\documentclass[1p]{elsarticle}

\usepackage[T1]{fontenc}
\usepackage[pdftex, svgnames, dvipsnames, table]{xcolor}
\usepackage[hyphens]{url}
\usepackage[hidelinks]{hyperref}
\usepackage{bbold}
\usepackage{amsmath}
\usepackage{amsfonts}
\usepackage{amssymb}
\usepackage{graphicx}
\usepackage{minted}
\usepackage{svg}
\usepackage{multirow}
\usepackage{booktabs}
\usepackage[para,online,flushleft]{threeparttable}
\usepackage{tablefootnote}
\usepackage{tabto}
\usepackage[toc,page]{appendix}
\newcommand*\xor{\oplus}

\usepackage{diagbox}
\usepackage{xspace}
\newcommand{\ie}{i.e.\@\xspace}  \newcommand{\eg}{e.g.\@\xspace}  \newcommand{\etal}{\textit{~et~al.\@}\xspace}

\usepackage{subcaption}
\usepackage{lipsum} 
\usepackage{bbding}

\usepackage{soul}
\newcommand{\changed}[1]{{#1}}

\usepackage{picins}

\usepackage{algorithm}
\usepackage{amsmath}
\usepackage{algpseudocode}

\newcommand{\DumpSeq}{D}
\newcommand{\Dump}{d}
\newcommand{\DumpSet}{D_s}

\newcommand{\LocSet}{L}

\newcommand{\TrueLoc}{l^\star}

\newcommand{\Score}{q}
\newcommand{\Ranking}{\pi}
\newcommand{\TrueRank}{\rho}

\newcommand{\GreedyPrune}{U}
\newcommand{\StatPrune}{V}

\newcommand{\StratInfo}{H}

\newcommand{\CandSet}{C}
\newcommand{\NumCand}{N}
\newcommand{\Recall}{R}

\newcommand{\InspectRule}{\kappa}

\usepackage{tikz}
\usetikzlibrary{arrows.meta, positioning, shapes.geometric} 
\journal{Computers \& Security}

\begin{document}

\begin{frontmatter}
\title{Statistical Effort Modelling\\ of Game Resource Localisation Attacks}

\author[1]{Alessandro Sanna\fnref{fn1}}
\ead{alessandro.sanna96@unica.it}
\author[2]{Waldo Verstraete\fnref{fn1}}
\ead{waldo.verstraete@ugent.be}
\author[1]{Leonardo Regano\corref{cor1}}
\ead{leonardo.regano@unica.it}
\author[1]{Davide Maiorca}
\ead{davide.maiorca@unica.it}
\author[2]{Bjorn De Sutter}
\ead{bjorn.desutter@ugent.be}
\affiliation[1]
{
organization={Dipartimento di Ingegneria Elettrica e Elettronica, Universit\`a di Cagliari},
addressline={Via Marengo 3},
postcode={09123},
city={Cagliari},
country={Italy}
}
\affiliation[2]
{
organization={Computer Systems Lab, Ghent University},
addressline={Technologiepark-Zwijnaarde 126},
postcode={9052},
city={Gent},
country={Belgium}
}
\cortext[cor1]{Corresponding author}
\fntext[fn1]{Alessandro Sanna and Waldo Verstraete share dual first authorship.}

\begin{abstract}
Evidence on the effectiveness of Man-At-The-End (MATE) software protections, such as code obfuscation, has mainly come from limited empirical research. Recently, however, an automatable method was proposed to obtain statistical models of the required effort to attack (protected) software. The proposed method was sketched for a number of attack strategies but not instantiated, evaluated, or validated for those that require human interaction with the attacked software. 

In this paper, we present a full instantiation of the method to obtain statistical effort models for game resource localisation attacks, which represent a major step towards creating game cheats, a prime example of MATE attacks. We discuss in detail all relevant aspects of our instantiation and the results obtained for two game use cases. Our results confirm the feasibility of the proposed method and its utility for decision support for users of software protection tools. These results open up a new avenue for obtaining models of the impact of software protections on reverse engineering attacks, which will scale much better than empirical research involving human participants. 

\end{abstract}

\end{frontmatter}

\section{Introduction}
\label{sec:intro}

The Man-At-The-End (MATE) attack model concerns adversaries with full control over machines seeking unauthorised access to software assets (e.g., secret keys, licence managers, game logic) through execution, inspection, reverse engineering, and manipulation of the software using static and dynamic analysis techniques~\cite{collbergbook}. Software Protections (SPs) such as obfuscations aim to safeguard the assets' confidentiality and integrity. Since full protection is unachievable in this attack model, SPs typically aim to reduce attack return on investment to unviable levels~\cite{collbergbook}.

Evaluating MATE SP strength lacks standardised methodologies~\cite{dagstuhl,desutter2024evaluation}, partly due to diverse attack techniques and attacker goals~\cite{survey2016}, and because of the fuzzy nature of SP, which tries to delay attacks rather than completely prevent them~\cite{collbergbook}.

Faingnaert et al.\ recently \changed{pitched} a simulation-based method to obtain statistical models of attack effort~\cite{checkmate24}, potentially enabling evaluation against real-world attacks rather than artificial metrics~\cite{desutter2024evaluation}, integration into risk management/decision support~\cite{Basile23}, and better scalability than expensive empirical studies. \changed{Their method is based on a four-line meta-model of which they claim that it can be instantiated to model and simulate a wide range of concrete, probabilistic reverse engineering attacks. They instantiated and validated a model for a fully automated cryptographic key localization attack~\cite{k-hunt++}, but only provided a brief sketch of how the meta-model might be instantiated to simulate attacks that include probabilistic manual human activities. They did not implement, present, or evaluate any such instantiation, thus offering no empirical validation of their claims regarding partially manual, probabilistic reverse engineering attacks.}   

\changed{In this paper, we fill this gap by presenting and evaluating the first instantiation of their meta-model and simulation method for such attacks. We do so for the case study of game resource localisation attacks, the initial step in many game cheating attacks~\cite{game_hacking} and a key reverse engineering attack category~\cite{survey2016}. Attackers locate memory addresses storing player resources (gold, coins, health) using memory scanning tools like CheatEngine~\cite{cheatengine}, then tamper with values. We deployed scanning tools and strategies on games with various data obfuscations (e.g., XOR-masking~\cite{xor}) and evaluated those obfuscations' impact on attack effort.}
\changed{Faingnaert et al.~\cite{checkmate24} already mentioned this use case and briefly sketched how to simulate such attacks, but as stated above they did not present, evaluate, or validate any concrete instantiation.  In other words, prior work provided the methodological skeleton, but not an end-to-end, empirically evaluated methodology for this attack class.}
\changed{To fill this gap, this paper offers} the following major contributions:
\begin{itemize}\setlength{\itemsep}{2pt}
    \item \changed{We formalize the game resource localisation attack.} 
    \item On two game use cases, we instantiate the method of Faingnaert et al.\ with various defences and attack strategies.
    \item We demonstrate the feasibility and utility of the method to obtain statistical models of the expected attack effort with minimal defender effort, including for probabilistic attack steps that require dynamic interaction with the software.
    \item We demonstrate how this provides useful, tailored decision support inputs and insights for defenders who have to select and reason about SPs.
    \item \changed{We present a small human experiment that validates obtained models.} 
    \item We open source all of our models and code for others to work on.  
\end{itemize}

With this work, we provide the first statistical models capturing SP impact on actual attack effort, complementing scarce and non-convincing empirical evidence typically provided in the SP literature~\cite{desutter2024evaluation}.

Section~\ref{sec:background} presents background on obfuscations, localisation strategies, and Faingnaert et al.'s method~\cite{checkmate24}. \changed{Section~\ref{sec:formalisation} formalizes the attack strategies, before} Section~\ref{sec:simulation} discusses our simulation approach. Section~\ref{sec:eval} presents the experimental evaluation on two games, \changed{and Section~\ref{sec:empirical-validation} presents an empirical validation by means of a human experiment.} Sections~\ref{sec:discussion}--\ref{sec:conclusions} cover limitations, related work, and conclusions.

 \section{Background}
\label{sec:background}

\subsection{Data Obfuscation}

To hide program values in memory, data obfuscation can be used. Collberg et al.\ define \emph{storage transformations} as obfuscations using non-conventional memory layouts~\cite{taxonomy}. For example, variable splitting divides a 32-bit integer into four 8-bit variables spread across memory.

\emph{Encoding transformations} use unnatural encodings for common data types. Instead of storing actual values, encoded forms are stored, and, if possible, computations operate on these encoded values (minimising decoding/re-encoding). For instance, Boolean values can be encoded as integers where even/odd numbers represent True/False. Table~\ref{tab:simple_encoding} lists encoding techniques used in practice, including by malware authors~\cite{Cannell2013}. Encodings can be static or dynamic. Static encodings do not change over time; XOR-masking with a constant mask value $p$ is static and primarily hides hard-coded constants in executables. To hinder resource data localisation attacks, defenders may use dynamic encodings, where the encoding changes during execution, such as varying the XOR-mask value $p$.

\begin{table}[t]
\begin{center}
\caption{Encoding transformations} \label{tab:simple_encoding}
\resizebox{\columnwidth}{!}{
\begin{tabular}{l l l}
\textbf{Method} & \textbf{Original} & \textbf{Encoded} \\ \hline \hline
 XOR masking & $x$ & $x \oplus p$ \\ \hline
ROT13 & $x \in \{ $`A', ..., `Z'\} & $chr(mod(asc(x)-65+13,65)+65)$ \\ \hline
 BASE64 & $x$ &  BASE64($x$) as described in RFC4648 \cite{rfc4648} \\ \hline
\end{tabular}
}
\end{center}
\end{table}

\emph{Residue number coding (RNC)}~\cite{RNCorig,RNC} combines storage and encoding transformations. With $m_1,m_2,...,m_u \in \mathbb{Z}$ where $gcd(m_i,m_j) = 1$ if $i \not = j$,  and $n = m_1 \cdot m_2 \cdot ...\cdot m_u$, any value $x \in [0,n-1]$ can be encoded as an array $[x\mod m_1, x\mod m_2, ... , x\mod m_u]$, of which the elements can be distributed throughout memory. The original value is recoverable using Euclid's extended algorithm for computing the GCD~\cite{Garner59}.

\subsection{Game Resource Hack Strategies}
\label{sec:background:subsec:hackstrategies}

In game cheats, resource hacks modify game behaviour/state regarding player resources like lives, coins, health, ammunition, gold, and energy. Resource hacks on (unprotected) games can be implemented in four steps~\cite{game_hacking}:
\begin{enumerate}
\item \emph{Resource data location pruning.} 
Tools like CheatEngine~\cite{cheatengine} or scanmem~\cite{scanmem} identify
memory locations holding resource data. The cheater plays, halts the process, and uses their tool to scan\footnote{\changed{Depending on the tool being used and how the pruning logic is implemented, an attacker might technically collect memory dumps that are later scanned, or directly scan the game's current address space itself. Any attack strategy relying on direct scans can trivially be implemented with dumps as well, but vice versa, that might not be the case. For example, when a pruning logic checks whether certain invariants hold between the consecutive values in memory locations (e.g., they evolve in the same manner as the on-screen values, or they are XOR-ed versions thereof with the same mask), more information is needed than is available in the current memory space of the game.}}
memory for values matching on-screen resources. This typically yields multiple locations (some
coincidental). To narrow candidates, they iteratively repeat gameplay and scanning after the value changes.
The number of iterations can vary, and pruning often converges to a few addresses rather than
one, so they may stop once the count stabilises even if more than one remains.

\item \emph{Resource data location validation.} To validate the correct location, attackers modify stored values at the remaining candidate location(s) and check if on-screen values change accordingly. With multiple candidates, changing values individually can further prune the search space. However, incorrect changes risk crashes or state changes preventing attack continuation, so this cannot be brute-forced on many candidates. Attackers hence typically begin this step only if the number of candidates has converged sufficiently low in the previous step.

\item \emph{Resource code location.} After attaching a debugger and setting a watchpoint on the found location, code fragments accessing/updating that data are easily identified.

\item \emph{Code tampering.} Found code fragments are reverse-engineered and patched using tools like Ghidra~\cite{ghidra} or IDA Pro~\cite{idapro} to persistently alter game behaviour, such as preventing ammunition reduction. These fragments already have pointers to resource data that patched code can reuse.
\end{enumerate}

These steps only work on unprotected programmes lacking memory scanning protections (step 1), anti-debugging measures~\cite{selfdebugging,circulardebugging} (step 3), or anti-tampering protections detecting code changes~\cite{viticchie2016reactive} (step 4). In games with anti-tampering, code fragments accessing resource data cannot be directly edited. However, step 1 remains relevant as it enables more advanced cheat techniques, like pointer chains for out-of-process attacks~\cite{game_hacking}. This work focuses exclusively on this resource data localisation step.
Through encoding transformations, possibly with storage transformations, defenders can make memory-stored resource data differ from on-screen values, complicating pruning and validation in early attack steps. Simple searches for on-screen values fail; attackers must then instead correlate changes in known on-screen bytes with changes in unknown memory bytes during gameplay.

In this paper, we model the process of finding these correlations, i.e., the effort an attacker is required to invest to do so, and the quality of the outcome (i.e., the number of candidate locations returned by step 1). Interestingly, the outcome quality of step 1 also affects the total effort, because the number of candidates from step 1 directly impacts the step 2 effort.

\changed{Importantly, step 1 is a probabilistic rather than deterministic. One reason is that attackers lack the required precision and motivation to play the gameplay deterministically. Often they only care about certain aspects of the gameplay in between scans, such as shooting exactly once, collecting exactly one coin, or moving in a specific direction; without caring about the other aspects, such as the precise location in which they do so or the exact time at which they do so. Moreover, the game itself and its internal state, as reflected in the values in a scan, might be non-deterministic, such was when the state depends on other players' actions, or when the representation of the internal state in memory is randomized, e.g., through security measures such as address space layout randomization (ASLR)~\cite{ASLR}. In short, every time attackers execute a location pruning strategy, they can get a different outcome. 
}

\subsection{Statistical Attack Effort Estimation}
\label{sec:metamodel}
Faingnaert et al.\ proposed to model reverse engineering effort---for any type of reverse engineering, not only for game cheats---by instantiating the following meta-model~\cite{checkmate24}: 
\begin{minted}[fontsize=\small,frame=none,linenos=false]{c}
  while (true)
    activity, artefacts = decision(knowledge, totalEffort)
    knowledge, activityEffort = execute(activity, artefacts, knowledge)
    totalEffort += activityEffort
\end{minted}

In this meta-model pseudo-code, the functions \mintinline{c}{decision} and \mintinline{c}{execute} implement the attacker's decision logic and the execution of individual attack steps, respectively. Each loop iteration corresponds to one attack step. The \mintinline{c}{decision} function considers gained knowledge and invested effort, returning the next attack step and the artefacts/knowledge on which to execute that step. The \mintinline{c}{execute} function performs it, returning updated knowledge and estimated effort, including halting when successful or when excessive effort leads to abandonment.

When executing a concrete attack strategy, attackers instantiate the meta-model with concrete \mintinline{c}{decision} and \mintinline{c}{execute} functions operating on targeted software artefacts and exploiting domain knowledge. Each instantiation corresponds to an attack strategy; each execution corresponds to a sequence of decisions and attack steps.

Both functions can be probabilistic. For example, attackers may randomly choose among equally-prioritised activities or artefacts as part of their decision making. Execution effort can also be probabilistic, such as when manually analysing the functionality of code fragments. Thus, the effort required for attack strategy $A$
on programme $P$ version $v$ (e.g., one variation of different protection combinations) should be modelled with statistical distribution $\phi_A^{P_v}$.
Since attacks can also fail probabilistically---ending without obtaining targeted information or achieving only partial outcomes---success rates should similarly be modelled with a distribution $\sigma_A^{P_v}$.

In the attacks discussed in the previous section, attackers execute strategies while playing: they decide how to alter resources through gameplay and when to scan memory with their tools. 
Probabilistic aspects include timing and choice of gaming actions and scans, plus the similarity between resource values and other values in the address space at scan times. Search space pruning per scan depends on matching value patterns in memory, which will vary between executions if the game is not fully deterministic.

To estimate the expected effort for any attack strategy, not just game resource hacks, the core idea behind the method is to script the strategy by implementing the \mintinline{c}{decision} and \mintinline{c}{execute} functions in scripts that simulate those two aspects of the strategy, using random number generators where probabilistic processes take place and statistical models of the expected effort are required for each individual attack step. Those models can be based, e.g., on software complexity metrics~\cite{halstead1977elements,mccabe1976complexity}, or on actual running times of analysis tools. We refer to Faingnaert et al.~\cite{checkmate24} for a more extensive discussion of this aspect.  

Each time the attack is simulated by executing the simulation scripts, a total effort estimate and some outcome will be obtained. Multiple simulations, each potentially yielding a different effort estimate and different outcome, will hence produce a statistical distribution $\hat{\phi}_A^{P_v}$ that approximates $\phi_A^{P_v}$ and from which the expected total attack effort can be estimated. Similarly, a success rate estimation $\hat{\sigma}_A^{P_v}$ approximates $\sigma_A^{P_v}$.

Crucially, ground-truth information about deployed protections and attacked assets is assumed available during simulation. Unlike real attacks where attackers lack ground truth, simulations are executed by modellers, i.e., defenders or researchers, to obtain statistical effort models. In the simulation scripts, the ground truth is not used for modelling attacker decision-making or code analysis results. Instead, it is only used for probabilistically estimating modelled attack step effort and for determining goal achievement.

Faingnaert et al.\ observed that when simulated attack steps require programme interaction, modellers cannot be asked to repeatedly perform those interactions for each simulation~\cite{checkmate24}. For example, in the case of a game resource localisation attack, the modeller cannot be asked to play the game over and over again for repeated simulations of the attack steps. 
Instead, Faingnaert et al.\ proposed a two-phase approach: modellers play the game once and collect a large set of memory dumps (more than a cheater would take), then each attack step simulation samples a random dump subset on which to perform scans.

We validated this method, obtaining the first empirically determined statistical effort models for MATE attacks on protected code. Section~\ref{sec:simulation} presents how the overall method is adopted for game resource localisation attacks; Section~\ref{sec:eval} presents our instantiation and results on two game use cases.

 \changed{
\section{Game Resource Localisation Formalisation} 
\label{sec:formalisation}
Before discussing how to use simulation to estimate the required effort of game resource localisation attacks, this section formalises such attacks. More precisely,  this section formalises the first attack step presented in Section~\ref{sec:background:subsec:hackstrategies}. \changed{We will denote the deployed strategy for this step $A_1$.}

\subsection{Attack process}
The attacks we consider can be formalised, for a fixed protected program version
$P_v$ and attack strategy $A_1$, as a discrete-time stochastic state-evolution process.
In a concrete attack execution, the attacker observes a sequence $D=(d_1,\dots,d_n)$
of memory dumps,\footnote{From here on, we use the term "dump" to refer to a stored snapshot of the game's memory, and the term "scan" to refer to the act of iterating over the data in one or more such dumps.} from which their knowledge state and invested effort evolve over the
course of the attack.

We distinguish between acquisition times, at which dumps are taken, and scans, at which the attacker processes the newly collected dumps and updates the current
knowledge state. Let
$i_0=0, 1\le i_1<i_2<\dots<i_m = n$
denote boundary indices in the dump sequence associated with the successive scans. Here, we define $i_0=0$ by convention, to denote the point immediately before the
first dump is collected, while $i_1,\dots,i_m$ denote the dump indices reached after
each successive scan. Then, for each $s\in\{1,\dots,m\}$, scan $s$
inspects the data in dumps $d_{i_{s-1}+1},\dots,d_{i_s}$.

\subsection{Attacker state}
Let $L$ denote the set of all candidate memory locations considered by the attacker, and
let $\ell^\star\in L$ denote the ground-truth target location. For scan $s$,
let $X_s=(K_s,E_s)$ denote the attacker state after incorporating all dumps up to $d_{i_s}$, where $K_s$ is
the attacker's knowledge state and $E_s$ the cumulative effort invested so far. The
attack execution therefore induces the state sequence $(X_1,\dots,X_m)$.

\paragraph{Greedy strategies}
For greedy strategies, the knowledge state consists of the surviving candidate set
together with any auxiliary information retained for those candidates from previous
dumps. We write $K_s=(C_s,H_s)$ where $C_s\subseteq L$ is the surviving candidate set after scan $s$, and $H_s$ denotes any strategy-specific information retained for those candidates. This
auxiliary information may encode, for example, parameter hypotheses, intermediate
consistency relations, or any other data needed to evaluate subsequent dump batches.
Initially, before any scans have been performed, all candidate locations are still
possible, so $C_0=L$.

At scan $s$, let $g_s(\ell,h)\in\{0,1\}$ 
denote whether a surviving location $\ell\in C_{s-1}$, together with its currently
retained auxiliary information $h\in H_{s-1}$, remains consistent with the pruning logic
when the newly collected dumps are taken into account in that scan. The new scan 
therefore updates both the surviving candidate set and the associated auxiliary
information. 
Abstractly, one may write
$(C_s,H_s)=U_s(C_{s-1},H_{s-1},d_{i_{s-1}+1},\dots,d_{i_s})$, 
where $U_s$ denotes the strategy-specific update rule for scan $s$. Given the
previous greedy-state components $(C_{s-1},H_{s-1})$ and the newly incorporated scans
$d_{i_{s-1}+1},\dots,d_{i_s}$, it returns the updated components $(C_s,H_s)$. In other words,
$U_s$ abstracts the concrete pruning logic of the attack, including both the elimination
of inconsistent candidates and the update of any retained per-candidate information.

The attacker hence does not need to re-scan the full location set $L$ after every batch of
dumps; instead, each new scan is applied only to the surviving candidates together with
the information retained for them from previous scans. The corresponding
localisation quality at scan $s$ is characterised by
\[
N_s=|C_s|,
\qquad
R_s=
\begin{cases}
1 & \text{if } \ell^\star\in C_s,\\
0 & \text{otherwise.}
\end{cases}
\]
where $N_s$ is the number of remaining candidates and $R_s$ indicates whether the
ground-truth location is still retained.

\paragraph{Statistical strategies}
For statistical strategies, no candidates are discarded. Instead, the knowledge state
consists of the current candidate scores together with any auxiliary information retained
from previous dumps that is needed to update those scores. We write
$K_s=(q_s,H_s)$, where $q_s(\ell)$ denotes the score assigned to candidate location $\ell\in L$ after
scan $s$, and $H_s$ denotes any strategy-specific information retained for
score updating. This auxiliary information may encode, for example, previously observed
encoded values, intermediate consistency relations, or other per-candidate summaries
needed to compare newly incorporated scans with earlier ones.

Abstractly, one may write $(q_s,H_s)=V_s(q_{s-1},H_{s-1},d_{i_{s-1}+1},\dots,d_{i_s})$, 
where $V_s$ denotes the strategy-specific score-update rule for scan $s$. Given
the previous score state $q_{s-1}$, the previously retained auxiliary information
$H_{s-1}$, and the newly incorporated scans $d_{i_{s-1}+1},\dots,d_{i_s}$, it returns
the updated score state $q_s$ and updated auxiliary information $H_s$. Thus, $V_s$
abstracts the concrete statistical pruning logic of the attack, including both the
accumulation of score evidence and the update of any retained per-candidate information
required for later scans.

The resulting scores induce a ranking $\pi_s$ over $L$. In that case, the primary
localisation quality is characterised by the rank of the ground-truth location,
$\rho_s=\mathrm{rank}_{\pi_s}(\ell^\star)$, 
which corresponds to the number of ranked locations the attacker would have to inspect,
in order, before reaching the ground-truth location. Following the discussion in
Section~\ref{subsec:result_gathering}, recall for statistical strategies can
additionally be defined relative to an assumed inspection criterion $\kappa$, which
determines how far down the ranking the attacker inspects. If $\kappa(\pi_s)$ denotes
the number of top-ranked locations inspected, then the corresponding recall indicator is
\[
R_s^\kappa=
\begin{cases}
1 & \text{if } \rho_s\le \kappa(\pi_s),\\
0 & \text{otherwise.}
\end{cases}
\]

\paragraph{Effort model}
Let $b_s=i_s-i_{s-1}$ denote the number of dumps inspected in scan $s$. Let $a_s$ denote the
effort invested between scans $s-1$ and $s$, including both the collection of
those new dumps and their inspection. Then the cumulative effort after scan $s$
is $E_s=\sum_{u=1}^{s} a_u$.

The model leaves $a_s$ generic, but it may naturally depend both on the number of newly
incorporated dumps and on the current knowledge state. In the example formulas below,
$\alpha$ denotes the average effort required to collect one additional dump, $\beta$ the
average effort required to apply the pruning logic on one candidate location for one newly incorporated dump, and $\gamma$ the average effort required to inspect one resulting candidate or ranked
location.

In particular, for greedy strategies, a natural assumption is that the scanning effort
decreases as the surviving candidate set shrinks. For example, the effort can be modeled as 
\[
a_s=\alpha\, b_s+\beta\, b_s\,N_{s-1}+\gamma\,N_s,
\]
where the first term captures the effort of collecting the new scans, the second the
effort of applying the pruning logic to the previously surviving candidates, and the
third the effort of inspecting the updated candidate set.

Analogously, for statistical strategies, the effort can be modeled as
\[
a_s=\alpha\, b_s+\beta\, b_s\,|L|+\gamma\,\kappa(\pi_s),
\]
where the first term again captures the effort of collecting the new scans, the second
the effort of updating candidate scores using the new batch, and the third the effort of
inspecting the top-ranked candidates selected under the inspection criterion. In
evaluation terms, the rank $\rho_s$ represents the realised number of candidates that
would need to be inspected before reaching the ground-truth location, and thus plays the
same downstream-effort role for statistical strategies as $N_s$ does for greedy ones.

\subsection{Stopping Criterion}
When an attacker has observed some sequence $D$ that ends with a dump $d_n$ and scan $m$, this implies that their stopping criterion was triggered on the basis of knowledge state $X_m$. Formally, let $\Gamma$ denote a stopping criterion
that, given the current attacker state, decides whether the attacker would continue or
stop. For a sequence of acquisition and scans to occur in an attack, the following condition needs to be met:
\[
\Gamma(X_s) =
\begin{cases}
1 & \text{if } s=m,\\
0 & \text{otherwise.}
\end{cases}
\]

For greedy strategies, $\Gamma$ may, for example, model criteria based on the number of remaining candidates, such as stopping when that number has become sufficiently low or stopping when it has (seemingly) converged to a stable number for a number of dumps. The latter obviously requires the attacker's knowledge state $K_s$ to comprise more than the surviving candidate set $C_s$. It is, among others, for that reason that we earlier included $H_s$ as the strategy-specific part of the knowledge state $K_s$.

For statistical strategies, $\Gamma$ may instead model criteria based on the current ranking or score profile, or simply a prescribed scan budget.

\paragraph{Induced distributions}
Over the possible executions of strategy $A_1$ on the protected program version $P_v$, these
random variables induce the actual effort distribution $\phi^{P_v}_{A_1}$ and the success distribution $\theta^{P_v}_{A_1}$ introduced in the previous section. Indeed, because the attack execution is probabilistic as discussed in earlier sections, each execution can yield a different sequence $D$ of memory dumps, and hence a different total effort estimate $E_m$, and different localisation outcome qualities $N_s$ (for greedy strategies) and $\pi_m$ and $R_m^\kappa$ (for statistical strategies). The different outcomes $E_m$ form the effort distribution $\phi^{P_v}_{A_1}$, and depending on which form of strategy $A_1$ is, one of the three outcomes $N_s$, $\pi_m$, and $R_m^\kappa$ yields the distribution $\theta^{P_v}_{A_1}$.

 }
\section{Resource Localisation Attack Simulation}
\label{sec:simulation}

To investigate and demonstrate the utility of the method, we select the use case of game resource localisation, i.e., the first attack step presented in Section~\ref{sec:background:subsec:hackstrategies}. \changed{In particular, we turn the generic meta-model of Faingnaert\etal~\cite{checkmate24} into an executable methodology for this use case by making explicit the attack-specific design choices that their framework leaves open: how dumps are collected, how valid attack traces are sampled from them, how pruning is simulated, and how success and effort are aggregated into distributions useful for defenders.}

We assume the game developer is the defender who must decide which software protections to deploy to mitigate potential resource hacks. The defender has source code access and can produce protected versions of the game, by implementing protections manually or by invoking protection tools such as Tigress~\cite{tigress2025}. It is for such protected versions that the defender seeks a statistical model of the required attack effort for a set of known attack strategies. In other words, the defender is the attack \emph{modeller}. With source code access, the modeller can instrument the game code to make it output the location of the resource in memory. The modeller hence knows the ground truth about the resource location when they play the game to collect memory dumps. 

As for the attack strategies to be modelled, we focus on attackers performing the resource data location pruning step (step 1 in Section~\ref{sec:background:subsec:hackstrategies}) until the number of candidate locations has converged to some low number, after which they will proceed to step 2. For such strategies, the defender may want to estimate the following statistical distributions:
\begin{itemize}
\item After how many memory scans is the pruning expected to converge? This relates directly to the expected effort $\phi_{A_1}^{P_v}$ that an attacker needs to invest in step 1.
\item To how many candidate locations does that pruning converge? This relates directly to the expected effort $\phi_{A_2}^{P_v}$ that an attacker will need to invest in step 2, namely to check those locations by writing data into them. 
\item What is the expected recall of the pruning? In other words, what expectation can the attacker have that the actual resource location is included in the remaining candidate locations? This relates directly to the success rate $\sigma_{A}^{P_v}$ of step 1. 
\end{itemize}

Alternatively, the defender may want to estimate the following distributions, which convey similar information: 
\begin{itemize}
\item After using $n$ memory scans for pruning, what is the expected number of remaining candidate locations? This relates directly to the expected effort $\phi_{A_{2,n}}^{P_v}$ that the attacker would still have to invest in step 2 after having invested the effort for performing $n$ scans in step 1.
\item  After using $n$ memory scans for pruning, what is the expected recall? This relates directly to the success rate $\sigma_{A_n}^{P_v}$ when step 1 is ended after $n$ dumps.
\end{itemize}

\begin{figure}[t]
    \centering
    \includesvg[width=\linewidth]{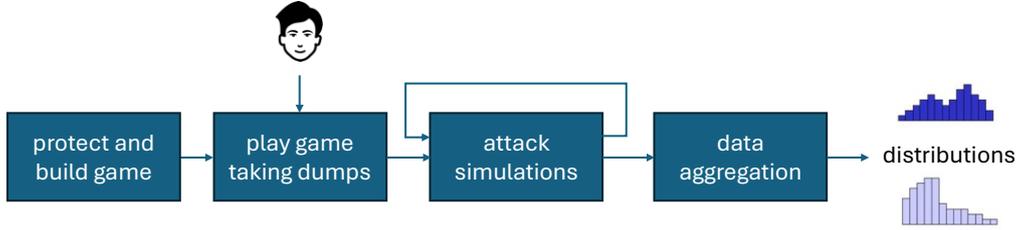}
    \caption{The simulation method of the game resource localisation attack. The defender needs to invest time once in playing the game while dumps are taken. Based on that one playing session, multiple attack executions can then be simulated, of which the results can be aggregated into statistical distributions that model the required attack effort.}
    \label{fig:flow}
\end{figure}

To obtain the estimates $\hat{\phi}_{A_1}^{P_v}$, $\hat{\phi}_{A_2}^{P_v}$, $\hat{\sigma}_{A}^{P_v}$, $\hat{\phi}_{A_{2,n}}^{P_v}$, and $\hat{\sigma}_{A_n}^{P_v}$, the defender/modeller will perform the next steps of the method, of which the overall flow is depicted in Figure~\ref{fig:flow}. In summary, on each considered protected version of the game, the defender will
\begin{enumerate}
    \item play the game taking many memory dumps; 
    \item run many attack simulations for each of the considered attack strategies;
    \item aggregate the results obtained from all simulations to obtain statistical distributions of the required effort. 
\end{enumerate}

Only in one stage of the defender's modelling work, which only has to be executed once per considered protected version, is there hence a need for human interaction. \changed{Figure~\ref{fig:attacksimul} provides a UML-like formalisation of the last two steps in which attacks are simulated and results are aggregated.}

\begin{figure}[t]
    \centering
\begin{tikzpicture}[
    node distance=8mm,
    >=Latex,
    line/.style={-Latex, thick},
    block/.style={
        rectangle,
        draw,
        rounded corners,
        align=center,
        minimum width=4.7cm,
        minimum height=9mm
    },
    decision/.style={
        diamond,
        draw,
        align=center,
        aspect=2.2,
        inner sep=1pt
    },
    startstop/.style={
        ellipse,
        draw,
        align=center,
        minimum width=2.2cm,
        minimum height=8mm
    }
]

\node[startstop] (start) {Start};
\node[block, below=of start] (init) {Capture dump sequence $\DumpSeq$};
\node[block, below=of init] (select) {Store ground-truth location $\TrueLoc$};
\node[block, below=of select] (obs) {Sample dump sequence $\DumpSet$ from $\DumpSeq$};
\node[block, below=of obs] (update) {Perform attack simulation on $\DumpSet$ (see Algorithm~\ref{alg:localisation-attack})};
\node[block, right=of update] (eff) {Update distributions};
\node[decision, above=of eff] (stop) {More possible\\samplings\\$\DumpSet$ over $\DumpSeq$};
\node[block, above=of stop] (out) {Return distributions};
\node[startstop, above=of out] (end) {End};

\draw[line] (start) -- (init);
\draw[line] (init) -- (select);
\draw[line] (select) -- (obs);
\draw[line] (obs) -- (update);
\draw[line] (update) -- (eff);
\draw[line] (eff) -- (stop);
\draw[line] (stop) -- node[right] {no} (out);
\draw[line] (out) -- (end);
\draw[line] (stop.west) -| ++(-1, 0, 0) |- node[right] {yes} (obs.east);

\end{tikzpicture}     \caption{\changed{Logic of the attack simulations process.}}
    \label{fig:attacksimul}
\end{figure}

\subsection{Step 1: Playing the game and collecting dumps}
\label{sec:simulation:step1}
After building a protected, instrumented game version, the defender plays it and uses memory analysis tools (CheatEngine, scanmem, debuggers) to capture memory dumps between actions. Each dump is also tagged with the game state (on-screen resource values), recent actions, and ground-truth data about resource storage locations.

Real attacks are non-deterministic, in the sense that scan timing, game actions, randomised events, and multiplayer activities will vary across attacks and attackers. Therefore, each real attack execution includes a different scan sequence $D_i$ with varying effort and outcomes. To cover sufficient variations of real attacks, the defender captures a much larger dump sequence $D$ than the number of scans a real attacker would typically use.

For the simulation method to accurately approximate the real attack strategy $S$'s distributions $\phi_A^{P_v}$ and $\sigma_A^{P_v}$ with $\hat{\phi}_A^{P_v}$ and $\hat{\sigma}_A^{P_v}$, $D$ must contain sufficiently diverse subsets $D_s$ that represent potential real attack sequences $D_i$. Formally: $\forall D_i: 
\exists D_s : ( attack(D_s) \equiv attack(D_i) ) \wedge (D_s \subset D$), where $attack(a) \equiv attack(b)$ means that sequences $a$ and $b$  yield the same attack decisions and outcomes.

No universal threshold exists for dump quantity or timing relative to gameplay. It is up to the defender to determine an appropriate set $D$ based on their knowledge of the game, deployed protections, implementation parameters, attack strategies, and subsequent analysis logic (see Section~\ref{sec:simul:subsec:dumpselection}).

\subsection{Step 2: Simulations of an Attack}
\label{sec:simul}

Next, the defender will run a script multiple times, each time simulating an attack strategy execution. This script consists of three pieces of logic that can easily be implemented in scripting languages such as Python. 

\subsubsection{Dump selection logic\label{sec:simul:subsec:dumpselection}} 
From the sequence $D$ of all memory dumps, a subsequence $D_s$ is selected that could correspond to a sequence $D_i$ of memory scans invoked during one actual execution of the attack. 
As a simple example, consider a localisation strategy in which an attacker would take at most one memory dump per different number of coins shown on screen. Then the sequence $D_s$ should not contain two dumps annotated with the same number of coins. 

Similarly to how the defender is responsible for selecting the dumps that form sequence $D$ in step 1, the defender is now responsible for implementing the code that will randomly select a relevant subset $D_s$ for each simulation. 
This selection defines the simulated attack strategy to a large degree. Specifically, it defines (part of) the \mintinline{c}{decision} function from the meta-model. If the chosen sets $D_s$ do not reflect real practice, the end result will be representative of attacks that make similar selections $D_i$, not real-world ones.

Obviously, steps 1 and 2 need to be co-designed by the modeller, such that the way the total sequence $D$ is assembled and the way subsequences $D_s$ are selected from it together model the envisioned attack strategy as accurately as desired. The defender has quite some leeway in this regard. For example, instead of picking subsequences $D_s$ randomly for each individual simulation of an attack strategy, the top-level simulation script could be adapted to exhaustively perform simulations for all subsequences $D_s$ of $D$ that meet certain criteria, without requiring randomised selection logic. 

In summary, modellers can freely choose the method used to obtain the sequences $D_s$ to simulate the localisation attack strategy. The statistical distributions $\hat{\phi}_A^{P_v}$ of the required attack effort and $\hat{\sigma}_A^{P_v}$ of the success ratio obtained from the simulation will then model attackers who apply the very same method to choose corresponding scan sequences $D_i$. Hence, dump selection defines the simulated attack strategy, rather than approximating it.

\subsubsection{Location pruning logic}
\label{sec:simulation:subsec:location-pruning}
On the dumps $D_s$, the logic is simulated to show how the attacker would try to prune the search space of possible resource locations during their own scans $D_i$. 
For example, consider attackers that would search for the memory addresses that hold, in each scan in $D_i$, the exact value shown on screen at the time of the scan. Such attackers would then discard locations for which certain scans do not contain the value shown on screen. To simulate such attackers, the defender should execute that exact same pruning logic on the dumps in $D_s$.

Importantly, the location pruning logic used in the simulation does not include the stop criterion that the attacker might use. For example, it neglects that an attacker might stop the pruning after the number of locations that are still candidate matches has been reduced to some low number other than one.  Instead, the defender will always execute the logic on the whole sequence $D_s$ to collect simulation results that cover a range of stop criteria.

Furthermore, attackers can use their pruning logic in two ways: greedily or statistically. In a \emph{greedy} strategy, the attacker discards memory locations irrevocably and as soon as they do not conform to the pruning logic. After each scan, the result is then the remaining number of candidate locations. In a \emph{statistical} strategy, no discarding of locations takes place. Instead, the attacker keeps track of statistics computed over all already considered scans. Specifically, for each location, they compute the fraction of the scans in which the location conforms to the pruning logic. Based on the gathered statistics, they then rank the memory locations. Thus, after each scan, they get an updated ranking of all fragments. 

For static encodings, a statistical attack does not provide any advantage over a greedy attack:
at a given timestamp, the set of remaining candidate locations in the greedy attack is identical to the set of memory locations with a perfect score of $1.0$ in the statistical attack.
For dynamic encodings ---where the parameters of the encoding change throughout the execution of the process, at times not necessarily known to the attacker--- that attacker might not be able to use certain pruning logics with a greedy attack if the logic builds on the assumption that the used encoding has static parameters. In that case, the attacker can use the statistical attack instead. 
If they can take scans sufficiently more frequently than the rate at which the parameters of the encoding vary, then the correct location should still end up higher in the ranking than other memory locations.

\changed{Algorithm~\ref{alg:localisation-attack} provides pseudo code for the simulation of an attack, using the formalization introduced in Section~\ref{sec:formalisation}. Lines 2--13 implement greedy strategies, and lines 15--27 implement statistical strategies.}

\begin{algorithm}[hbtp]
\caption{\changed{Simulation of a localisation attack}}
\label{alg:localisation-attack}
\begin{algorithmic}[1]
\Require dump sequence $\DumpSet$, greedy pruning rule $\GreedyPrune$ (or statistical pruning rule $\StatPrune$, inspection criterion $\InspectRule$), ground truth location $\TrueLoc$, candidate location set $\LocSet$
\Ensure number of candidate locations $\NumCand$ (or ground-truth rank $\TrueRank$), recall $\Recall$ (or recall indicator $\Recall^\InspectRule$)
\If{$\GreedyPrune$ is defined}
 \State $\CandSet_0 \gets \LocSet, \Recall_0 \gets 1, \StratInfo_0 \gets \emptyset$ \For{\(i \gets 1\) to \(|\DumpSet|\)}
\State Select next dump $\Dump_i$
  \State $(\CandSet_i, \StratInfo_i) \gets \GreedyPrune(\CandSet_{i-1}, \StratInfo_{i-1}, \Dump_i)$
  \State $\NumCand_i \gets |\CandSet_i|$
  \If {$\TrueLoc \in \CandSet_i$}
  \State $\Recall_i \gets 1$
  \Else
  \State $\Recall_i \gets 0$
  \EndIf
 \EndFor
 \State \Return $\NumCand, \Recall$
\ElsIf{$\StatPrune$ is defined}
 \State $\Score_0 \gets \LocSet, \TrueRank_0 \gets 1, \StratInfo_0 \gets \emptyset$ \For{\(i \gets 1\) to \(|\DumpSet|\)}
  \State Select next dump $\Dump_i$
  \State $(\Score_i, \StratInfo_i) \gets \StatPrune(\Score_{i-1}, \StratInfo_{i-1}, \Dump_i)$
  \State Compute ranking $\Ranking_i$ from $\Score_i$
  \State $\TrueRank_i \gets rank(\Ranking_i)(\TrueLoc)$
  \If {$\TrueRank_i \leq \InspectRule(\Ranking_i)$}
  \State $\Recall_i^\InspectRule \gets 1$
  \Else
  \State $\Recall_i^\InspectRule \gets 0$
  \EndIf
 \EndFor
 \State \Return $\TrueRank, \Recall^\InspectRule$
\Else
 \State \Return $\emptyset$
\EndIf

\end{algorithmic}
\end{algorithm}
 
\subsubsection{Result gathering} 
\label{subsec:result_gathering}
To compute the results of the simulation, the ground-truth information about the targeted location is used, as well as the conformance of all considered memory locations to the pruning logic in the sequence $D_s$ of dumps $d_0,d_1,...,d_n$. 
\\

For greedy attack strategies, the results include the following for each $d_i$:
\begin{itemize}
\item how many memory locations have not yet been pruned after considering dumps $d_0$ up to $d_i$, i.e., how many locations in memory are still being considered as potential candidates of where the resource data might be stored;
\item whether the ground-truth location is still included in those locations. 
\end{itemize}
For each dump $d_i$ in the sequence, these two pieces of information capture how effective the greedy pruning strategy has been up to that point, e.g., in terms of the precision and recall that an attacker would have achieved with the simulated attack up to dump $d_i$.  

For statistical strategies, for each dump $d_i$ the result consists solely of the rank of the ground-truth location in the list of all locations that were ordered based on the fraction of the considered dumps $d_0,...,d_i$ in which the location's value conformed to the pruning logic.
This corresponds to the number of memory locations an attacker would have to inspect before reaching the correct one when considering all locations in this sorted order. 

Since \emph{all} locations are ranked in statistical strategies, there is at first sight no notion of recall for such strategies. However, we can introduce that notion by introducing a criterion that the attacker can be assumed to use for determining how many locations in the list to inspect. For example, they might limit the inspection to locations with a score above a certain threshold, or to a certain number of locations, or to those locations ranked above the point where a first large drop in scores (i.e., a delta above a certain threshold) takes place. If the ground-truth location is not inspected given the assumed criterion, the attack up to that dump is considered not to have recalled the targeted location.

\subsection{Step 3: Data Aggregation}

The above three-part simulation is executed multiple times, each time for a different $D_s$. From all simulations, the data can then be aggregated into distributions $\hat{\phi}_{A_1}^{P_v}$, $\hat{\phi}_{A_2}^{P_v}$, $\hat{\sigma}_{A}^{P_v}$, $\phi_{A_{2,n}}^{P_v}$, and $\sigma_{A_n}^{P_v}$.
In this aggregation, the modeller has one final chance to ``tune'' the attack strategy for which they estimate distributions, namely by choosing the weights of all simulations, i.e., the weight of the outcomes obtained for each simulated $D_s$.

\subsection{\changed{Extension to Multiple Game Plays}}
\label{sec:extension_multiple_game_plays}
\changed{So far, we assumed, and put forward the advantage of, having to perform only one game play to collect the dump set $D$. However, for some games it might be impossible to cover all possibly relevant real-world dump sequences $D_i$ with one such set of dumps $D$. In such as case, the simulation method can readily be extended to include multiple game plays. During those plays, multiple dump sequences $D^{[1]},...,D^{[n]}$ will then be collected as described in Section~\ref{sec:simulation:step1}. The simulations described in Section~\ref{sec:simul} will then be executed separately for each sequence $D^{[i]}$, and the data aggregation discussed in the previous section will then consider all data obtained from all those simulations, without any additional complications.}

 \section{Instantiation and Experimental Evaluation}
\label{sec:eval}
We validated the feasibility of the simulation method for estimating the required attack effort distribution on two games, which we protected with a range of data obfuscations, and on which we simulated corresponding localisation strategies.
We group our experiments in two sets, one of which explores different attacks on various static encodings (Section~\ref{sec:evaluation_static}), while the other set of experiments explores multiple attacks on a single dynamic encoding (Section~\ref{sec:evaluation_dynamic}). Section~\ref{sec:results:subsec:duration} evaluates the run times of our experiments. 

Before discussing the results of those experiments, we describe, mostly at a conceptual level, the two games (Section~\ref{sec:games}), the used obfuscation strategies (Section~\ref{sec:obf-strategies}), the dump collection strategies we evaluated (Section~\ref{sec:atk-strategies:subsec:dump-collection}), and the resource localisation heuristics (Section~\ref{sec:atk-strategies}). The appendices provide concrete parameter values used in the experiments.

\subsection{Games}
\label{sec:games}
We evaluated our method on two open-source games: SuperTux\footnote{\url{https://www.supertux.org/} We forked from the main branch after commit \texttt{15dfac1}.} and AssaultCube\footnote{\url{https://assault.cubers.net/} We forked from the main branch after commit \texttt{13f0d8e}.}. 

In SuperTux, the player controls a penguin that collects coins by moving and interacting with on-screen objects. In the modelled attacks, the attacker seeks the memory location of the coin counter to allow cheating by altering the counter. This value is stored within a player status object alongside other state information. The attacker must therefore locate the coin value in the running process memory.

Interestingly, the coin count is stored twice in process memory. The value in the player status object  is the targeted value with which cheaters aim to tamper. Additionally, the heads-up display code retrieves this value and stores a duplicate in a buffer. No tampering with the duplicate is needed for a cheat, so it is not targeted. Still, it can distract the attacker. 
Moreover, the game also tracks the number of coins collected in the current level. Although this value differs from the total coin count, it evolves similarly and could therefore be mistaken for it by an attacker who does not know the used encoding. Across all our experiments, we obfuscate only the original total coin value.

In AssaultCube, the player explores a 3D first-person shooter environment, using various weapons to eliminate enemies. Only one weapon can be used at a time, and each has a limited magazine that can be fired and reloaded during gameplay. We model attacks in which the attacker attempts to locate the memory value representing the number of bullets, \eg to enable infinite ammunition. Bullet counts are stored in a player-status object that also contains other information (\eg player name, available weapons).

In AssaultCube, the targeted value of the bullet count is only stored once; no copy is stored in a buffer. However, a related value stored in memory tracks the number of fired shots. Unless the magazine gets reloaded with extra ammunition, this number evolves in the opposite direction of the bullet count. So also this value can distract attackers that do not know the used encodings.

In both games, the resource (coin count or bullet count for the active weapon) is continuously displayed on screen; thus, the attacker always knows the target value, but not how it is encoded or stored.

We opted for these open-source games because source code access allows us to implement various data obfuscations on the targeted resources, and because it makes obtaining ground-truth information trivial.
Being open-source games, AssaultCube and SuperTux are, of course, not games that cheaters and builders of cheats would reverse engineer starting from a binary executable. They are good representatives for other games that are the target of such cheats, however, because like most other games, the player's resources are stored in a dynamically allocated data structure that the cheater needs to locate in memory for the purpose of tampering with it~\cite{game_hacking}. The attack strategies we evaluate hence all apply to commercial games as well.

\subsection{Obfuscation Strategies}
\label{sec:obf-strategies}

We evaluated our method for estimating the resource locating attack effort on eight versions of the SuperTux and AssaultCube binaries. Each version employs a different encoding to store the protected resource amount (\ie coins in SuperTux or bullets in AssaultCube) in memory. Six versions use static encodings, where any secrets remain unchanged during execution, and constitute our first experimental set. The remaining two use dynamic encodings and form the second set. In the remainder of this paper, dynamic encodings are explicitly identified in their descriptions, otherwise encodings are static. For all encodings, the resource memory location remains constant over time. Parameters used for each protection are listed in Appendix~\ref{app:obf_parameters}.

\paragraph{Base Encoding}
With the \emph{base} encoding, we introduce no protection measures to hide the resource amount from the attacker.
It is stored unobfuscated in memory, i.e., in the default two's complement encoding commonly used for storing integers.

\paragraph{$+$-Encoding}
Using the $+$-encoding, we do not store the resource amount $A$ in memory directly, but rather obfuscate it by storing $A+O$ in memory, with $O$ being a secret. Ideally, the value of $O$ is such that all bytes of the stored value change. 

\paragraph{$\xor$-Encoding}
The $\xor$-encoding hides the resource amount $A$ using XOR masking.
The value $X$ in memory is obfuscated by computing the bitwise $\xor$ using a secret mask $M$: $X=A \oplus M$.

\paragraph{$+\xor$-Encoding}
The $+\xor$-encoding combines the $+$- and $\xor$-encodings by first adding a secret offset $O$ to the resource amount and then using the secret mask $M$ in a bitwise XOR-masking operation: $X = (A+O) \oplus M$.

\paragraph{$\xor +$-Encoding}
Similar to the $+\xor$-encoding, the $\xor +$-encoding combines XOR-masking with an offset.
It does so in the reverse order, however, and first applies a bitwise XOR mask before adding secret offset $O$ to the result: $X = (A \oplus M) + O$.

\paragraph{RNC encoding}
With \emph{RNC} encoding, the resource amount is stored as an array of $n$ values $X_1 = A\mod m_1, ... ,\ X_n = A\mod m_n$ where all $m_k$ are constant integer values co-prime with each other. In our implementation, $n=3$. 

\paragraph{Dynamic $\xor$-Encoding}
In the second experiment set, we use two dynamic variants of the $\xor$-encoding. The \emph{Update on Write} (UoW) version randomly changes mask $M$ with probability $p_{u,w}$ on every asset value update, while the Update on Read (UoR) version updates the mask with probability $p_{u,r}$ every time the game reads the asset value from memory.

\subsection{Collecting Dumps}
\label{sec:atk-strategies:subsec:dump-collection}
Attack strategies differ not only in the \emph{location pruning logic} used to narrow candidate locations, but also in how attackers decide when to perform the next pruning step (i.e., next scan) and which game actions to execute between steps. Consequently, when simulating attacks, the defender must carefully select both the dump collection logic and dump selection approach. For our experimental evaluation, we use two dump collection approaches, each tailored to different attack strategies. \changed{Each of these approaches rely on a single game play, so they do not involve the extension discussed in Section~\ref{sec:extension_multiple_game_plays}.}

For each dump collection approach, we use a script that launches a game and leverages scanmem to capture process memory at regular intervals\footnote{The interval between the memory dumps corresponds to in-game time. Taking the dump itself also takes some time, but this is not counted as part of the interval, because the game process is suspended.} during gameplay. The approaches differ only in the interval and in how the resource value is changed between subsequent dumps (coins in SuperTux, bullet count in AssaultCube). In all experiments, we play the games to record dumps with the following resource range: in SuperTux from 100 coins up to 107, in AssaultCube from 20 bullets\footnote{In AssaultCube, the user can use different weapons (gun, assault rifle, grenades, etc.). The magazines for each weapon are stored in a single array, thus we opted to protect the whole array, \ie the magazines for all weapons. However, we performed the simulation only on the default weapon, the assault rifle.} down to 13.

For both SuperTux and AssaultCube, our script uses OCR to automatically annotate each dump with the on-screen number of coins or bullets. Because this step may fail, we manually inspected all screenshots, correcting the annotations when necessary.

\hypertarget{sec:atk-strategies:subsec:dump-collection:par:paced}{\paragraph{Paced Dump Collection}}
For the first experiment set, we use the \emph{paced} dump collection approach, modelling an attacker who changes the resource value between each memory scan. Specifically, we increment or decrement the targeted resource by 1 after every 3 dumps, collecting a coin in SuperTux or firing a shot in AssaultCube. In SuperTux, we take an additional dump before starting the level, as the coin count is loaded into memory beforehand. In contrast, in AssaultCube the first dump is taken once the level is already running.

\hypertarget{sec:atk-strategies:subsec:dump-collection:par:fast}{\paragraph{Fast Dump Collection}}

In the second experiment set, we use the \emph{fast} dump collection approach to model an attacker who presumes a dynamic encoding, where encoding parameters change during execution, and thus tries to perform as many memory scans as possible between encoding updates. We collect a dump every 0.5 seconds and change the resource value every 6 dumps. To support frequent coin collection in SuperTux, we created a custom level with easily accessible coins, whereas AssaultCube requires no such modification. Additionally, we automated parts of the controls to trigger coin collection (or firing bullets) at the appropriate times.

In our proof-of-concept implementation of our simulation method, the defender must collect dumps for each protected game version, requiring repeated gameplay for every protection. Future work will investigate ways to reduce this effort. In particular, when multiple protection variants share the same data storage layout for the resource, a single dump sequence could be reused by patching it with the encodings of other protections. This would eliminate the need to replay the game and collect separate dumps for each protection version.

\subsection{Localisation Attack Strategy Simulation}\label{sec:atk-strategies}

In the first experiment set, we simulated nine greedy localisation attack strategies. In the second set, we simulated three strategies: one statistical strategy, and two greedy strategies also used in the first set. Each simulation required implementing the three components of the simulation script described in Section~\ref{sec:simul}. We detail these implementations in the following three sections.

\subsubsection{Dump Selection Logic\label{sec:atk-strategies:subsec:dumpselection}}
For our experiments, we have implemented four dump selection approaches.
Each of these dump selection logics is appropriate for use with one or more of the pruning logics outlined in Section~\ref{sec:atk-strategies:subec:location-pruning}.

\hypertarget{sec:atk-strategies:subsec:dumpselection:par:binned}{\paragraph{Binned Dump Selection}}
In the \emph{binned} dump selection approach, we select sequences of dumps $D_s$ such that no dump shares the same resource amount with another dump in that sequence.

\hypertarget{sec:atk-strategies:subsec:dumpselection:par:incremental}{\paragraph{Incremental Dump Selection}}
The \emph{incremental} dump selection approach further narrows down the results of the \emph{binned} dump selection approach by allowing only sequences where the value of the resource decreases or increases by one at every step.

\hypertarget{sec:atk-strategies:subsec:dumpselection:par:fully-random}{\paragraph{Fully Random Dump Selection}}
As its name suggests, the \emph{fully random} dump selection approach creates sequences by selecting dumps at random.

\hypertarget{sec:atk-strategies:subsec:dumpselection:par:rapid}{\paragraph{Rapid Dump Selection}}
The \emph{rapid} dump selection approach chooses random dump subsets where consecutive dumps are at most $t$ seconds apart, with $t$ as a configurable parameter. This models an attacker who performs pruning steps quickly, for instance assuming a dynamic encoding and aiming to maximise the number of dumps collected between encoding parameter changes.
\\

When combining these dump selection approaches with the \emph{paced} dump collection approach in the first experiment set, we exhaustively simulate all selected dump sequences, excluding those too short for a given localisation strategy. In contrast, the \emph{fast} dump collection approach used in the other experiments yields substantially more dumps and, consequently, many more valid sequences. We hence cap the number of selected sequences at 1000 per simulated attack, game version, and sequence length.

\subsubsection{Location pruning logic}\label{sec:atk-strategies:subec:location-pruning}

We simulate nine different location pruning logics, some tailored to the encoding obfuscations described in Section~\ref{sec:obf-strategies}, and others more generally applicable. While the implementations differ in their encoding assumptions, all assume 32-bit aligned storage; these assumptions are not fundamental and can be easily relaxed.

\paragraph{Base Logic}
The \emph{base} location pruning logic assumes that no obfuscation has been applied to the data. It searches for an exact match in each memory dump, i.e., for the exact resource amount as shown on screen. 

\paragraph{$+$-Logic}
The $+$-logic expects that the data $X_i$ in memory dump $i$ does not match the value $A_i$ shown on screen, but that it has been incremented with a secret offset $O$ before being stored in memory.
This parameter $O$ is presumed constant across all dumps.

Under these assumptions, one cannot directly search the memory without knowledge of the used secret.
Instead, the logic hence compares values across dumps and requires that candidate locations adhere to $X_j - X_i = A_j - A_i$, exploiting the fact that the difference between displayed resource amounts during different scans should equal the difference between the values stored in memory.

\paragraph{$\xor$-Logic}
The $\xor$-logic is similar in nature to the $+$-logic, but rather than assuming the encoding of the resource value $A_i$ with an offset, this logic assumes that value $X_i$ in memory was obfuscated using XOR-masking with a secret mask $M$.
Analogous to the $+$-logic, we compare two values across dumps and require that they adhere to $X_j \oplus X_i = A_j \oplus A_i$, circumventing the need for knowledge of the secret mask $M$.

\paragraph{$+\xor$-Logic}
The $+\xor$-logic targets the similarly-named $+\xor$-encoding, where the encoded value $X_i$ in memory is given by encoding value $A_i$ as $(A_i + O) \oplus M$, with $M$ and $O$  secret values. The logic that we evaluate only targets dump sequences in which the different values $A_{i}$ increase (or decrease) monotonically with a constant stride 1, as this is the stride occurring in the simulated game scenarios.
In that scenario, the value $X_{i+1} \oplus X_i$ will equal $p \oplus (p + 1)$ for some value $p$. 

The function $\chi : p \mapsto p \oplus (p+1)$ gives an easily recognisable pattern where each value $\chi (p)$ equals $\mathbb{1}_{\zeta(p)}$, with $\mathbb{1}_{k}$ being a shorthand notation for $2^{k}-1$ and $\zeta : \mathbb{N} \to \mathbb{N}$ is the \emph{find first zero} function that gives the position of the least significant zero bit in the two's complement representation of its argument.
\autoref{tab:chi-pattern} lists the values for the $\chi$ function for the lowest natural numbers. In every considered dump, this logic prunes candidate memory locations if their values no longer adhere to this pattern.

Interestingly, this logic will not only yield memory locations of which the value increments by 1, but also those of which it decrements by 1. Indeed, any sequence of values $X_i$ stored in memory that matches the pattern can be the result of an incrementing sequence $A_i$ encoded with mask $M$ and offset $O$, as well as of a decrementing sequence $A'_i$ encoded with a "complementary" mask $M'$ and offset $O'$. In other words, the approach cannot distinguish between decrementing and incrementing sequences.

\begin{table}[t]
    \centering
    \caption{The first 9 non-negative values for the $\chi : p \mapsto p \oplus (p+1)$ function used in the $\xor +$-logic.}
    \label{tab:chi-pattern}
    \setlength{\tabcolsep}{5pt}
    \renewcommand{\arraystretch}{1.2}
    \begin{tabular}{c|c c c c c c c c c}
        $p$     & 0 & 1 & 2 & 3 & 4 & 5 & 6 & 7 & 8 \\
        \hline
        $\chi(p)$ & 1 & 3 & 1 & 7 & 1 & 3 & 1 & 15 & 1 \\
    \end{tabular}
\end{table}

\paragraph{$\xor +$-Logic}
\phantomsection
\label{par:xor_add_strat}

The $\xor +$-logic is the counterpart of the $\xor +$-encoding, which encodes $X_i$ as $(A_i \oplus M) + O$.
We again limit our approach to consecutive resource amounts, which allows us to detect patterns more easily.

In this location pruning logic, we inspect the value $X_{i+1} - X_i$.
Since $O$ is presumed static, we expect that $X_{i+1} - X_i = (A_{i+1} \oplus M) - (A_i \oplus M)$.
Further, because $A_{i+1} = A_i + 1$, we know that $A_{i+1}$ and $A_i$ will differ only in their $\zeta(A_i)$ least significant bits.
This allows us to derive that $X_{i+1} - X_i = \mathbb{1}_{\zeta(A_i)} - 2\cdot(\mathbb{1}_{\zeta(A_i) - 1} \oplus M_{[\zeta(A_i):0]})$, where $M_{[k:0]} \triangleq \mathbb{1}_k \land M$ contains the $k$ least significant bits of $M$.
This value will be positive, will always be odd, and only its $\zeta(A_i)$ least significant bits can be non-zero.
For every candidate memory location, we thus check whether it meets these criteria, otherwise we discard it.
For the correct memory location, this value also allows us to determine the $\zeta(A_i)$ least significant bits of the secret $M$.
We derive this value for each candidate location at every time step.
If such an update contradicts with the previously determined bits of the hypothesis for $M$ for that candidate, then we can also dispose of that candidate memory location.

Similarly to the $+ \xor$-logic, this $\xor +$-logic will identify not only values incrementing with 1, but also values decrementing with 1. 

\paragraph{RNC-logic}
Using RNC, the encoding of the value $A$ of the asset is stored as multiple numbers $X_k$ each with their corresponding modulus $m_k$, such that $X_k = A \mod m_k$.
Consequently, for a memory location $k$ in memory dump $i$, the value $X_{k,i}$ will always be less than or equal to the value $A_i$ of the resource at that time.
Moreover, the difference $A_i - X_{k,i}$ will always be a non-negative multiple of $m_k$.
For every memory location, we thus compute $A_i - X_{k,i}$ for all considered dumps $i$.
We then compute the greatest common divisor (GCD) of all these differences.
If this GCD is $1$, then this means that there exists no modulus $m_k$ for which this memory location would be a valid part of an RNC encoding\footnote{We consider $GCD(0, 0)$ to be $0$. Further, note that $GCD(0, a) = a$.}.
The location thus does not constitute a valid candidate; we discard it.

Note that the RNC-logic only tries to identify memory locations where the stored values consistently correspond to the result of a modulo operation with some fixed modulus $m_k$.
The logic does not leverage the fact that the moduli for the different memory locations should be pairwise coprime, nor that $n=\prod_{k}m_k$ should be greater than all values $A$ that the defender wants to represent using the encoding.

\paragraph{Increase/Decrease Logic}
The \emph{increase/decrease} logic is a weaker version of the $+$-logic.
It assumes that the attacker knows when the encoded value in memory will increase, decrease or stay unaltered, compared to the previously observed encoded value.
However, unlike the $+$-logic, this pruning logic does not assume to know the amount by which the encoded value will change.

\paragraph{Change/No Change Logic}
The \emph{change/no change} logic is not tailored for a specific encoding.
It only assumes that the attacker knows when the encoded value in memory changes and when it does not, compared to the previously observed encoded value.
The logic discards memory locations if the values stored in these locations change when they are assumed to stay unaltered, and vice versa.

\paragraph{Change Logic}
The \emph{change} logic assumes that the attacker knows when the value stored in memory will change, but assumes that it cannot guarantee when this value will stay unaltered.\\

While some of the above location pruning logics are tailored for an assumed encoding, others are more general.
Some of the logics use a subset of the information that other logics use.
The increase/decrease logic, for example, is a weakened version of the $+$-logic.
We say that the $+$-logic is more \emph{specific} than the increase/decrease logic.
An attack logic $A$ is more \emph{specific} than a logic $B$ ($A > B$) if $A$ leverages more information in an attack.
A more specific logic will thus allow to narrow down the number of candidate memory locations faster when executing a location pruning attack, but can be used against fewer obfuscating encodings.

\begin{figure}[t]
    \centering
    \includesvg[height=5cm,keepaspectratio]{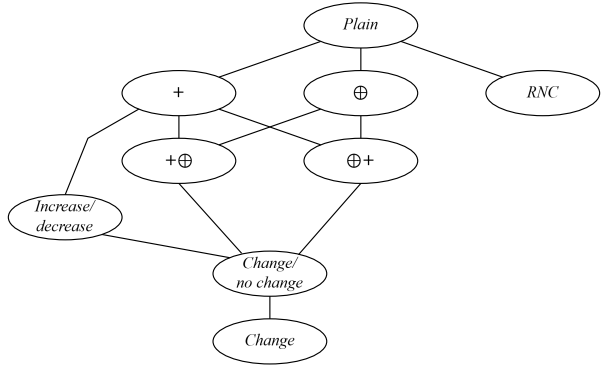}
    \caption{Hasse diagram for the partially ordered set of pruning logics used in the experimental evaluation in this paper, induced by the \emph{specificity} relation}
    \label{fig:attack-strategies-po}
\end{figure}

The specificity relation induces a partial ordering over the considered pruning logics, of which \autoref{fig:attack-strategies-po} shows the Hasse diagram. 
It thus allows a qualitative comparison of the attacker effort for those logics between which a partial ordering exists.
However, it does not allow for a quantitative comparison, nor does it allow for comparing the attacker effort for logics that are incomparable with respect to this \emph{specificity} relation.

As explained in Section \ref{sec:simulation:subsec:location-pruning}, an attacker can use every pruning logic in a greedy attack strategy as well as in a statistical attack strategy.
The latter type of strategy can prove useful on binaries where a dynamic encoding is used for the resource of interest.
In our first set of experiments, we use each of the above pruning logics in a greedy attack on all the versions of the considered games that use static encodings.
In our second set of experiments, we have created two versions each of both considered games that use a dynamic $\xor$-encoding, as described in Section \ref{sec:obf-strategies}. 
In these experiments, we compare the properties of running a statistical $\xor$ attack on these versions with those of running a greedy \emph{change} or \emph{change/no change} attack. We opted for this combination of pruning logics and attack strategies because these are the only effective ones on assets protected with dynamic encodings. In practice, the change/no change logic is one of the most popular ones among users of CheatEngine~\cite{cheatengine}.

\subsubsection{Data aggregation}
We evaluated each attack on many different combinations of dumps. For sequences consisting of a single dump (which can only produce results for the base attack logic), the number of simulations is limited to the number of dumps taken, which can be as small as 16 for some game versions. For longer sequences of length $n$, for each version $v$ of each game $P$ and for each pruning logic $A$, the number of simulations depends on the number of dump combinations of length $n$ that can be sampled from $D_s$ that meet the constraints of $A$. For example, our $+ \xor$-logic requires the resource value to increment or decrement with exactly 1 between consecutive dumps, while the RNC-logic and change/no change logics have no such constraints. Fewer constraints obviously result in more combinations being available from a similar number of dumps. For each data point, the number of simulated combinations hence ranged from 63 (2 dumps with some attack logics) up to 25515 (when the relatively small number of dumps in $D_s$ allows for a large number of subsequences of length $n$) and back down to 1 (when all dumps in $D_s$ were used, as is possible for, e.g., the change/no change logic).
As mentioned in Section~\ref{sec:atk-strategies:subsec:dumpselection}, when using the \emph{fast} dump collection approach in our second set of experiments, the total number of possible combinations becomes unfeasibly large; we therefore cap the number of considered combinations for every data point.

For each data point, i.e., for each version $v$ of each game $P$, for each pruning logic $A$, and for each number of used dumps $n$, we aggregate the number of remaining candidate locations obtained with the different simulations into a distribution $\hat{\phi}_{A_{2,n}}^{P_v}$
 and the success rate into a distribution $\hat{\sigma}_{A_n}^{P_v}$. We opted to aggregate the data into these distributions instead of into $\hat{\phi}_{A_1}^{P_v}$, $\hat{\phi}_{A_2}^{P_v}$, and $\hat{\sigma}_{A}^{P_v}$, because, in our view, plots such as those in Figure~\ref{fig:base_obfuscation_results} of percentiles of $\hat{\phi}_{A_{2,n}}^{P_v}$ and mean of $\hat{\sigma}_{A_n}^{P_v}$ for increasing numbers of dumps $n$ convey more useful information than plots based on the other distributions.

\subsection{Statistical Models of Effort and Outcomes for Attacks on Static Encodings}
\label{sec:evaluation_static}
This section presents a selection of the results obtained with our simulations of different pruning logics deployed against static protection strategies.

\subsubsection{Greedy Targeted Attack Strategies on Unobfuscated Games}
\begin{figure*}[t]
    \centering
\scriptsize
    \begin{subfigure}[b]{0.08\textwidth}
        \centering
        \includegraphics[trim={0cm 1.6cm 9.55cm 0.75cm},clip,height=3.3cm]{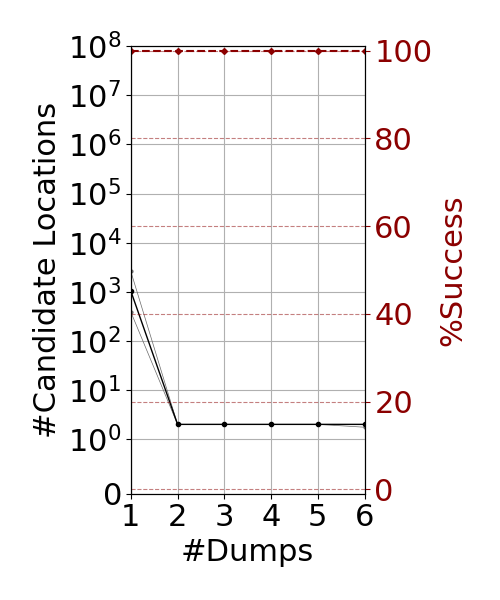}
        \vspace{0.5cm}
    \end{subfigure}
    \hspace{-0.35cm}
    \begin{subfigure}[b]{0.12\textwidth}
        \centering
        \includegraphics[trim={3.2cm 1.6cm 3.2cm 0.75cm},clip,height=3.3cm]{st_R_LinePlot_base-attack_on_base-obfuscation.png}
        \caption{Base}
    \end{subfigure}
    \hfill
    \begin{subfigure}[b]{0.12\textwidth}
        \centering
        \includegraphics[trim={3.2cm 1.6cm 3.2cm 0.75cm},clip,height=3.3cm]{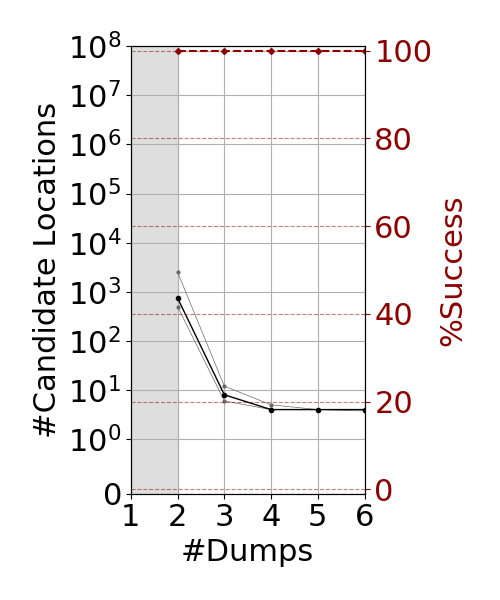}
        \caption{$+$}
    \end{subfigure}
    \hfill
    \begin{subfigure}[b]{0.12\textwidth}
        \centering
        \includegraphics[trim={3.2cm 1.6cm 3.2cm 0.75cm},clip,height=3.3cm]{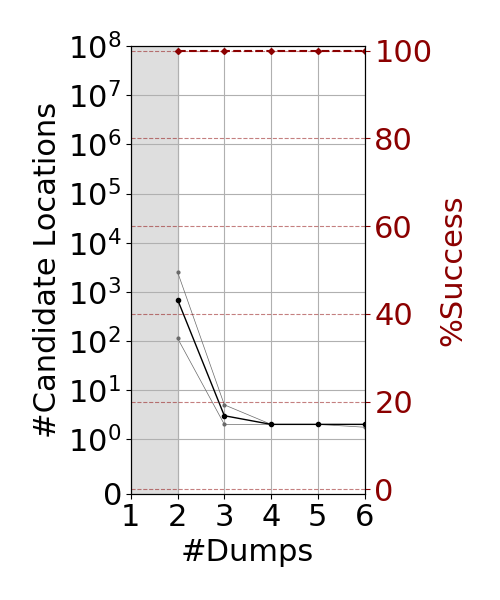}
        \caption{$\xor$}
    \end{subfigure}
    \hfill
    \begin{subfigure}[b]{0.12\textwidth}
        \centering
        \includegraphics[trim={3.2cm 1.6cm 3.2cm 0.75cm},clip,height=3.3cm]{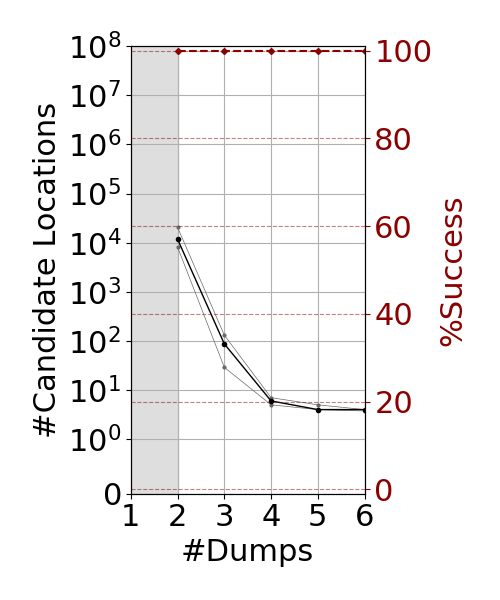}
        \caption{$+ \xor$}
    \end{subfigure}
    \hfill
    \begin{subfigure}[b]{0.12\textwidth}
        \centering
        \includegraphics[trim={3.2cm 1.6cm 3.2cm 0.75cm},clip,height=3.3cm]{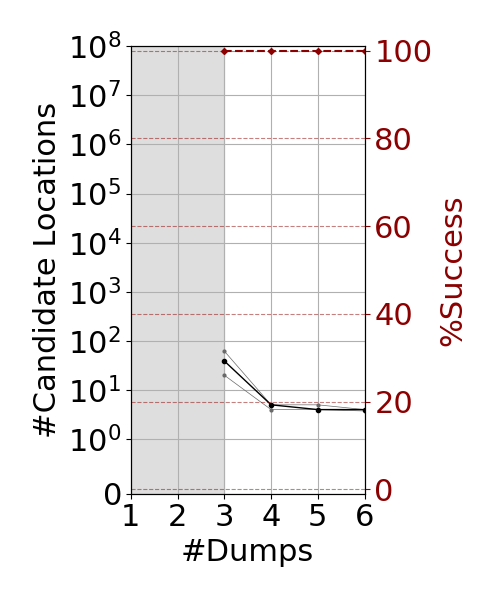}
        \caption{$\xor +$}
    \end{subfigure}
    \hfill
    \begin{subfigure}[b]{0.12\textwidth}
        \centering
        \includegraphics[trim={3.2cm 1.6cm 3.2cm 0.75cm},clip,height=3.3cm]{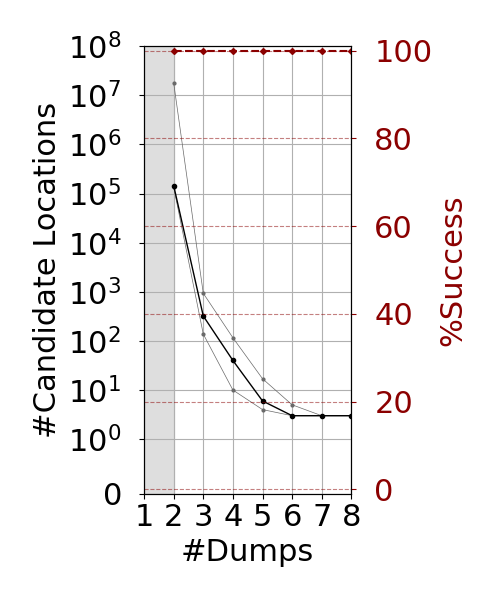}
        \caption{RNC}
    \end{subfigure}
    \hspace{-0.25cm}
    \begin{subfigure}[b]{0.08\textwidth}
        \centering
        \includegraphics[trim={9.55cm 1.6cm 0cm 0.75cm},clip,height=3.3cm]{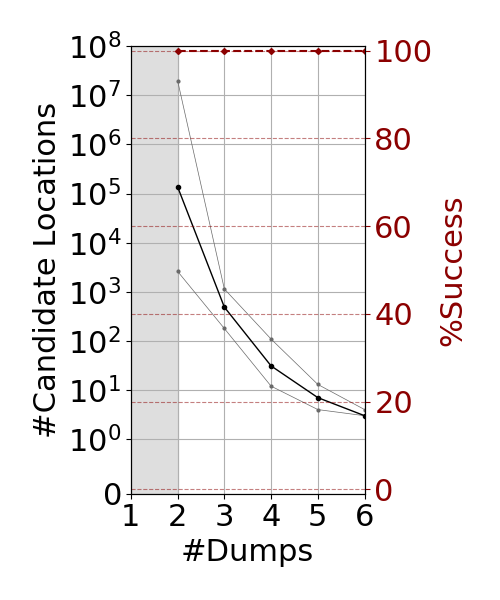}
        \vspace{0.48cm}
    \end{subfigure}

    \vspace{0.2cm}

    $\uparrow$: SuperTux \quad\quad $\downarrow$: AssaultCube

    \vspace{0.2cm}

    \begin{subfigure}[b]{0.08\textwidth}
        \centering
        \includegraphics[trim={0cm 1.6cm 9.55cm 0.75cm},clip,height=3.3cm]{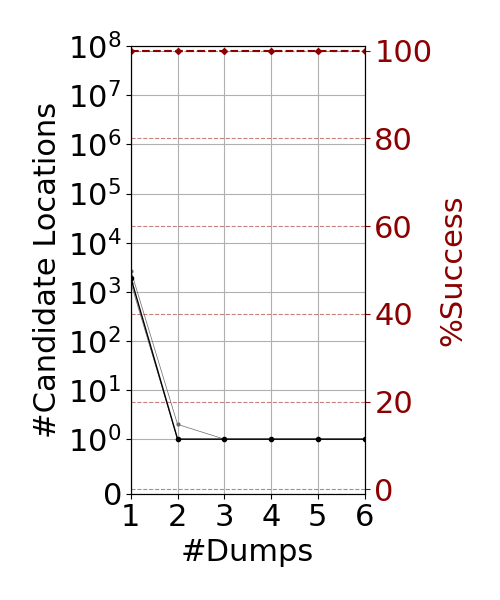}
        \vspace{0.5cm}
    \end{subfigure}
    \hspace{-0.35cm}
    \begin{subfigure}[b]{0.12\textwidth}
        \centering
        \includegraphics[trim={3.2cm 1.6cm 3.2cm 0.75cm},clip,height=3.3cm]{ac_R_LinePlot_base-attack_on_base-obfuscation.png}
        \caption{Base}
    \end{subfigure}
    \hfill
    \begin{subfigure}[b]{0.12\textwidth}
        \centering
        \includegraphics[trim={3.2cm 1.6cm 3.2cm 0.75cm},clip,height=3.3cm]{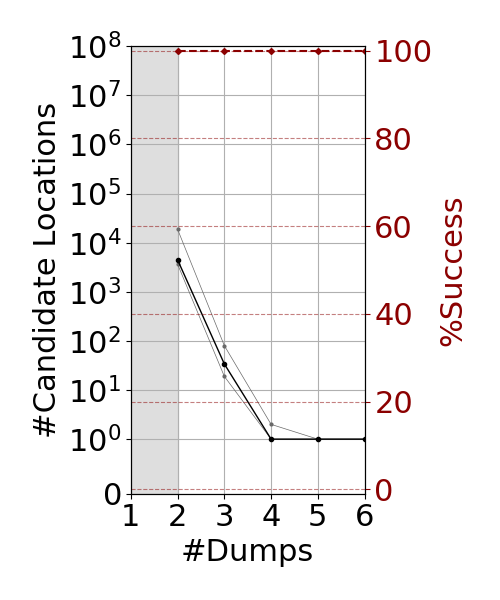}
        \caption{$+$}
    \end{subfigure}
    \hfill
    \begin{subfigure}[b]{0.12\textwidth}
        \centering
        \includegraphics[trim={3.2cm 1.6cm 3.2cm 0.75cm},clip,height=3.3cm]{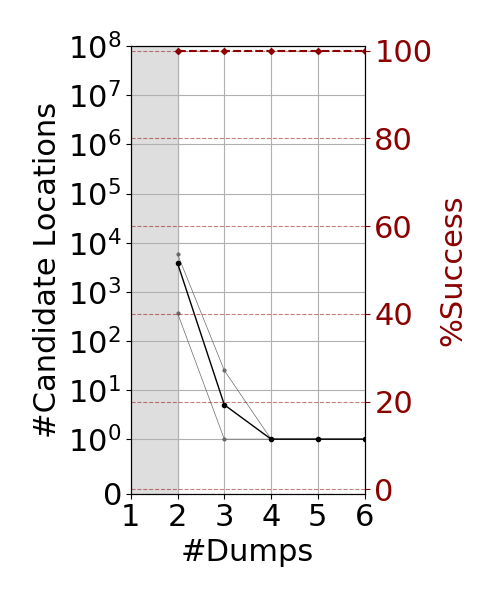}
        \caption{$\xor$}
    \end{subfigure}
    \hfill
    \begin{subfigure}[b]{0.12\textwidth}
        \centering
        \includegraphics[trim={3.2cm 1.6cm 3.2cm 0.75cm},clip,height=3.3cm]{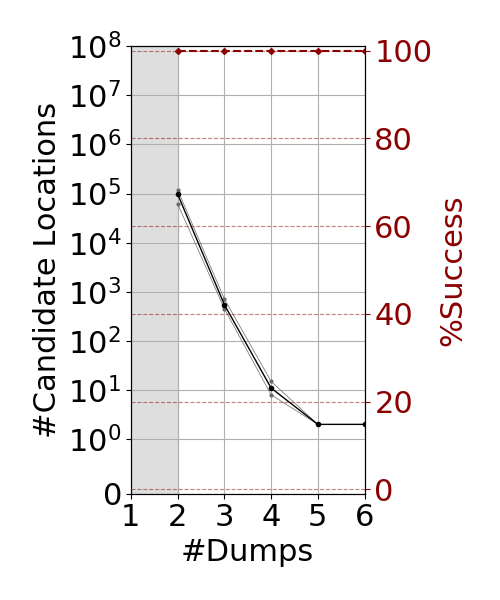}
        \caption{$+ \xor$}
    \end{subfigure}
    \hfill
    \begin{subfigure}[b]{0.12\textwidth}
        \centering
        \includegraphics[trim={3.2cm 1.6cm 3.2cm 0.75cm},clip,height=3.3cm]{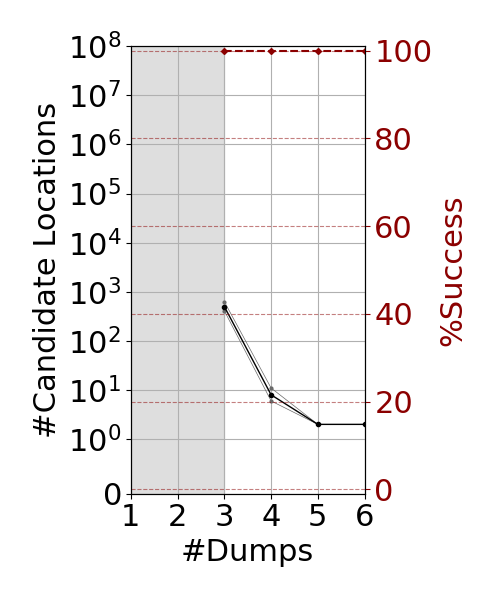}
        \caption{$\xor +$}
    \end{subfigure}
    \hfill
    \begin{subfigure}[b]{0.12\textwidth}
        \centering
        \includegraphics[trim={3.2cm 1.6cm 3.2cm 0.75cm},clip,height=3.3cm]{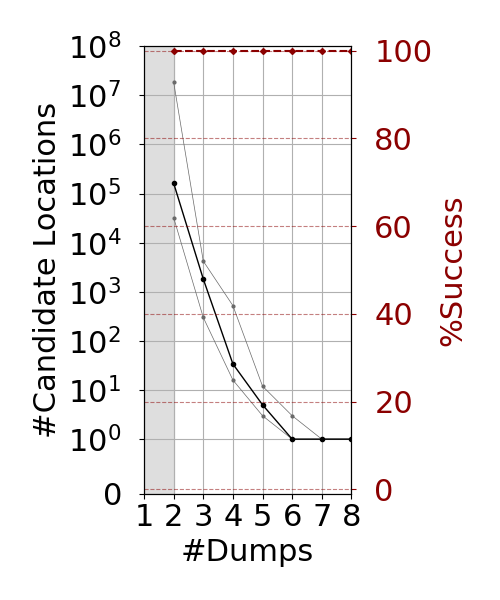}
        \caption{RNC}
    \end{subfigure}
    \hspace{-0.25cm}
    \begin{subfigure}[b]{0.08\textwidth}
        \centering
        \includegraphics[trim={9.55cm 1.6cm 0cm 0.75cm},clip,height=3.3cm]{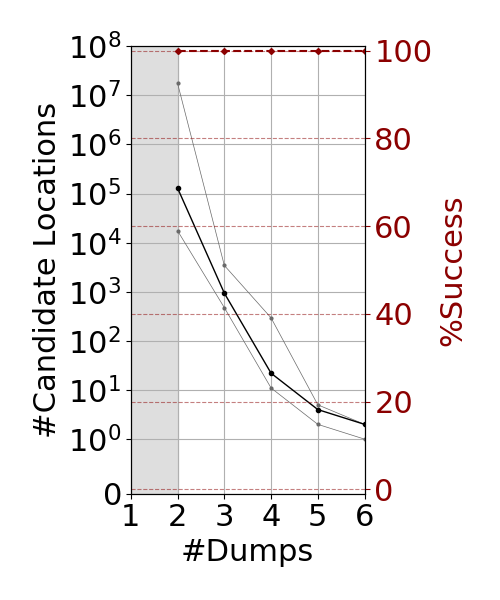}
        \vspace{0.48cm}
    \end{subfigure}
    
    \caption{Outcomes of six targeted pruning logics in greedy attack strategies on unobfuscated versions of the two games. (a)--(f) show the results on SuperTux, (g)--(l) show the results on AssaultCube. The X-axis shows the number of dumps $n$ taken by the modeller.     
    In red (right Y-axis), the mean success rate is plotted, i.e., the mean value of $\hat{\sigma}_{A_n}^{P_v}$. In black and gray, the $P_{25}$, $P_{50}$, and $P_{75}$ percentiles of the remaining number of candidate locations are plotted, i.e., of $\hat{\phi}_{A_{2,n}}^{P_v}$. The red line in graph (b), e.g., shows that with a $+$-attack using two or more dumps, the mean success rate is 100\%, meaning that remaining candidate locations to be considered will highly likely always include the actual location the attacker is after. The black line shows that with only two dumps, half of the attacks had already pruned the search space for the attacker to less than about 900 candidate locations. The gray lines show that 75\% had pruned the search space to less than 2000 locations, and 25\% had already pruned it to less than 700 locations. Finally, the grayed area of each plot marks the zone where the attack cannot yet achieve any useful pruning of the search space, because the number of used dumps is simply too low.}
    \label{fig:base_obfuscation_results}
\end{figure*}

The top row of Figure~\ref{fig:base_obfuscation_results} (charts (a)–-(f)) presents results for six targeted pruning logics applied in a greedy strategy to an unobfuscated version of SuperTux. All six logics achieve a 100\% success rate (\ie recall), as indicated by the red line, once a sufficient number of dumps is available: they prune the search space without excluding the targeted location. The base and $\xor$ logics are the \emph{most effective}, reducing candidates to two locations, which is the minimum achievable because the value is duplicated in memory for on-screen display. The $+$, $+\xor$, and $\xor+$ logics yield four candidates: the correct location, the location storing the number of coins in the current level (Section~\ref{sec:games}), and an additional spurious value that happened to evolve similarly in the observed dumps.

The base logic is also the \emph{most efficient}, reaching its optimal result with only 2 dumps, whereas the other logics require at least 4.

The bottom row of Figure~\ref{fig:base_obfuscation_results} shows analogous results for AssaultCube. Although absolute numbers differ, the same trends appear. In particular, both the $+\xor$ and $\xor+$ logics identify not only the bullet count but also the location tracking the number of fired shots, which evolves with the opposite stride of $-1$.

Overall, these expected results confirm that our method effectively models both the efficiency and effectiveness of localisation strategies on unobfuscated games.

This includes identifying other data that can confound certain localisation pruning logics, because such data behaves similarly when observed through the lens of those logics. In other words, the method can reveal data that can be exploited to create a (small) anonymity set in which the data to be protected can be hidden to some extent. We consider this an important benefit of our method. In our study, as software protection researchers, we invested substantial effort to analyse such other data to validate and explain the obtained results, which required substantial insight into the games’ internal operations. From conversations with major software protection vendors, we know that this level of insight cannot be expected from the game vendor employees or their consultants who need to deploy protection tools on their games, typically late in the software development life cycle when time-to-market dominates. In other words, in practice those configuring the tools and choosing protections are often largely steering blind. Our method can then help them gather relevant insights more quickly.

\pagebreak

\subsubsection{Greedy Targeted Attack Strategies on Obfuscated Games}
\label{sec:evaluation-greedy}

\begin{figure*}[t]
    \centering
\scriptsize
    \begin{subfigure}[b]{0.08\textwidth}
        \centering
        \includegraphics[trim={0cm 1.6cm 9.55cm 0.75cm},clip,height=3.3cm]{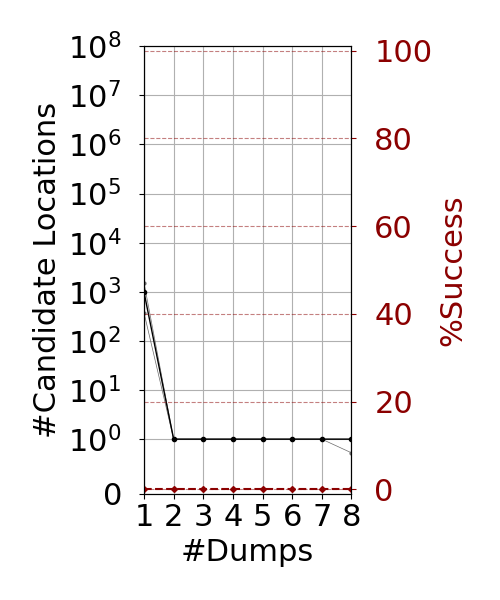}
        \vspace{0.5cm}
    \end{subfigure}
    \hspace{-0.35cm}
    \begin{subfigure}[b]{0.12\textwidth}
        \centering
        \includegraphics[trim={3.2cm 1.6cm 3.2cm 0.75cm},clip,height=3.3cm]{st_R_LinePlot_base-attack_on_rnc-obfuscation.png}
        \caption{Base}
    \end{subfigure}
    \hfill
    \begin{subfigure}[b]{0.12\textwidth}
        \centering
        \includegraphics[trim={3.2cm 1.6cm 3.2cm 0.75cm},clip,height=3.3cm]{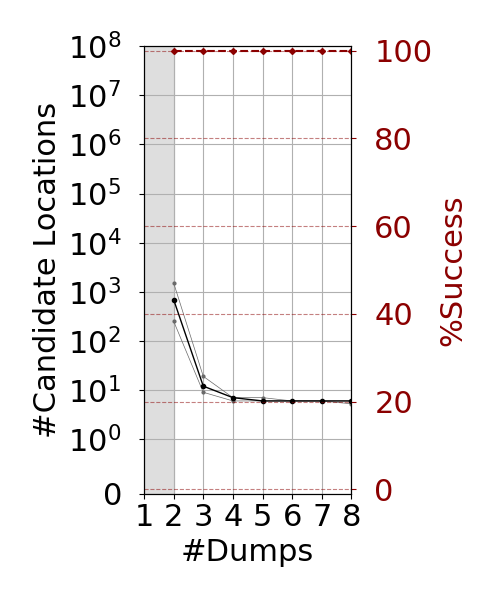}
        \caption{$+$}
    \end{subfigure}
    \hfill
    \begin{subfigure}[b]{0.12\textwidth}
        \centering
        \includegraphics[trim={3.2cm 1.6cm 3.2cm 0.75cm},clip,height=3.3cm]{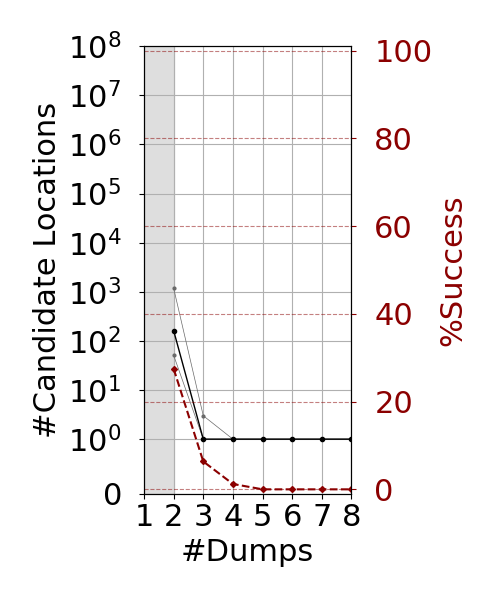}
        \caption{$\xor$}
    \end{subfigure}
    \hfill
    \begin{subfigure}[b]{0.12\textwidth}
        \centering
        \includegraphics[trim={3.2cm 1.6cm 3.2cm 0.75cm},clip,height=3.3cm]{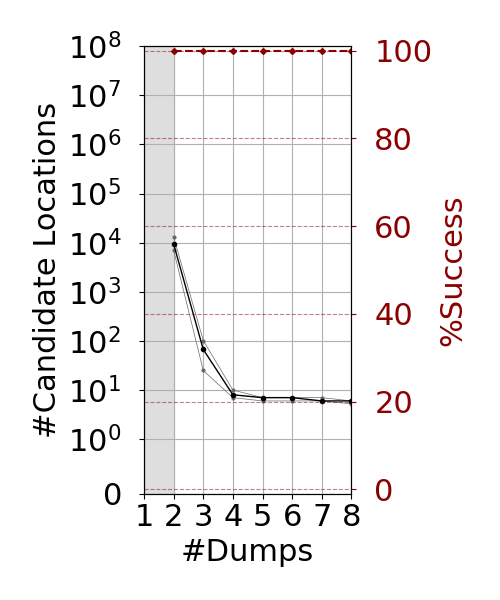}
        \caption{$+ \xor$}
    \end{subfigure}
    \hfill
    \begin{subfigure}[b]{0.12\textwidth}
        \centering
        \includegraphics[trim={3.2cm 1.6cm 3.2cm 0.75cm},clip,height=3.3cm]{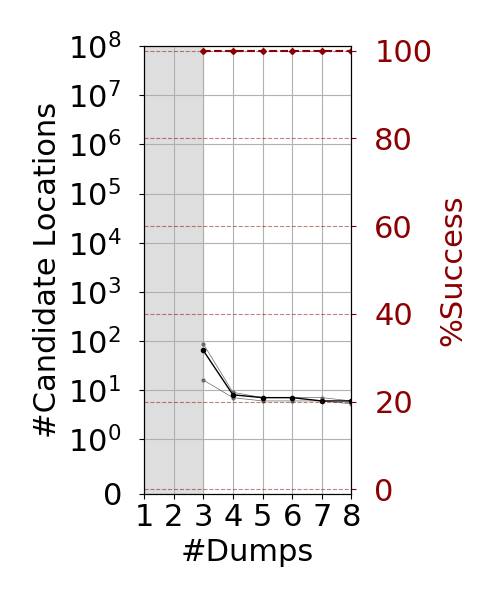}
        \caption{$\xor +$}
    \end{subfigure}
    \hfill
    \begin{subfigure}[b]{0.12\textwidth}
        \centering
        \includegraphics[trim={3.2cm 1.6cm 3.2cm 0.75cm},clip,height=3.3cm]{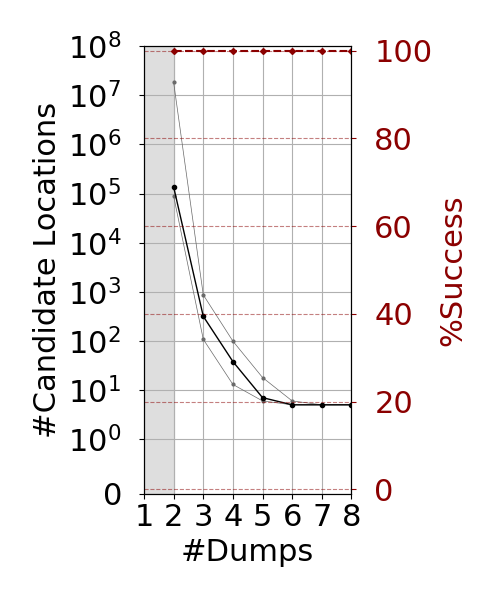}
        \caption{RNC}
    \end{subfigure}
    \hspace{-0.25cm}
    \begin{subfigure}[b]{0.08\textwidth}
        \centering
        \includegraphics[trim={9.55cm 1.6cm 0cm 0.75cm},clip,height=3.3cm]{st_R_LinePlot_rnc-attack_on_rnc-obfuscation.png}
        \vspace{0.48cm}
    \end{subfigure}

    \vspace{0.2cm}

    $\uparrow$: SuperTux \quad\quad $\downarrow$: AssaultCube

    \vspace{0.2cm}

    \begin{subfigure}[b]{0.08\textwidth}
        \centering
        \includegraphics[trim={0cm 1.6cm 9.55cm 0.75cm},clip,height=3.3cm]{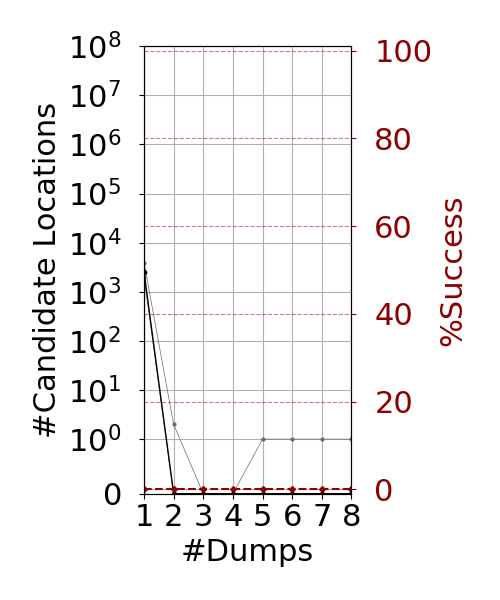}
        \vspace{0.5cm}
    \end{subfigure}
    \hspace{-0.35cm}
    \begin{subfigure}[b]{0.12\textwidth}
        \centering
        \includegraphics[trim={3.2cm 1.6cm 3.2cm 0.75cm},clip,height=3.3cm]{ac_R_LinePlot_base-attack_on_rnc-obfuscation.png}
        \caption{Base}
    \end{subfigure}
    \hfill
    \begin{subfigure}[b]{0.12\textwidth}
        \centering
        \includegraphics[trim={3.2cm 1.6cm 3.2cm 0.75cm},clip,height=3.3cm]{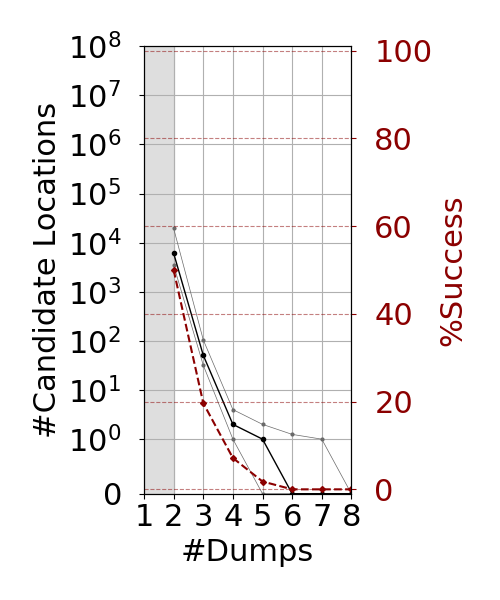}
        \caption{$+$}
    \end{subfigure}
    \hfill
    \begin{subfigure}[b]{0.12\textwidth}
        \centering
        \includegraphics[trim={3.2cm 1.6cm 3.2cm 0.75cm},clip,height=3.3cm]{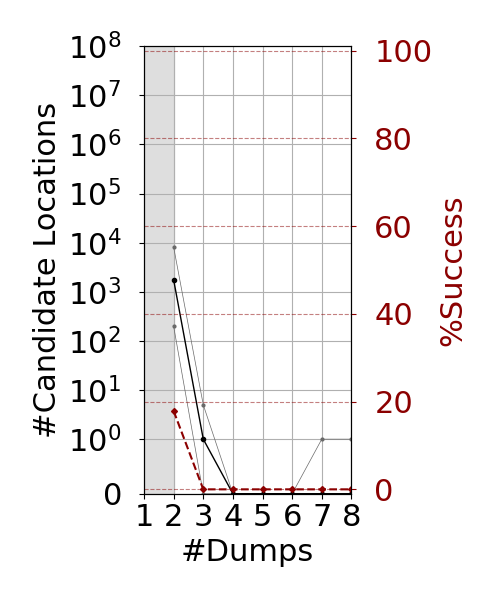}
        \caption{$\xor$}
    \end{subfigure}
    \hfill
    \begin{subfigure}[b]{0.12\textwidth}
        \centering
        \includegraphics[trim={3.2cm 1.6cm 3.2cm 0.75cm},clip,height=3.3cm]{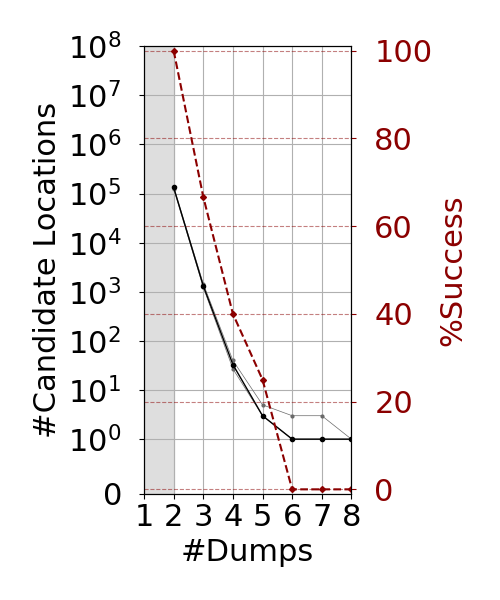}
        \caption{$+ \xor$}
    \end{subfigure}
    \hfill
    \begin{subfigure}[b]{0.12\textwidth}
        \centering
        \includegraphics[trim={3.2cm 1.6cm 3.2cm 0.75cm},clip,height=3.3cm]{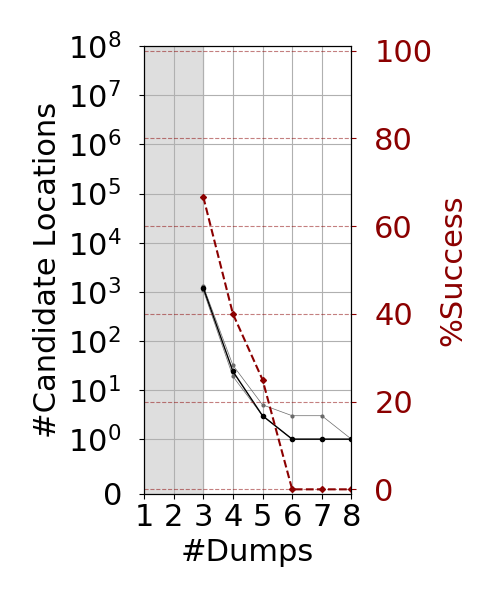}
        \caption{$\xor +$}
    \end{subfigure}
    \hfill
    \begin{subfigure}[b]{0.12\textwidth}
        \centering
        \includegraphics[trim={3.2cm 1.6cm 3.2cm 0.75cm},clip,height=3.3cm]{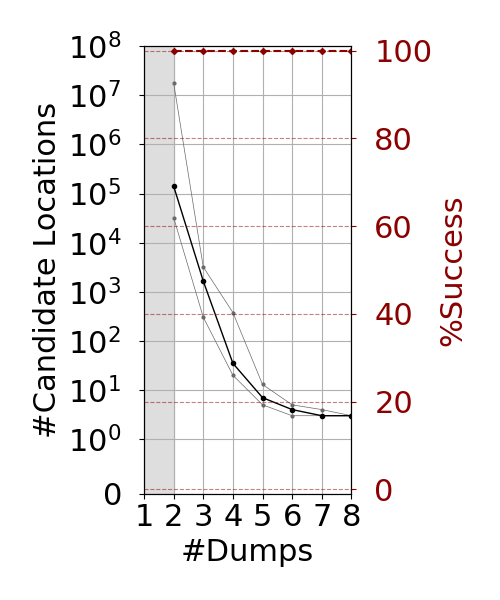}
        \caption{RNC}
    \end{subfigure}
    \hspace{-0.25cm}
    \begin{subfigure}[b]{0.08\textwidth}
        \centering
        \includegraphics[trim={9.55cm 1.6cm 0cm 0.75cm},clip,height=3.3cm]{ac_R_LinePlot_rnc-attack_on_rnc-obfuscation.png}
        \vspace{0.48cm}
    \end{subfigure}
    
    \caption{Outcomes of six targeted pruning logics applied greedily on RNC-protected games. (a)--(f) show results on SuperTux, (g)--(l) show results on AssaultCube. The legend is the same as in Figure~\ref{fig:base_obfuscation_results}. }
    \label{fig:rnc_obfuscation_results}
\end{figure*}

Figure~\ref{fig:rnc_obfuscation_results} presents results for the same greedily applied, targeted pruning logics, now on RNC-protected games. For AssaultCube, the outcome matches expectations: only the RNC-specific attack succeeds, while simpler logics fail.\footnote{For some attacks, the success rate initially seems high but then quickly drops as more dumps are considered. The reason for the high initial success rate is, of course, that the candidate locations have initially hardly been pruned yet, so the correct location has not yet been pruned. With more dumps considered, and hence more pruning, the correct location reliably gets pruned with the failing techniques.}

For SuperTux, however, results differ markedly from expectations. Besides the RNC logic, the $+$, $+\xor$, and $\xor+$ logics also succeed and do so more efficiently. This behaviour results from the interaction between the coin values observed in the dumps (100--107) and the RNC moduli ($m_1=89, m_2=93, m_3=97$), which prevents value wrapping. For example, when the number of coins evolves from 101 to 107 in unit steps, the first stored value (mod 89) evolves from 12 to 18, also in unit steps. Consequently, the stored values evolve exactly as under a $+$-encoding with offset $-89$, allowing offset-based logics to succeed as well. Our method correctly captures this interaction.

These findings lead to two observations. First, our method helps identify subtle interactions between defence parameters, on the one hand, and pruning logics and attack strategies, on the other hand, that can significantly weaken a protection. Second, the method is highly sensitive to defence and attack parameters and their interplay. If the SuperTux dumps had by chance included only cases when the values wrapped around the moduli, the results would likely resemble those of AssaultCube, making the parameter weakness less apparent. Note that this does not imply that outcome of the experiment was entirely accidental and due to pure luck. At the start of the game, the number of coins is 100. So it makes sense to expect that rational attackers will start making dumps in the range we targeted during our simulations, i.e., to model resource localisation attacks that focus on the start of the game.

\begin{figure*}[t]
    \centering
\scriptsize
    \begin{subfigure}[b]{0.08\textwidth}
        \centering
        \includegraphics[trim={0cm 1.6cm 9.55cm 0.75cm},clip,height=3.3cm]{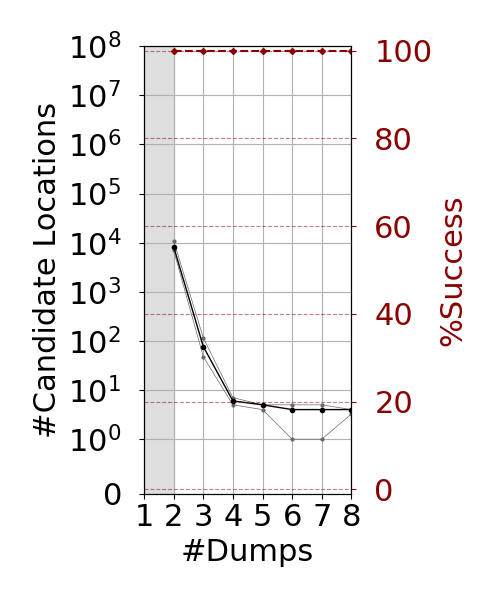}
        \vspace{0.77cm} \end{subfigure}
    \hspace{-0.60cm} \begin{subfigure}[b]{0.16\textwidth}
        \centering
        \includegraphics[trim={3.2cm 1.6cm 3.2cm 0.75cm},clip,height=3.3cm]{st_Results_LinePlot_add_xor-attack_on_add_xor-obfuscation.png}
        \caption{$+\xor$ attack on $+\xor$ obfuscation}
    \end{subfigure}
    \hfill
    \begin{subfigure}[b]{0.16\textwidth}
        \centering
        \includegraphics[trim={3.2cm 1.6cm 3.2cm 0.75cm},clip,height=3.3cm]{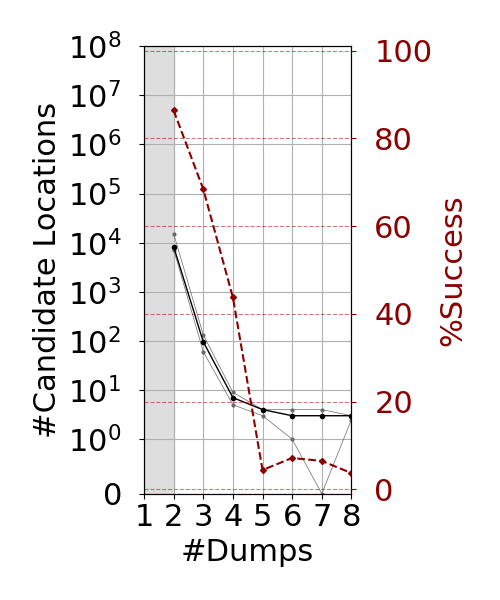}
        \caption{$+\xor$ attack on $\xor+$ obfuscation}
    \end{subfigure}
    \hfill
    \begin{subfigure}[b]{0.16\textwidth}
        \centering
        \includegraphics[trim={3.2cm 1.6cm 3.2cm 0.75cm},clip,height=3.3cm]{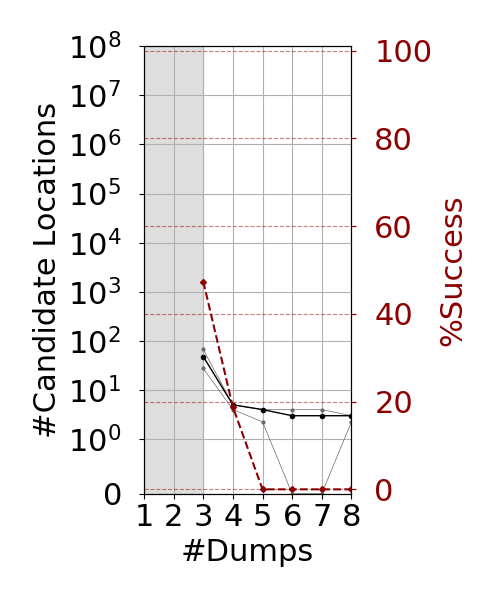}
        \caption{$\xor+$ attack on $+\xor$ obfuscation}
    \end{subfigure}
    \hfill
    \begin{subfigure}[b]{0.16\textwidth}
        \centering
        \includegraphics[trim={3.2cm 1.6cm 3.2cm 0.75cm},clip,height=3.3cm]{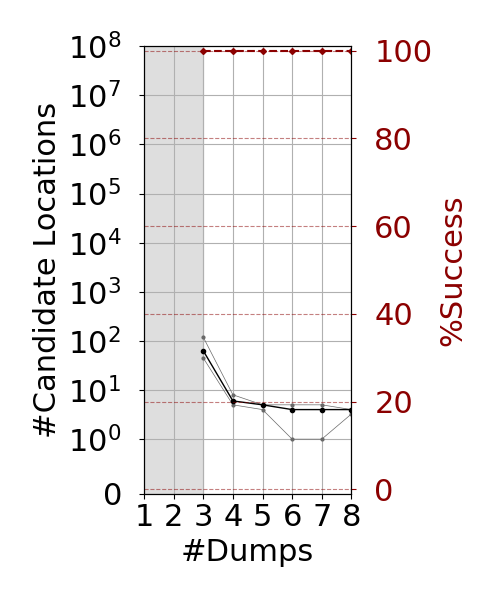}
        \caption{$\xor+$ attack on $\xor+$ obfuscation}
    \end{subfigure}
    \hspace{-0.25cm}
    \begin{subfigure}[b]{0.08\textwidth}
        \centering
        \includegraphics[trim={9.55cm 1.6cm 0cm 0.75cm},clip,height=3.3cm]{st_Results_LinePlot_xor_add-attack_on_xor_add-obfuscation.png}
        \vspace{0.48cm}
    \end{subfigure}

    \vspace{0.2cm}

    $\uparrow$: SuperTux \quad\quad $\downarrow$: AssaultCube

    \vspace{0.2cm}

    \begin{subfigure}[b]{0.08\textwidth}
        \centering
        \includegraphics[trim={0cm 1.6cm 9.55cm 0.75cm},clip,height=3.3cm]{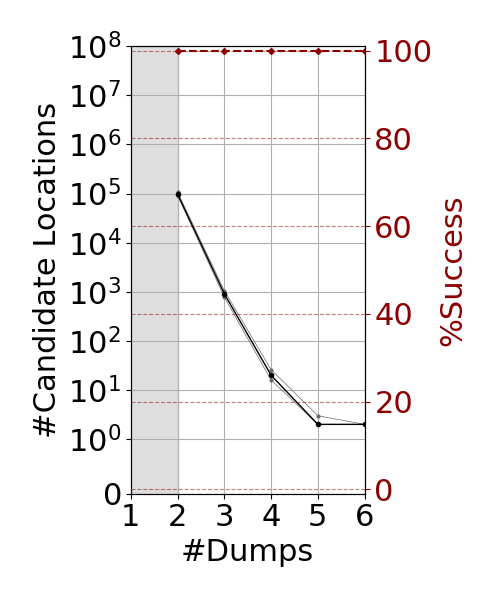}
        \vspace{0.77cm} \end{subfigure}
    \hspace{-0.60cm} \begin{subfigure}[b]{0.16\textwidth}
        \centering
        \includegraphics[trim={3.2cm 1.6cm 3.2cm 0.75cm},clip,height=3.3cm]{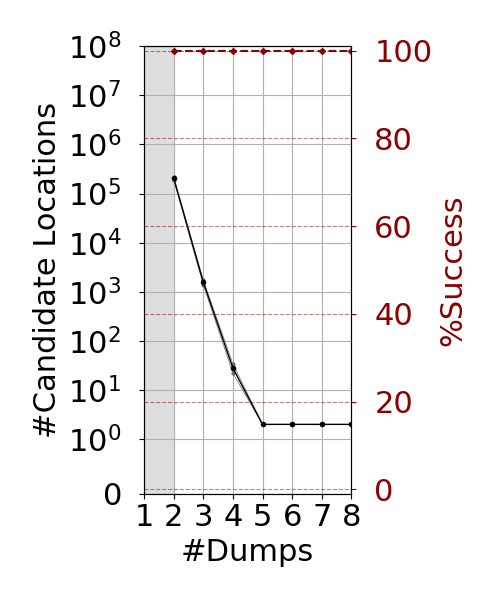}
        \caption{$+\xor$ attack on $+\xor$ obfuscation}
    \end{subfigure}
    \hfill
    \begin{subfigure}[b]{0.16\textwidth}
        \centering
        \includegraphics[trim={3.2cm 1.6cm 3.2cm 0.75cm},clip,height=3.3cm]{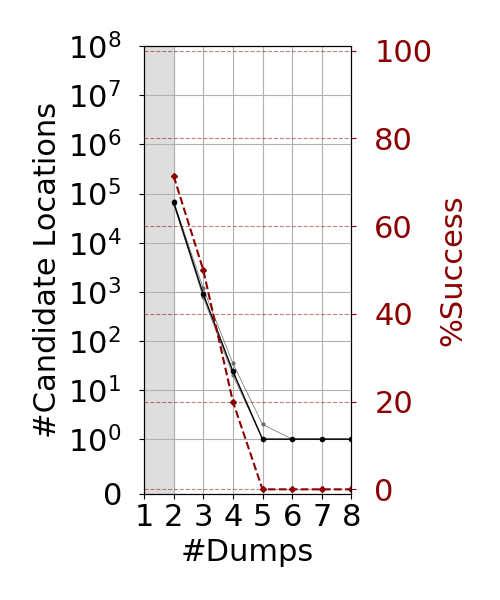}
        \caption{$+\xor$ attack on $\xor+$ obfuscation}
    \end{subfigure}
    \hfill
    \begin{subfigure}[b]{0.16\textwidth}
        \centering
        \includegraphics[trim={3.2cm 1.6cm 3.2cm 0.75cm},clip,height=3.3cm]{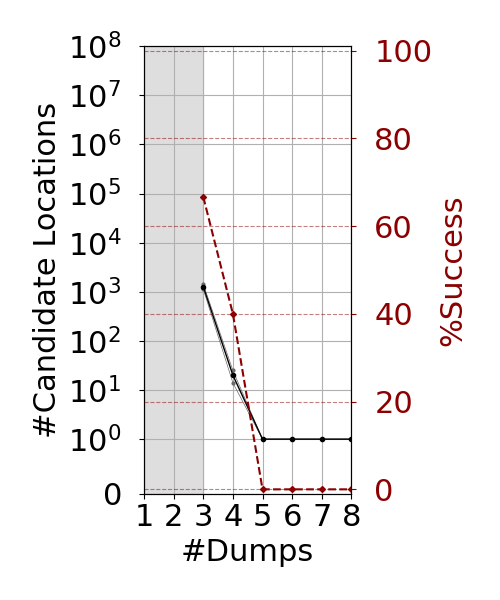}
        \caption{$\xor+$ attack on $+\xor$ obfuscation}
    \end{subfigure}
    \hfill
    \begin{subfigure}[b]{0.16\textwidth}
        \centering
        \includegraphics[trim={3.2cm 1.6cm 3.2cm 0.75cm},clip,height=3.3cm]{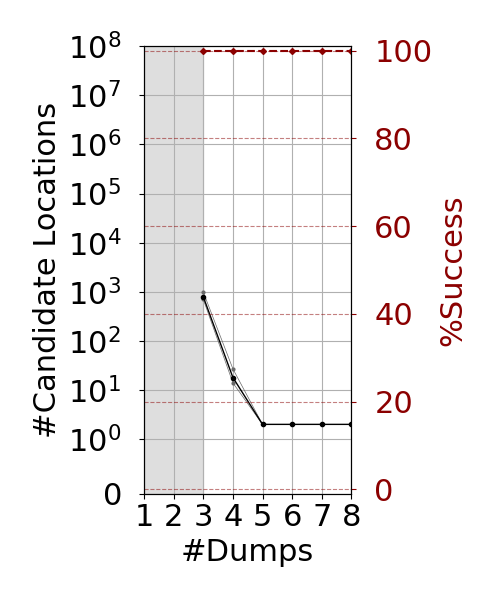}
        \caption{$\xor+$ attack on $\xor+$ obfuscation}
    \end{subfigure}
    \hspace{-0.25cm}
    \begin{subfigure}[b]{0.08\textwidth}
        \centering
        \includegraphics[trim={9.55cm 1.6cm 0cm 0.75cm},clip,height=3.3cm]{ac_Results_LinePlot_xor_add-attack_on_xor_add-obfuscation.png}
        \vspace{0.48cm}
    \end{subfigure}

    \caption{Outcomes of two related pruning logics applied greedily on two versions of the games. (a)--(d) show results on SuperTux, (e)--(h) show results on AssaultCube. The legend is the same as in Figure~\ref{fig:base_obfuscation_results}.}
    \label{fig:switching_strategy_results}
\end{figure*}

Figure~\ref{fig:switching_strategy_results} presents results for the related $\xor+$ and $+\xor$ pruning logics applied to both corresponding obfuscations. These results are notable for three reasons. First, they confirm that attack effectiveness is highly sensitive to the applied defence: for both programs, each attack succeeds only against its targeted obfuscation and fails on the variant. Second, for each logic and game, the evolution of the number of candidate locations is very similar regardless of whether the attack is successful, preventing attackers from inferring correctness during their attack based on pruning behaviour alone. Third, because the games store related values not encoded in a way that breaks the attempted logic, failing attacks do not prune to zero locations. As a result, failures produce both false negatives and false positives, forcing attackers to invest additional effort to validate candidates, \ie to discriminate between true and false positives.

We conclude that our method provides useful models for comparing the efficiency and effectiveness of pruning logics across different encodings, particularly in scenarios where defence–attack interactions depend sensitively on encoding configuration choices. This is especially valuable for users of software protection tools, who otherwise must operate largely without guidance.

\subsubsection{Comparing Worst-Case Scenarios on Different Protections}

\begin{figure*}[t]
    \centering
\scriptsize
    \begin{subfigure}[b]{0.08\textwidth}
        \centering
        \includegraphics[trim={0cm 1.6cm 9.55cm 0.75cm},clip,height=3.3cm]{st_R_LinePlot_base-attack_on_base-obfuscation.png}
        \vspace{0.5cm}
    \end{subfigure}
    \hspace{-0.35cm}
    \begin{subfigure}[b]{0.12\textwidth}
        \centering
        \includegraphics[trim={3.2cm 1.6cm 3.2cm 0.75cm},clip,height=3.3cm]{st_R_LinePlot_base-attack_on_base-obfuscation.png}
        \caption{Base}
    \end{subfigure}
    \hfill
    \begin{subfigure}[b]{0.12\textwidth}
        \centering
        \includegraphics[trim={3.2cm 1.6cm 3.2cm 0.75cm},clip,height=3.3cm]{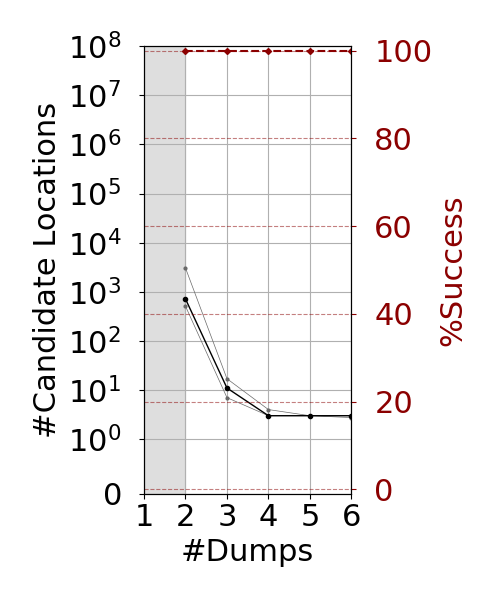}
        \caption{$+$}
    \end{subfigure}
    \hfill
    \begin{subfigure}[b]{0.12\textwidth}
        \centering
        \includegraphics[trim={3.2cm 1.6cm 3.2cm 0.75cm},clip,height=3.3cm]{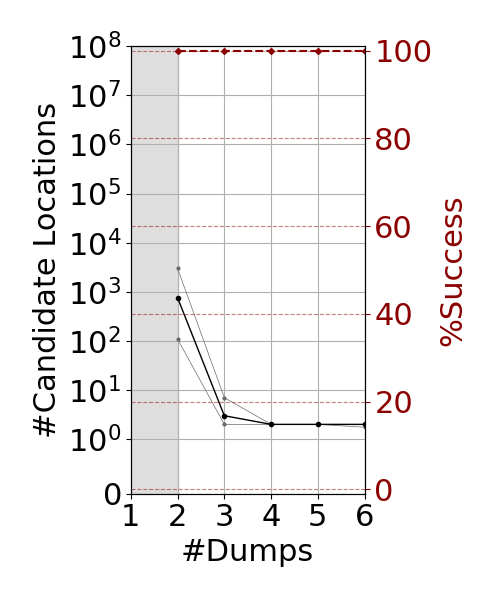}
        \caption{$\xor$}
    \end{subfigure}
    \hfill
    \begin{subfigure}[b]{0.12\textwidth}
        \centering
        \includegraphics[trim={3.2cm 1.6cm 3.2cm 0.75cm},clip,height=3.3cm]{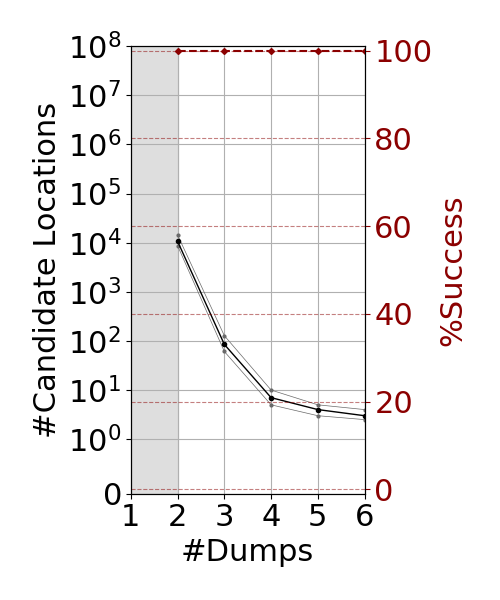}
        \caption{$+\xor$}
    \end{subfigure}
    \hfill
    \begin{subfigure}[b]{0.12\textwidth}
        \centering
        \includegraphics[trim={3.2cm 1.6cm 3.2cm 0.75cm},clip,height=3.3cm]{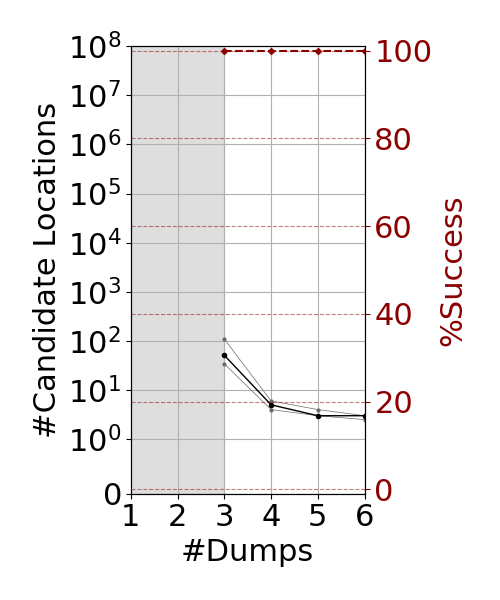}
        \caption{$\xor+$}
    \end{subfigure}
    \hfill
    \begin{subfigure}[b]{0.12\textwidth}
        \centering
        \includegraphics[trim={3.2cm 1.6cm 3.2cm 0.75cm},clip,height=3.3cm]{st_R_LinePlot_rnc-attack_on_rnc-obfuscation.png}
        \caption{RNC}
    \end{subfigure}
    \hspace{-0.25cm}
    \begin{subfigure}[b]{0.08\textwidth}
        \centering
        \includegraphics[trim={9.55cm 1.6cm 0cm 0.75cm},clip,height=3.3cm]{st_R_LinePlot_rnc-attack_on_rnc-obfuscation.png}
        \vspace{0.48cm}
    \end{subfigure}

    \vspace{0.2cm}

    $\uparrow$: SuperTux \quad\quad $\downarrow$: AssaultCube

    \vspace{0.2cm}

    \begin{subfigure}[b]{0.08\textwidth}
        \centering
        \includegraphics[trim={0cm 1.6cm 9.55cm 0.75cm},clip,height=3.3cm]{ac_R_LinePlot_base-attack_on_base-obfuscation.png}
        \vspace{0.5cm}
    \end{subfigure}
    \hspace{-0.35cm}
    \begin{subfigure}[b]{0.12\textwidth}
        \centering
        \includegraphics[trim={3.2cm 1.6cm 3.2cm 0.75cm},clip,height=3.3cm]{ac_R_LinePlot_base-attack_on_base-obfuscation.png}
        \caption{Base}
    \end{subfigure}
    \hfill
    \begin{subfigure}[b]{0.12\textwidth}
        \centering
        \includegraphics[trim={3.2cm 1.6cm 3.2cm 0.75cm},clip,height=3.3cm]{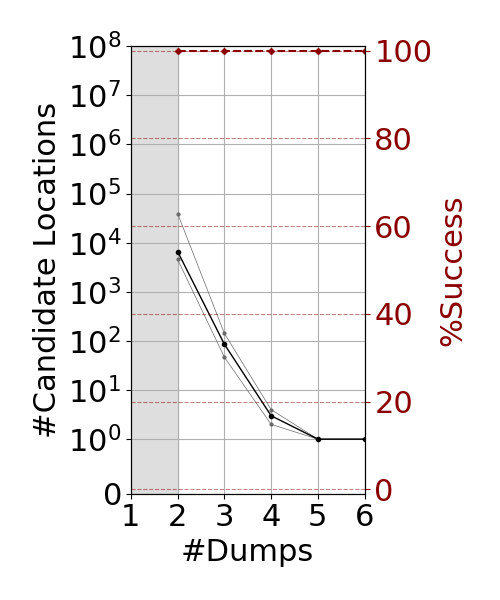}
        \caption{$+$}
    \end{subfigure}
    \hfill
    \begin{subfigure}[b]{0.12\textwidth}
        \centering
        \includegraphics[trim={3.2cm 1.6cm 3.2cm 0.75cm},clip,height=3.3cm]{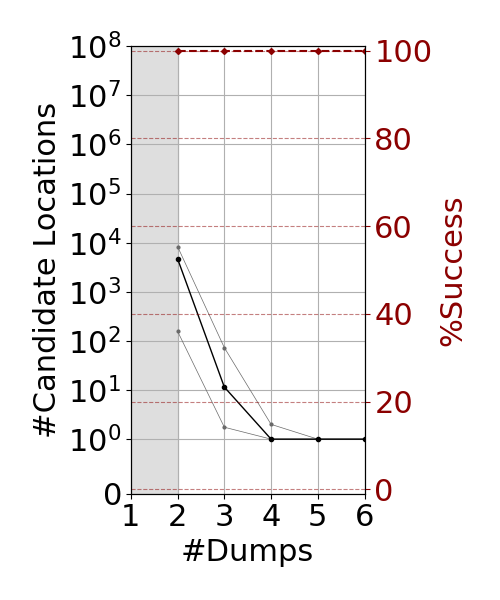}
        \caption{$\xor$}
    \end{subfigure}
    \hfill
    \begin{subfigure}[b]{0.12\textwidth}
        \centering
        \includegraphics[trim={3.2cm 1.6cm 3.2cm 0.75cm},clip,height=3.3cm]{ac_R_LinePlot_add_xor-attack_on_add_xor-obfuscation.png}
        \caption{$+\xor$}
    \end{subfigure}
    \hfill
    \begin{subfigure}[b]{0.12\textwidth}
        \centering
        \includegraphics[trim={3.2cm 1.6cm 3.2cm 0.75cm},clip,height=3.3cm]{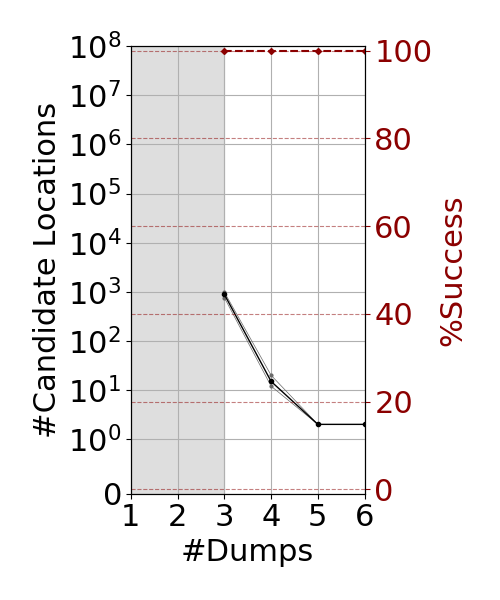}
        \caption{$\xor+$}
    \end{subfigure}
    \hfill
    \begin{subfigure}[b]{0.12\textwidth}
        \centering
        \includegraphics[trim={3.2cm 1.6cm 3.2cm 0.75cm},clip,height=3.3cm]{ac_R_LinePlot_rnc-attack_on_rnc-obfuscation.png}
        \caption{$RNC$}
    \end{subfigure}
    \hspace{-0.25cm}
    \begin{subfigure}[b]{0.08\textwidth}
        \centering
        \includegraphics[trim={9.55cm 1.6cm 0cm 0.75cm},clip,height=3.3cm]{ac_R_LinePlot_rnc-attack_on_rnc-obfuscation.png}
        \vspace{0.48cm}
    \end{subfigure}
    
    \caption{Outcomes of six pruning logics, each applied greedily to the game version protected with the exactly matching protection. (a)--(f) show the results on SuperTux, (g)--(l) show the results on AssaultCube. The legend is the same as in Figure~\ref{fig:base_obfuscation_results}.}
    \label{fig:corresponding_attack_results}
\end{figure*}

Figure~\ref{fig:corresponding_attack_results} shows results for the six targeted pruning logics applied to game versions protected with their corresponding defences. This allows comparing the worst-case strength of the defences, \ie how effectively they delay attackers who somehow know which protection is used and can therefore deploy the optimal attack logic.

On our use cases, these confirm that RNC encoding offers stronger protection than offset/XOR-based encodings alone. Although RNC’s strength has previously been argued descriptively~\cite{RNCorig,RNC}, our approach is, to the best of our knowledge, the first to validate this claim empirically.

\subsubsection{Greedy General Attack Strategies on Different Protections}
\begin{figure*}[t]
    \centering
\scriptsize
    \begin{subfigure}[b]{0.08\textwidth}
        \centering
        \includegraphics[trim={0cm 1.6cm 17.2cm 0.75cm},clip,height=3.3cm]{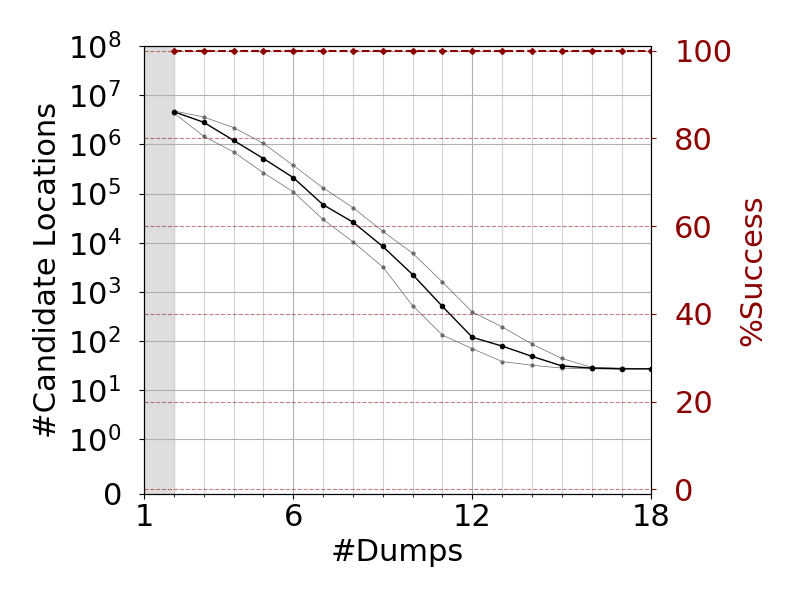}
        \vspace{0.5cm}
    \end{subfigure}
    \hspace{-0.35cm}
    \begin{subfigure}[b]{0.24\textwidth}
        \centering
        \includegraphics[trim={3.2cm 1.6cm 3.2cm 0.75cm},clip,height=3.3cm]{ac_R_LinePlot_change_no_change-attack_on_base-obfuscation.png}
        \caption{Base}
    \end{subfigure}
    \hfill
    \begin{subfigure}[b]{0.24\textwidth}
        \centering
        \includegraphics[trim={3.2cm 1.6cm 3.2cm 0.75cm},clip,height=3.3cm]{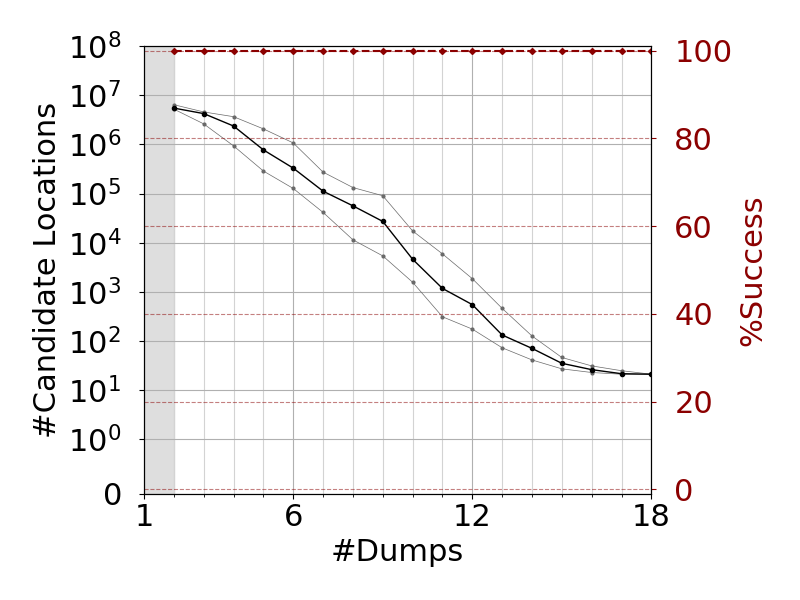}
        \caption{$+$}
    \end{subfigure}
    \hfill
    \begin{subfigure}[b]{0.24\textwidth}
        \centering
        \includegraphics[trim={3.2cm 1.6cm 3.2cm 0.75cm},clip,height=3.3cm]{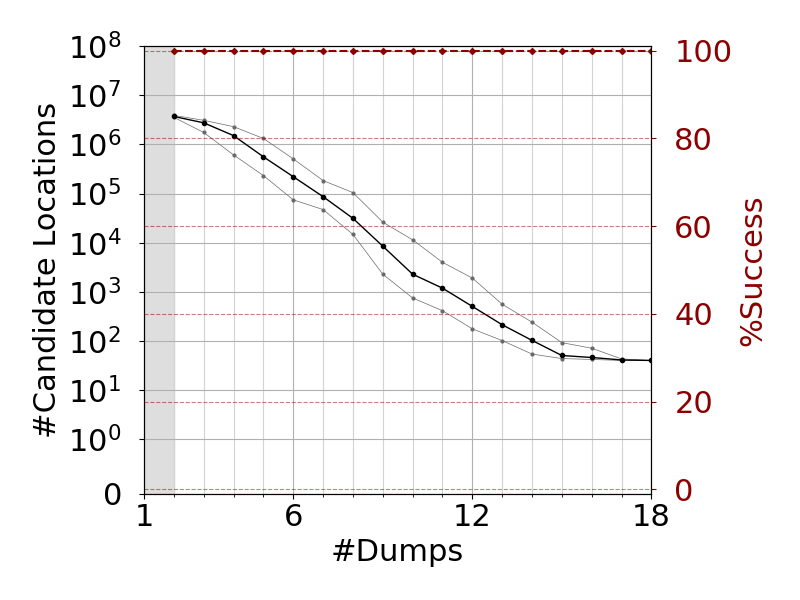}
        \caption{$\xor$}
    \end{subfigure}
    \hspace{0.05cm}
\begin{subfigure}[b]{0.08\textwidth}
        \centering
        \includegraphics[trim={17.1cm 1.6cm 0cm 0.75cm},clip,height=3.3cm]{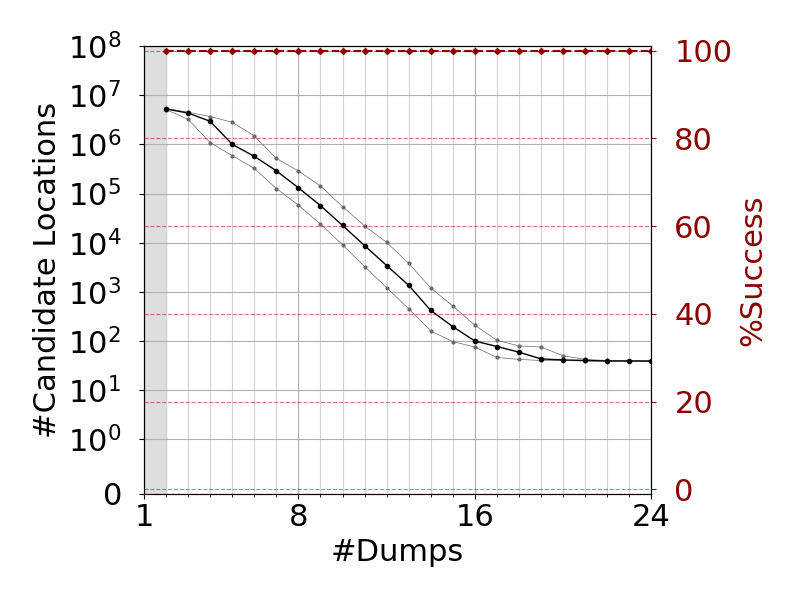}
        \vspace{0.49cm}
    \end{subfigure}

 \begin{subfigure}[b]{0.08\textwidth}
        \centering
        \includegraphics[trim={0cm 1.6cm 17.2cm 0.75cm},clip,height=3.3cm]{ac_R_LinePlot_change_no_change-attack_on_base-obfuscation.png}
        \vspace{0.5cm}
    \end{subfigure}
    \hspace{-0.35cm}
    \begin{subfigure}[b]{0.24\textwidth}
        \centering
        \includegraphics[trim={3.2cm 1.6cm 3.2cm 0.75cm},clip,height=3.3cm]{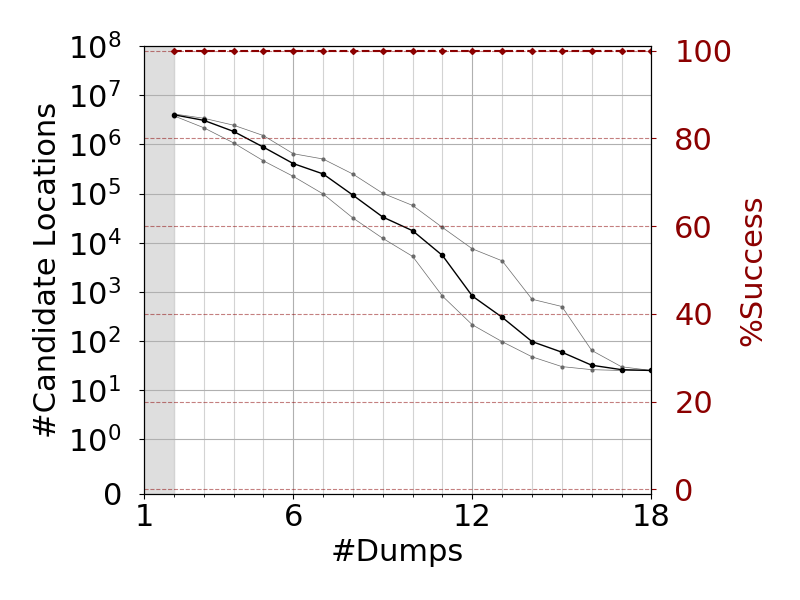}
        \caption{$+\xor$}
    \end{subfigure}
    \hfill
    \begin{subfigure}[b]{0.24\textwidth}
        \centering
        \includegraphics[trim={3.2cm 1.6cm 3.2cm 0.75cm},clip,height=3.3cm]{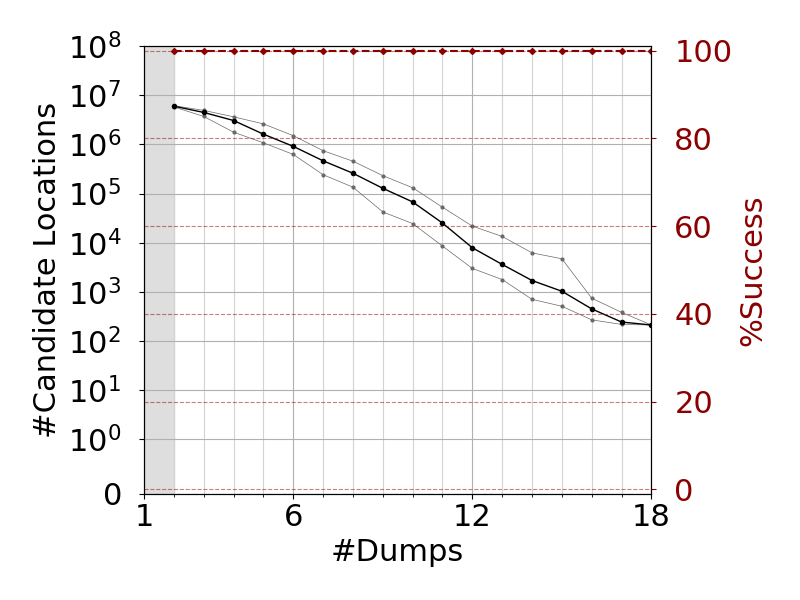}
        \caption{$\xor+$}
    \end{subfigure}
    \hfill
    \begin{subfigure}[b]{0.24\textwidth}
        \centering
        \includegraphics[trim={3.2cm 1.6cm 3.2cm 0.75cm},clip,height=3.3cm]{ac_R_LinePlot_change_no_change-attack_on_rnc-obfuscation.png}
        \caption{$RNC$}
    \end{subfigure}
    \hspace{0.05cm}
    \begin{subfigure}[b]{0.08\textwidth}
        \centering
        \includegraphics[trim={17.1cm 1.6cm 0cm 0.75cm},clip,height=3.3cm]{ac_R_LinePlot_change_no_change-attack_on_rnc-obfuscation.png}
        \vspace{0.49cm}
    \end{subfigure}
    
    \caption{Outcomes of the change/no change attack logic applied to AssaultCube protected with six different protections. The legend is the same as in Figure~\ref{fig:base_obfuscation_results}.}
    \label{fig:change_no_change_attack_results}
\end{figure*}

Figure~\ref{fig:change_no_change_attack_results} presents the models obtained by applying the change/no change attack logic to 6 statically encoded versions of AssaultCube. This is the only logic that consistently reduces the candidate set across all 6 versions and therefore the only one that works reliably for an attacker.\footnote{As shown in Figure~\ref{fig:attack-strategies-po}, the \emph{change} attack logic ---of which we do not show plots of the outcomes--- is less specific than the \emph{change/no change} logic, so its recall is also perfect for all 6 of the static encodings. However, its effectiveness is very low, converging to a number of candidate locations orders of magnitude larger than that of the \emph{change/no change} logic. It is hence not useful for an attacker.} The logic is nevertheless inefficient: candidate reduction is much slower than for the targeted attacks discussed above. Moreover, for some defences, such as the $\xor+$ variant, pruning converges to only about one order of magnitude above the actual number of locations. Thus, although recall is high, precision increases slowly with increased attack effort and may never become high for certain attacks.

Lastly, Figure~\ref{fig:change_no_change_attack_results} also illustrates the method’s dependency on gameplay variations. Since the AssaultCube versions differ only in encoding, we expect memory behaviour to be similar across versions except at the ground-truth location(s). Accordingly, one would expect similar curves for all versions, aside from the contribution of the ground-truth locations. However, noticeable differences appear, most prominently in plot (e). This reflects the influence of gameplay and in-game events on the resulting distributions, which vary between runs. This confirms that the obtained models depend on the defender’s enacted gameplay, underscoring the need to ensure sufficient coverage of possible attacker gameplays.

\subsection{Statistical Models of Effort and Outcomes for Attacks on Dynamic Encodings}
\label{sec:evaluation_dynamic}

This section presents selected results from our simulations of attack strategies deployed against dynamic protections; Appendix~\ref{app:dyn-experiments-combos} summarizes all evaluated encoding–attack combinations in this experimental set.

\begin{figure*}[t]
    \centering
\scriptsize
    \begin{subfigure}[b]{0.06\textwidth}
        \centering
        \includegraphics[trim={0.8cm 1.6cm 17.2cm 0.75cm},clip,height=3.2cm]{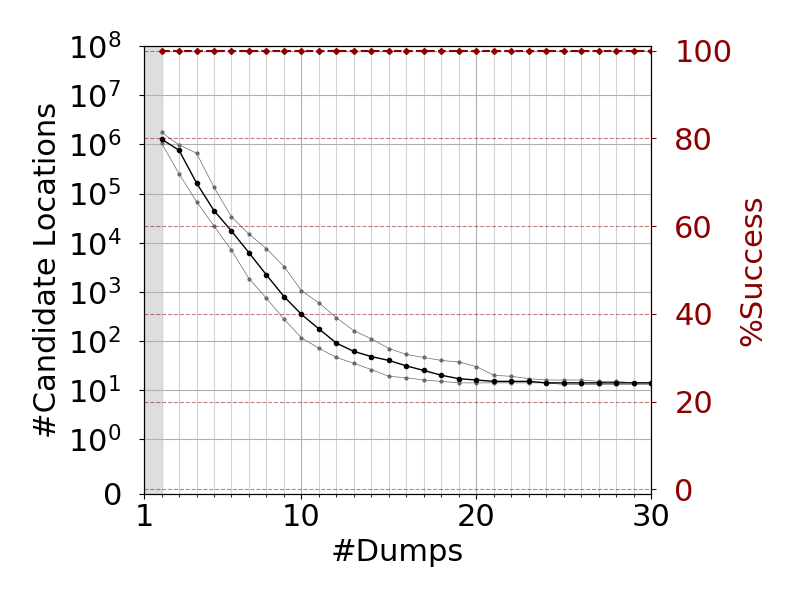}
        \vspace{0.5cm}
    \end{subfigure}
    \hspace{-0.35cm}
    \begin{subfigure}[b]{0.24\textwidth}
        \centering
        \includegraphics[trim={3.2cm 1.6cm 3.2cm 0.75cm},clip,height=3.2cm]{st_R_LinePlot_change_no_change-attack_on_xor_UoW-obfuscation.png}
        \caption{\textit{Change/no change}}
    \end{subfigure}
    \hspace{0.1cm}
    \begin{subfigure}[b]{0.24\textwidth}
        \centering
        \includegraphics[trim={3.2cm 1.6cm 3.2cm 0.75cm},clip,height=3.2cm]{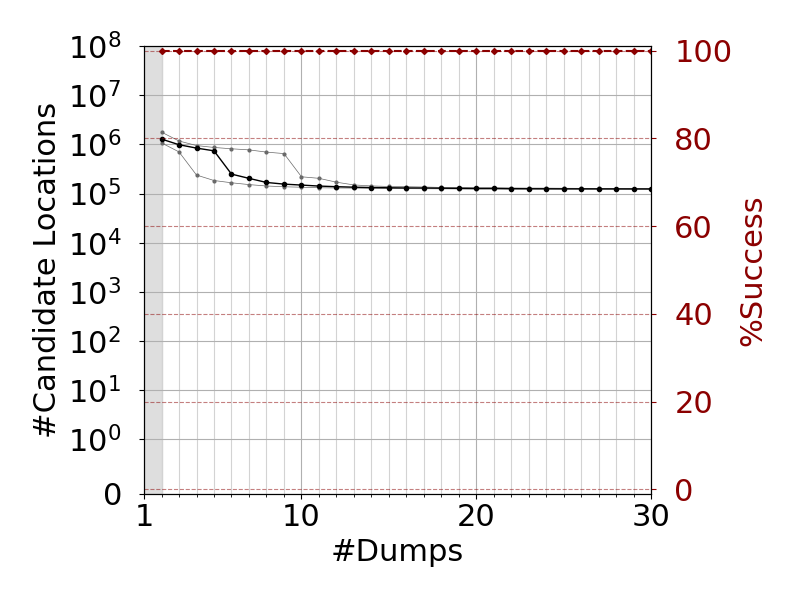}
        \caption{\textit{Change}}
    \end{subfigure}
    \hspace{-0.05cm}
    \begin{subfigure}[b]{0.08\textwidth}
        \centering
        \includegraphics[trim={17.1cm 1.6cm 0cm 1cm},clip,height=3.2cm]{st_R_LinePlot_change-attack_on_xor_UoW-obfuscation.png}
        \vspace{0.49cm}
    \end{subfigure}
\begin{subfigure}[b]{0.34\textwidth}
        \centering
        \includegraphics[trim={0.9cm 1.6cm 0.9cm 0.8cm},clip,height=3.2cm]{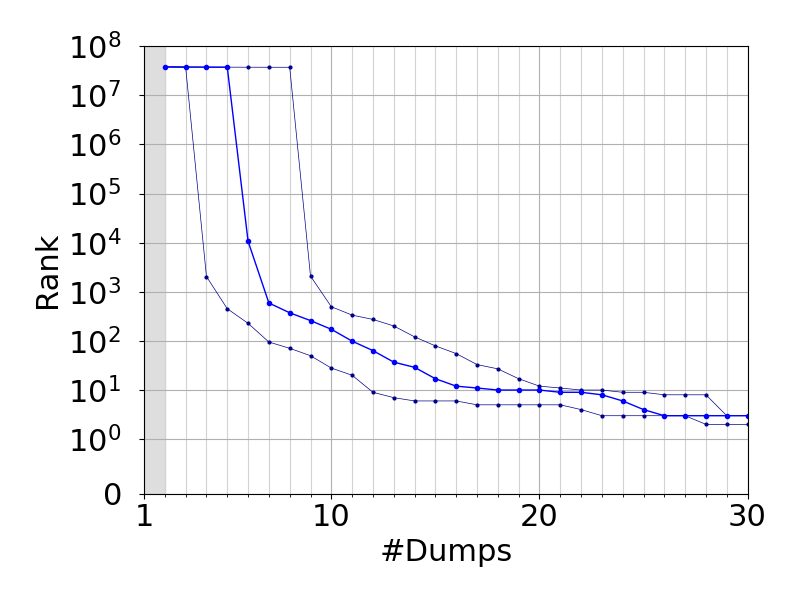}
        \caption{Statistical $\xor$}
    \end{subfigure}

    \vspace{0.2cm}

    $\uparrow$: Update on Write (UoW) \quad\quad $\downarrow$: Update on Read (UoR)

    \vspace{0.2cm}

    \begin{subfigure}[b]{0.06\textwidth}
        \centering
        \includegraphics[trim={0.8cm 1.6cm 17.2cm 0.75cm},clip,height=3.2cm]{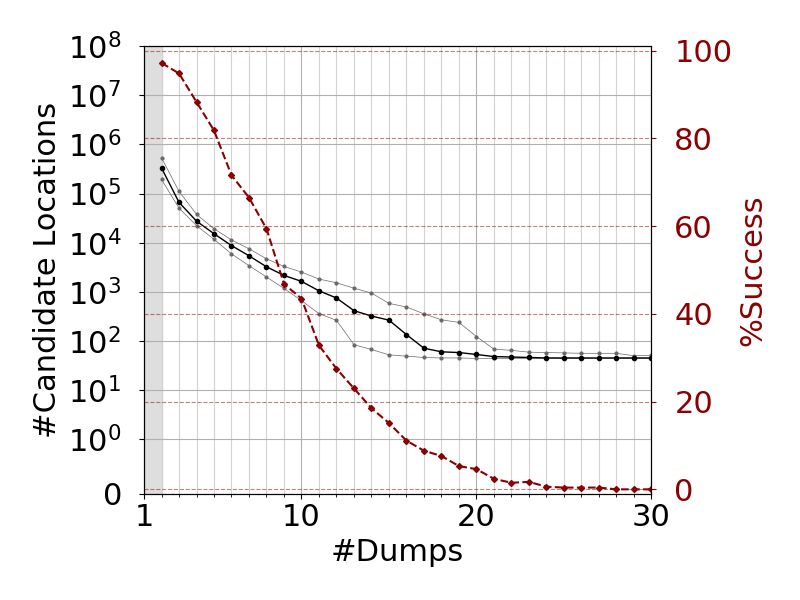}
        \vspace{0.5cm}
    \end{subfigure}
    \hspace{-0.35cm}
    \begin{subfigure}[b]{0.24\textwidth}
        \centering
        \includegraphics[trim={3.2cm 1.6cm 3.2cm 0.75cm},clip,height=3.2cm]{st_R_LinePlot_change_no_change-attack_on_xor_UoR-obfuscation.png}
        \caption{\textit{Change/no change}}
    \end{subfigure}
    \hspace{0.1cm}
    \begin{subfigure}[b]{0.24\textwidth}
        \centering
        \includegraphics[trim={3.2cm 1.6cm 3.2cm 0.75cm},clip,height=3.2cm]{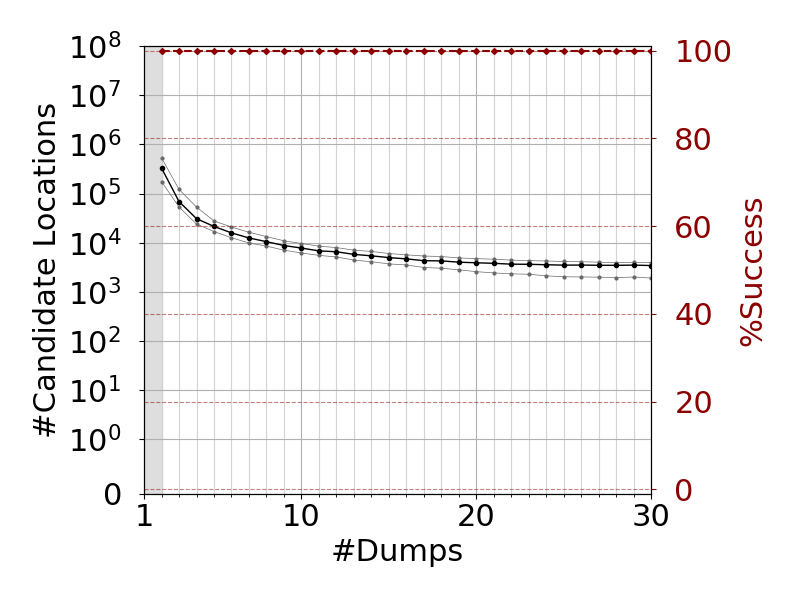}
        \caption{\textit{Change}}
    \end{subfigure}
    \hspace{-0.05cm}
    \begin{subfigure}[b]{0.08\textwidth}
        \centering
        \includegraphics[trim={17.1cm 1.6cm 0cm 1cm},clip,height=3.2cm]{st_R_LinePlot_change-attack_on_xor_UoR-obfuscation.png}
        \vspace{0.49cm}
    \end{subfigure}
\begin{subfigure}[b]{0.34\textwidth}
        \centering
        \includegraphics[trim={0.9cm 1.6cm 0.9cm 0.8cm},clip,height=3.2cm]{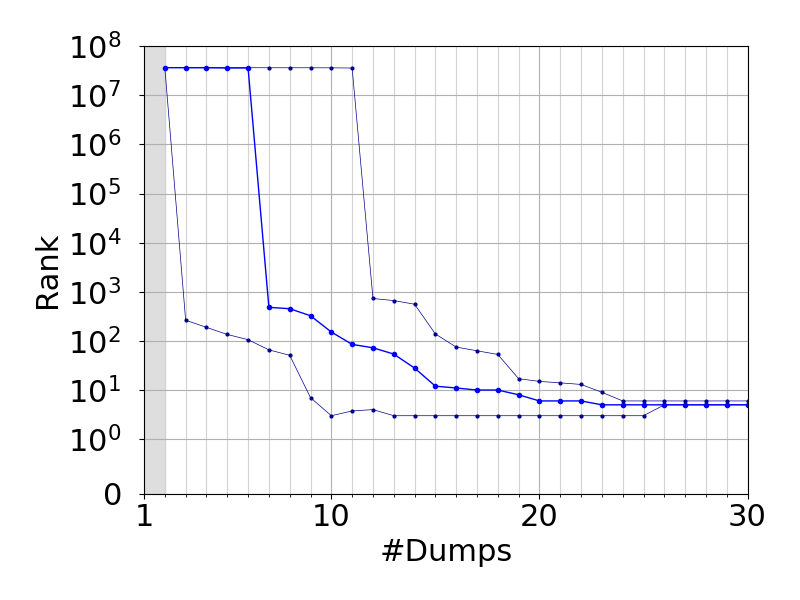}
        \caption{Statistical $\xor$}
    \end{subfigure}

    \caption{Outcomes of different relevant attack strategies applied to SuperTux protected with two dynamic versions of the $\xor$-encoding. For the greedy attacks, the legend is the same as in Figure~\ref{fig:base_obfuscation_results}. For the statistical attack the blue lines show the $P_{25}$, $P_{50}$, and $P_{75}$ percentiles of the rank of the ground-truth location, which is an estimation $\hat{\phi}_{A_{2,n}}^{P_v}$ of the expected effort that an attacker will have to invest in a second step of the attack, like the remaining number of candidates is for a greedy attack strategy.}
    \label{fig:dyn_xor_encoding_results}
\end{figure*}

Figure~\ref{fig:dyn_xor_encoding_results} shows distributions for greedy \emph{change} and \emph{change/no change} attacks and the \emph{statistical $\xor$}-attack on two SuperTux versions using dynamic-$\xor$ encoding with UoW and UoR policies. 
Note the different Y-axes for the statistical $\xor$ attack. As this is a statistical attack, these charts do not show remaining candidate counts after pruning, but rather the rank of the ground-truth location(s) among all memory locations ordered by conformity to the pruning logic. Assuming attackers inspect locations in increasing rank order, rank measures the required effort; assuming an attacker inspecting only top-ranked locations, it also reflects success, as discussed in Section~\ref{subsec:result_gathering}. Analogous to the candidates count in greedy attacks, the rank distribution after $n$ scans corresponds to $\hat{\phi}_{A_{2,n}}^{P_v}$. This demonstrates our method’s versatility, enabling comparison between fundamentally different attack modes, where reasoning about their trade-offs would otherwise be practically impossible.

While we avoid general recommendations in this paper about which encodings or attacks work best, Figure~\ref{fig:dyn_xor_encoding_results} yields several insights. First, plot (d) shows that the \emph{change/no change} attack fails against dynamic $\xor$ UoW, despite succeeding for all other considered encodings. This is expected, as updating encoding parameters while the in-game value remains unchanged causes memory values to change without screen changes, violating the attack’s assumptions. Second, the \emph{change} attack succeeds on both dynamic encodings (as for static ones), but remains impractically ineffective, leaving too many candidates to inspect. Third, the statistical $\xor$ attack becomes highly effective with sufficient scans, leading to a rank an order of magnitude lower than the candidate counts of \emph{change/no change}. This indicates that statistical attacks can be effective against certain dynamic encodings and should be modelled by defenders.

Overall, the ability to derive such insights by instantiating our method further demonstrates its utility for defender decision support.

\noindent

\subsection{Simulation Running Times}
\label{sec:results:subsec:duration}

\begin{figure}[t]

\begin{subfigure}[b]{0.72\linewidth}
        \centering
        \includegraphics[trim={0.48cm 1cm 1.88cm 0cm},clip,height=6.8cm,keepaspectratio]{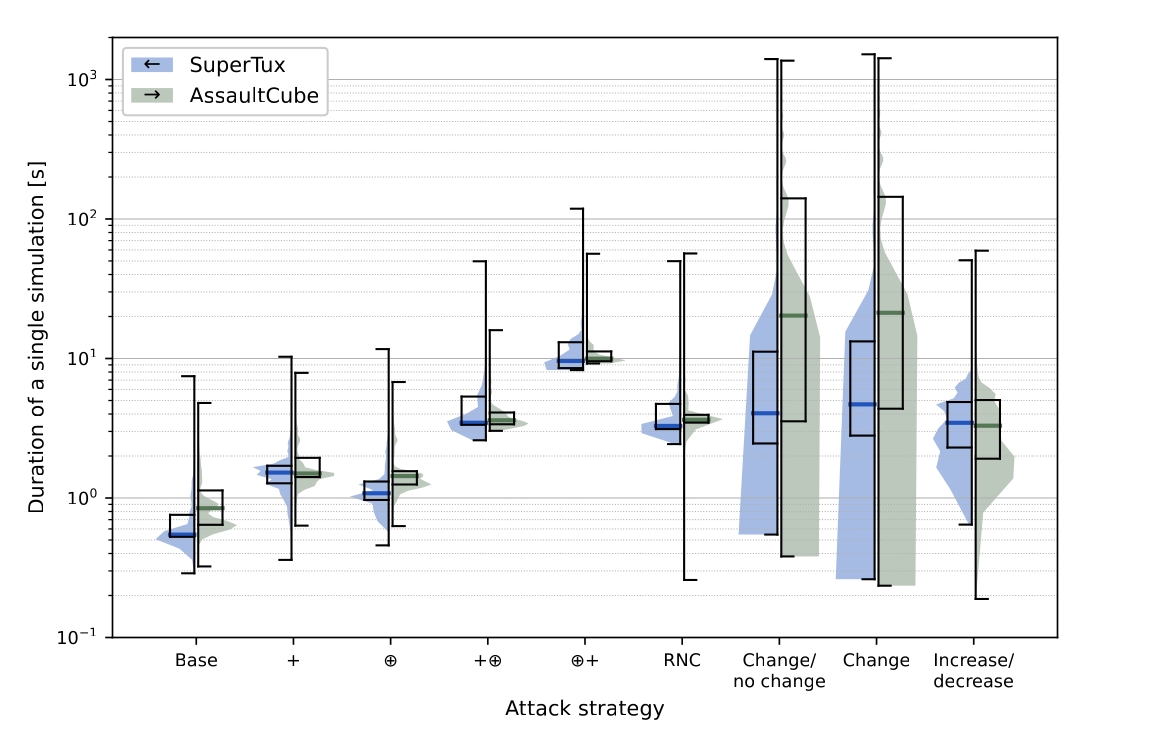}
        \caption{Static encodings}
        \label{fig:simulation_times_set_1}
    \end{subfigure}
    \hspace{0.05cm}
    \begin{subfigure}[b]{0.26\textwidth}
        \centering
        \includegraphics[trim={1.88cm 1cm 1.88cm 0cm},clip,height=6.8cm,keepaspectratio]{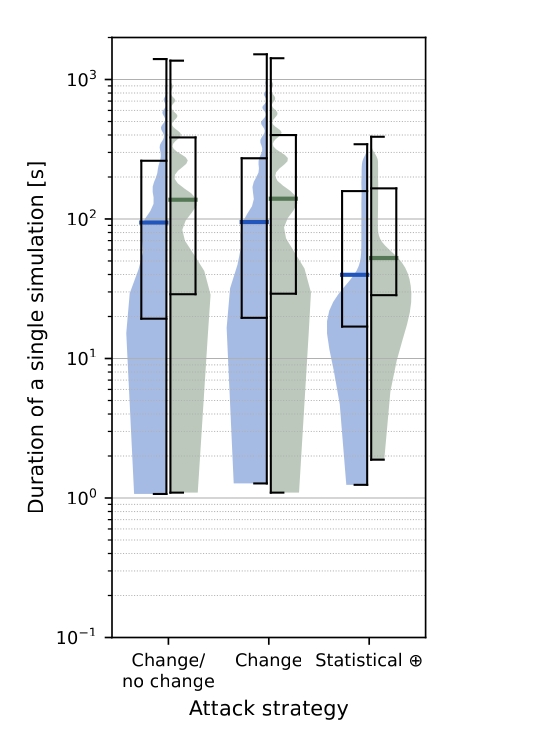}
        \caption{Dynamic encodings}
        \label{fig:simulation_times_set_2}
    \end{subfigure}

    \caption{The distribution of the time taken by each attack simulation for the different attack strategies and pruning logics in both sets of experiments. The whiskers of the boxplots correspond to $P_{0}$ and $P_{100}$.}
    \label{fig:simulation_times}
\end{figure}

For the sake of completeness, this section reports simulation running times. Our proof-of-concept implementation was rapid-prototyped in Python and is not optimised for performance. Each attack is simulated independently, requiring dumps to be loaded and scanned once per attack logic, so data loading---dominating execution time--- is not amortized. The reported times should therefore be interpreted as upper bounds.

For static encodings, we ran experiments on an Ubuntu 22.04 workstation with an Intel Xeon w9-3575X processor and 256 GB DDR5 RAM, without using GPU acceleration. Simulations were executed in parallel, with up to 54 concurrent runs, \ie the number of encoding–attack combinations.
Figure~\ref{fig:simulation_times_set_1} shows execution-time distributions per attack logic in this experimental set.

For dynamic encodings, experiments ran on an Ubuntu 24.04.03 workstation with an AMD Threadripper Pro 7985WX processor and 512 GB DDR5 RAM, again without GPU acceleration. Up to 128 simulations ran concurrently. Figure~\ref{fig:simulation_times_set_2} shows execution-time distributions for this experimental set.

For conciseness, each violin plot aggregates all simulations of a pruning logic, spanning different dump sequence lengths and versions of both games with different encodings.

Both in Figure~\ref{fig:simulation_times_set_1} and Figure~\ref{fig:simulation_times_set_2} the slowest simulations are orders of magnitude slower than the fastest simulations for the same pruning logic. This variance arises from (i) aggregating results from different sequence lengths, (ii) early termination when greedy attacks fail (\ie when no candidate locations remain), and (iii) large fluctuations in dump loading times due to resource contention among parallel simulations. Indeed, while parallelism reduced total wall-clock time, it also introduced substantial latency variability when multiple workers needed to load the same dumps concurrently.

 \changed{
\section{Empirical Validation of Statistical Models}
\label{sec:empirical-validation}

To assess whether these distributions obtained through simulation in the previous section reflect realistic attacker behaviour, we conducted a small empirical experiment with human participants. The goal of this experiment was to validate whether the distributions predicted by our simulation method are close to the distributions obtained when real users perform the same localisation task with the same automated support. The experiment was limited to SuperTux, to number of greedy strategies, and to static versions of the obfuscations, i.e., where the used masks, offsets, and moduli do not change dynamically. 

\subsection{Memhunt}
Because CheatEngine does not provide out-of-the-box support for the different pruning strategies that target the different data obfuscation strategies we evaluated in the previous section, we created our own CheatEngine like tool in Python, called \texttt{memhunt}. It provides functionality to (i) attach to a running game, (ii) take dumps from the game's heap and provide the on-screen values for them, (iii) perform on-demand scans of all already taken dumps using various greedy pruning logics (iv) show the resulting candidate locations of those logics, (v) overwrite data in memory locations among those candidates, and (vi) reset the whole dump and scan history. Whenever some pruning logic is no longer applicable, e.g., because the target resource value does not increase with constant stride 1 in the taken dumps, the tool emits a warning. 

Importantly, all available pruning logics are executed in parallel: the attacker does not need to choose which one they execute, but they have the freedom to choose which logics' outcomes they rely on or neglect to decide on their next action, such as when to take the next dump, for which resource value, and when to execute a new scan. The pruning is optimized in the sense that later scans only visit candidate locations remaining after previous scans. This allows to free parts of earlier dumps that are no longer of interest, thus avoiding out-of-memory issues. Moreover, it speeds up subsequent scans, thus making frequent scanning a perfectly viable option, rather than forcing attackers to take multiple dumps in between scans to optimize their time usage. 

For the sake of our experiment, the memhunt tool was instrumented to produce logs for post-experiment analysis. The game was also instrumented such that it logged the ground-truth location(s) of the coin value, enabling memhunt to compute and log the recall of each pruning strategy after each scan. Participants were instructed not to inspect those logs, and not to use any other static or dynamic analysis tools such as debuggers. This way, we placed them in the role of rational attackers who try to minimise their effort while relying only on the information normally available during the localisation phase.

The  pruning logics implemented by memhunt and enabled during the experiments were the following: base, $+$, $\xor$, $+\xor$, $\xor +$, and RNC. 

\subsection{Participants, Treatment, and Assignment}
Seven participants took part voluntarily in the experiment. All were students enrolled in a PhD programme in cybersecurity at one of our institutions, and as such employees of our institutions. No personal identifiable information was logged. 

Each participant worked on eleven binaries: one unobfuscated version and two versions for each of the five obfuscation families \(+\), \(\xor\), \(+\xor\), \(\xor+\), and RNC. The two binaries of each obfuscation family used different protection parameters, such as different addition offsets, XOR masks, or RNC moduli; the concrete values are reported in Appendix~\ref{app:empirical_obf_parameters}. Participants were instructed to analyse the binaries in a precise order, but a different order was assigned to each participant, to reduce ordering effects due to participants becoming more familiar with the tool or with the game over time. 

To avoid results from being biased by participants gradually learning basic properties of SuperTux or of the tool, participants were given the same general background information for all binaries. In particular, they were informed that SuperTux often stores the number of coins twice, as explained in Section~\ref{sec:games}. They were also told that only the actual gameplay value may be obfuscated, whereas the duplicate remains unobfuscated, and that the relevant values are aligned four-byte integers. This mirrors a scenario in which attackers have already analysed earlier versions of the same game and have learned general information about its memory layout, but not the exact layout, and not the concrete target location in a new execution.

In the empirical experiment, participants were asked to complete the localisation task rather than to stop after a predefined number of dumps, scans, or remaining candidate locations. They therefore decided autonomously when the candidate locations were sufficiently pruned to start testing candidate locations. In practice, this meant that they first used \texttt{memhunt} to reduce the candidate set, and then overwrote selected candidate locations with a new coin value to check whether the change was reflected on screen. The observed stopping points hence reflect participants' own assessment of when further pruning was no longer worth the additional effort relative to direct validation. 

\subsection{Data Analysis}
We compared the empirical distributions obtained from the participants' logs with the corresponding simulated distributions. As in the previous sections, we focus on two quantities: the number of remaining candidate locations, which estimates the effort left for the subsequent validation phase of the attack, and recall, which indicates whether the ground-truth location is still among the candidates. Since both quantities evolve with the number of dumps, we compare empirical and simulated distributions step by step, rather than only at the final step.

To summarise the difference between empirical and simulated distributions in a compact way, we use the Hellinger distance~\cite{beran1977minimum}. Given two discrete distributions \(P\) and \(Q\), this distance is defined as
\[
H(P,Q)=\frac{1}{\sqrt{2}}\sqrt{\sum_i\left(\sqrt{P_i}-\sqrt{Q_i}\right)^2}.
\]
It ranges from 0 to 1, where 0 means that the two distributions are identical and larger values indicate increasing disagreement.

\subsection{Results}

\begin{table}[t]
\centering
\caption{Hellinger Distance computed on the distributions of the number of remaining candidate locations at each step.}
\label{tab:hellinger-candidates}
\begin{tabular}{ccccccc}
\toprule
\multirow{2.5}{*}{\shortstack{\textbf{Obfuscation}\\\textbf{strategy}}}
&
\multicolumn{6}{c}{\textbf{Location pruning logic}} \\
\cmidrule(lr){2-7}
& Base & $+$ & $\xor$ & $+\xor$ & $\xor+$ & RNC \\
\midrule
Base & 0.000000 & 0.071694 & 0.027393 & 0.006506 & 0.046119 & 0.125275 \\
$+$ & 0.000000 & 0.015251 & 0.029551 & 0.020168 & 0.028360 & 0.158577 \\
$\xor$ & 0.000000 & 0.060104 & 0.042025 & 0.007645 & 0.086759 & 0.288250 \\
$+\xor$ & 0.000000 & 0.048150 & 0.045966 & 0.022095 & 0.192141 & 0.240640 \\
$\xor+$ & 0.000000 & 0.048117 & 0.035204 & 0.008383 & 0.082623 & 0.324555 \\
RNC & 0.117094 & 0.105341 & 0.093394 & 0.024827 & 0.149309 & 0.193407 \\
\bottomrule
\end{tabular}
\end{table}

Table~\ref{tab:hellinger-candidates} reports the Hellinger distances for the distributions of the number of remaining candidate locations. Overall, the distances are small. The average value over the 36 protection--attack combinations is 0.076, the median is 0.046, and 26 out of 36 values are below 0.1. This indicates that, in most cases, the empirical candidate-count distributions are close to the simulated ones. The agreement is especially strong for the base, \(+\), \(\xor\), and \(+\xor\) pruning logics, whose column averages are 0.02, 0.058, 0.046, and 0.015, respectively. The values in the base column are zero for all non-RNC obfuscations, meaning that the model accurately predicts the candidate-count behaviour of the base matcher in those cases. The diagonal of Table~\ref{tab:hellinger-candidates} is particularly relevant, as it compares each obfuscation with the attack logic tailored to it. These distances are low for all matched pairs except RNC. Thus, for five of the six matched protection--attack combinations, the empirical distributions are very close to the simulated ones. The RNC/RNC case shows a larger discrepancy, but it is still below the largest off-diagonal values and remains compatible with the qualitative trend predicted by the model. The largest deviations in Table~\ref{tab:hellinger-candidates} are concentrated in the RNC attack column, whose average distance is 0.222. The largest individual distances are obtained when applying the RNC pruning logic to \(\xor\)-, \(\xor+\)-, and \(+\xor\)-protected binaries. This suggests that the candidate-count behaviour of the RNC matcher is harder to predict than that of the other matchers. This is plausible, because RNC matching depends on arithmetic relations between the observed resource values and the selected moduli, so small differences in the concrete parameterisation or in the dump sequence can noticeably affect the remaining candidate set.

\begin{figure*}[t]
    \centering
\scriptsize
    \begin{subfigure}[b]{0.08\textwidth}
        \centering
        \includegraphics[trim={0cm 1.6cm 6.2cm 0.75cm},clip,height=3.3cm]{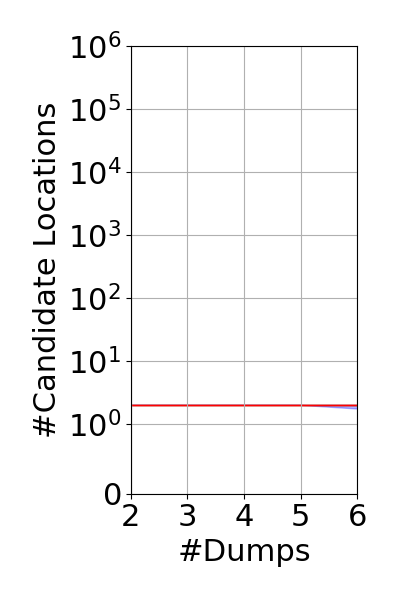}
        \vspace{0.21cm}
    \end{subfigure}
    \hspace{-0.45cm}
    \begin{subfigure}[b]{0.12\textwidth}
        \centering
        \includegraphics[trim={3.1cm 1.6cm 0.5cm 0.75cm},clip,height=3.3cm]{empirical_Exact_on_Base_empirical_no_p50s.png}
        \caption{Base}
    \end{subfigure}
    \hfill
    \begin{subfigure}[b]{0.12\textwidth}
        \centering
        \includegraphics[trim={3.1cm 1.6cm 0.5cm 0.75cm},clip,height=3.3cm]{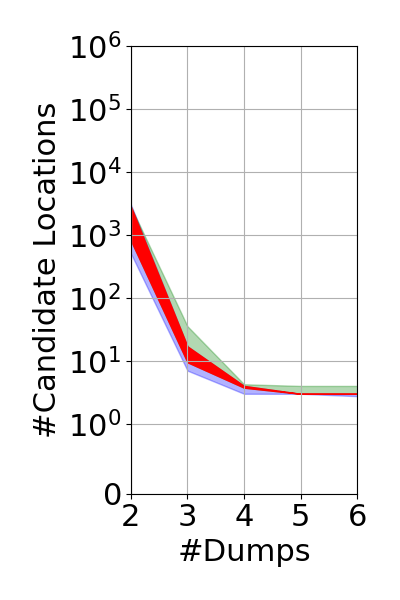}
        \caption{$+$}
    \end{subfigure}
    \hfill
    \begin{subfigure}[b]{0.12\textwidth}
        \centering
        \includegraphics[trim={3.1cm 1.6cm 0.5cm 0.75cm},clip,height=3.3cm]{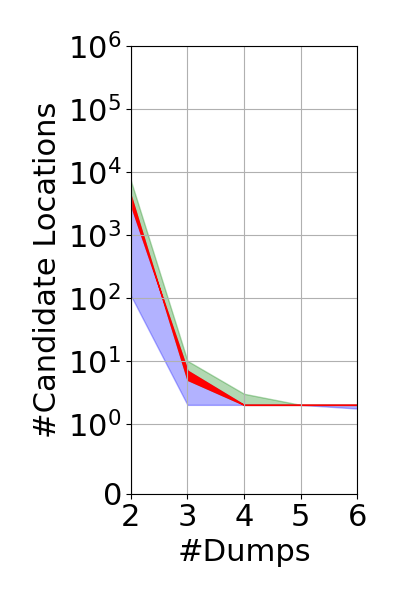}
        \caption{$\xor$}
    \end{subfigure}
    \hfill
    \begin{subfigure}[b]{0.12\textwidth}
        \centering
        \includegraphics[trim={3.1cm 1.6cm 0.5cm 0.75cm},clip,height=3.3cm]{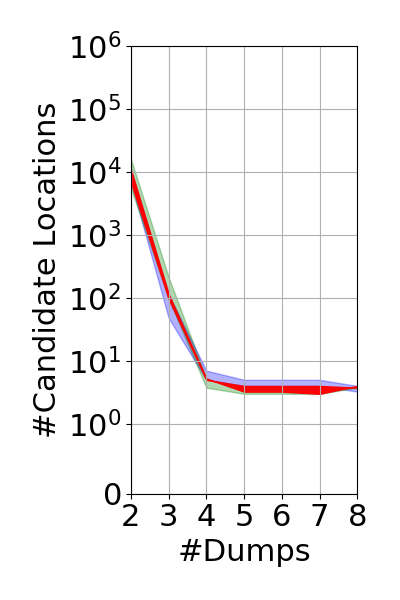}
        \caption{$+ \xor$}
    \end{subfigure}
    \hfill
    \begin{subfigure}[b]{0.12\textwidth}
        \centering
        \includegraphics[trim={3.1cm 1.6cm 0.5cm 0.75cm},clip,height=3.3cm]{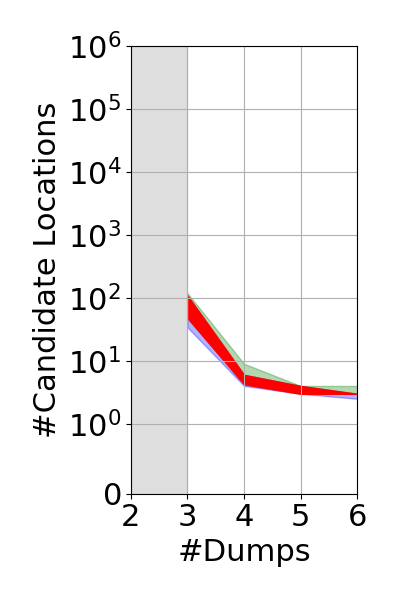}
        \caption{$\xor +$}
    \end{subfigure}
    \hfill
    \begin{subfigure}[b]{0.12\textwidth}
        \centering
        \includegraphics[trim={3.1cm 1.6cm 0.5cm 0.75cm},clip,height=3.3cm]{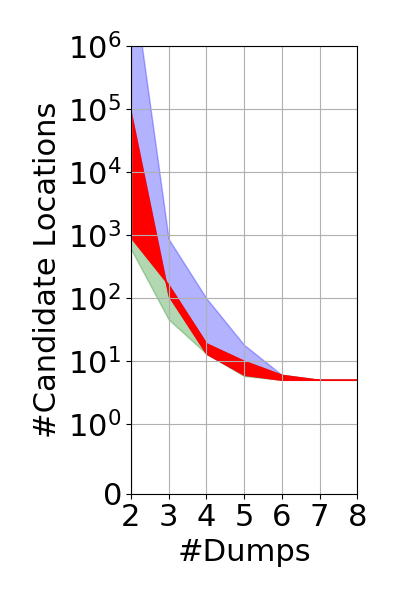}
        \caption{RNC}
    \end{subfigure}
    
    \caption{Outcomes of six pruning logics, each applied greedily to the game version protected with the exactly matching protection. The green region indicates the area between the p25 and the p75 values obtained by \emph{participants}, while the blue one indicates the area between the same values computed on our simulation. The region in red represents the intersection of the regions. The axes are the same of Figure~\ref{fig:base_obfuscation_results}.}
\label{fig:empirical_superimposition}
\end{figure*}

Figure~\ref{fig:empirical_superimposition} visualises the comparison for the six matched protection--attack combinations. The blue regions represent the interquartile ranges obtained from the simulations, the green regions the interquartile ranges obtained from the participants, and the red regions their intersections. The plots confirm the interpretation of Table~\ref{tab:hellinger-candidates}. For base, \(+\), \(\xor\), \(+\xor\), and \(\xor+\), the simulated and empirical interquartile ranges largely overlap. In all these cases, both the model and the participants exhibit a rapid decrease in the number of candidate locations after the first useful dumps, followed by convergence to a small candidate set. The RNC plot shows the largest difference, but still captures the same qualitative behaviour: candidate reduction is slower than for most other matched attacks, and more dumps are needed before the candidate set approaches its final range.

The interpretation of the RNC results should take into account the parameter sensitivity already discussed in Section~\ref{sec:evaluation-greedy}. In the simulation experiments presented there, the specific moduli used in the RNC-protected SuperTux binary and the observed coin values caused the residues to not wrap around the moduli. Consequently, $+$-compatible pruning logics could also succeed on that RNC-protected binary. In the empirical experiment, however, the two RNC-protected binaries used different moduli, that did not induce the same behaviour. The larger RNC-related distances therefore do not indicate a generic failure of the modelling approach. Rather, they reinforce the observation that RNC effectiveness depends on the concrete moduli and on the resource values sampled by the attacker.

\begin{table}[t]
\centering
\caption{Hellinger Distance computed on the distributions of the recalls at each step.}
\label{tab:hellinger-recall}
\begin{tabular}{ccccccc}
\toprule
\multirow{2.5}{*}{\shortstack{\textbf{Obfuscation}\\\textbf{strategy}}}
&
\multicolumn{6}{c}{\textbf{Location pruning logic}} \\
\cmidrule(lr){2-7}
& Base & $+$ & $\xor$ & $+\xor$ & $\xor+$ & RNC \\
\midrule
Base & 0.000000 & 0.000000 & 0.000000 & 0.000000 & 0.000000 & 0.000000 \\
$+$ & 0.000000 & 0.000000 & 0.113508 & 0.000000 & 0.000000 & 0.000000 \\
$\xor$ & 0.000000 & 0.000000 & 0.000000 & 0.000000 & 0.000000 & 0.000000 \\
$+\xor$ & 0.000000 & 0.409817 & 0.318250 & 0.000000 & 0.391900 & 0.000000 \\
$\xor+$ & 0.000000 & 0.000000 & 0.323467 & 0.564626 & 0.000000 & 0.000000 \\
RNC & 0.000000 & 0.436279 & 0.338559 & 0.436279 & 0.456725 & 0.000000 \\
\bottomrule
\end{tabular}
\end{table}

\begin{table}[t]
\centering
\caption{Percentage of samples that had recall equal to our expected recall for the chosen attack-obfuscation combination at the last step. As we discussed, the only case where this does not occur is the RNC protected files when attacked with an ADD compatible pruning logic. This is because the supertux version considered in the previous experiments used coprimes and coin values that made so that the moduli were monotonically increasing. This is not the case for the other binaries used in the empirical experiment.}
\label{tab:final-recall-match}
\begin{tabular}{ccccccc}
\toprule
\multirow{2.5}{*}{\shortstack{\textbf{Obfuscation}\\\textbf{strategy}}}
&
\multicolumn{6}{c}{\textbf{Location pruning logic}} \\
\cmidrule(lr){2-7}
& Base & $+$ & $\xor$ & $+\xor$ & $\xor+$ & RNC \\
\midrule
Base & 100\% & 100\% & 100\% & 100\% & 100\% & 100\% \\
$+$ & 100\% & 100\% & 100\% & 100\% & 100\% & 100\% \\
$\xor$ & 100\% & 100\% & 100\% & 100\% & 100\% & 100\% \\
$+\xor$ & 100\% & 100\% & 100\% & 100\% & 100\% & 100\% \\
$\xor+$ & 100\% & 100\% & 100\% & 100\% & 100\% & 100\% \\
RNC & 100\% & 0\% & 100\% & 0\% & 0\% & 100\% \\
\bottomrule
\end{tabular}
\end{table}

Tables~\ref{tab:hellinger-recall} and~\ref{tab:final-recall-match} report the corresponding validation results for recall. Table~\ref{tab:hellinger-recall} contains Hellinger distances computed on the step-wise recall distributions. Since recall is binary, these values measure whether the model and the empirical observations agree on the probability that the true location is retained after each number of dumps. The results are strong: 26 out of 36 values are exactly zero, and all diagonal values are zero. Thus, for every matched protection--attack combination, the simulated and empirical recall distributions coincide. This is an important result, because it means that for the attacks most relevant to evaluating each protection, the model correctly predicts not only the size of the candidate set reasonably well, but also whether the ground-truth location remains among the candidates. The non-zero values in Table~\ref{tab:hellinger-recall} are concentrated in mismatched protection--attack combinations, where the pruning logic makes assumptions that are not guaranteed by the deployed encoding. The largest discrepancy is observed for the \(\xor+\)-protected binaries attacked with the \(+\xor\) logic, with a distance of 0.565. Other non-zero values occur for \(+\xor\)-protected binaries attacked with \(+\), \(\xor\), or \(\xor+\) logic, and for RNC-protected binaries attacked with \(+\)-compatible or \(\xor\)-compatible logics. Table~\ref{tab:final-recall-match} complements this step-wise analysis by considering the final recall outcome. It reports the percentage of samples for which the empirical final recall matches the expected final recall for the corresponding protection--attack combination. The results show that final recall is correctly predicted in all cases except for RNC-protected binaries attacked with ADD-compatible pruning logics. This exception is again explained by the different RNC moduli used in the empirical experiment with respect to those used when producing the statistical models.

\subsection{Conclusions}

Overall, the empirical evaluation supports the external validity of the statistical models. The candidate-count distributions produced by the simulations are close to those obtained from human participants for most protection--attack combinations, and the matched protection--attack cases are particularly well predicted. Recall is even more robust: all matched protection--attack combinations have identical simulated and empirical recall distributions, and the final recall behaviour is correctly predicted except in the RNC cases affected by deliberate parameter variation. The observed discrepancies therefore do not contradict the modelling approach. Instead, they confirm one of the main observations of the previous sections: localisation outcomes can depend on subtle interactions between protection parameters, resource-value ranges, and dump-selection behaviour. The value of the simulation method is precisely that it makes such interactions visible and measurable.
}
 \section{Discussion}
\label{sec:discussion}

The evaluation demonstrated the simulation method's utility for defenders mitigating modelled resource localisation attacks, but it also has limitations and disadvantages.

First, as discussed previously, outcomes can be highly sensitive to interactions between program properties, attack parameters, and protection parameters, such as dump frequency versus XOR-mask change frequency, and resource values versus RNC-encoding moduli. The method hence requires cautious deployment.

The most important limitation is that  simulating multiple varying localisation attack executions based on a single game execution is limited to attack steps requiring only a single execution by attackers, and to scenarios where defenders can execute required interactions far more frequently than attackers without altering game state enough to impede further interactions or change outcomes. In the meta-model (Section~\ref{sec:metamodel}), the modelled execute function must be amenable to executing multiple step variations rather than implementing exact attack steps as attackers would devise for optimising their productivity.

This may not hold for other attack strategies or steps. For example, step 2 of complete resource cheating attacks (Section~\ref{sec:background:subsec:hackstrategies}) validates remaining candidate locations by overwriting values (\eg with a debugger) and observing effects. This likely cannot execute multiple times within one game execution, as overwriting wrong data likely destroys game state. While it might not destroy attacker-relevant state, and workarounds like check-pointing to roll back state may exist~\cite{rr}, modelling such steps with our method becomes much more complex, if feasible at all.

Although applicability is hence limited to certain attack step categories, we consider the method widely applicable. While we experimented with open-source games potentially differing from commercial games in architecture, design, and implementation, and ran them on Linux (unpopular among commercial game players), our method is applicable to commercial games. Indeed, the game hacking strategy from Section~\ref{sec:background:subsec:hackstrategies}, whose step 1 we modelled, is used by cheat creators on commercial games~\cite{game_hacking}, and CheatEngine~\cite{cheatengine} implements memory dump/scan functionality we simulate because cheat creators use it on commercial games. We did not study commercial games only because (i) as defenders we need ground truth, and (ii) commercial game licences prohibit ``attacking'' those games.

Moreover, we claim broader applicability than game cheats. The general attack meta-model (Section~\ref{sec:metamodel}) applies to all localisation strategies pruning search spaces for assets or iteratively navigating search spaces, not only those comparing memory-stored values with on-screen values. As Schrittwieser et al.~\cite{survey2016} noted, code and data localisation are two of the four prime reverse engineering goals when attacking protected software assets.

\changed{Game resource localisation is one instance of this pattern, and the one analysed in this paper. The hidden asset is the memory location storing the resource value, while the observable signal is the value shown on screen. Each newly collected dump adds a constraint on candidate locations, so that the candidate set is gradually pruned.

Closely related mechanisms appear in other domains. In dynamic cryptographic key extraction~\cite{k-hunt++}, repeated runtime evidence is used to narrow candidate key-related code blocks, operands, and buffers, until the memory locations holding those artefacts are identified. In malware unpacking~\cite{malwareMemory}, repeated dynamic observations of memory regions are used to identify the unpacked code body before further reverse engineering can proceed. In protocol reverse engineering~\cite{huangProtocolRE,tangProtocolRE}, network traces are compared iteratively to progressively constrain candidate field boundaries and message structure.

Our experiments validate the method only for game resource localisation attacks, and we therefore do not claim empirical validation for reverse-engineering scenarios in general. Nevertheless, }we conclude that the evaluated method has broader utility than game cheat prevention. \section{Related Work}
\label{sec:related}

MATE software protection effectiveness has been assessed through two main approaches: evaluating impact on software metrics of protected applications, and empirically assessing attack delays via controlled studies with students or professional hackers.

\subsection{Software metrics as a measure of software protection strength}

Collberg\etal\cite{taxonomy} first advocated using software metrics to assess protection effectiveness, introducing \textit{potency}—correlating increases in software engineering metrics (\eg Cyclomatic Complexity, Halstead Length) from applied protections with attack resistance. Anckaert\etal\cite{anckaert2007obfuscation} compared obfuscation techniques by potency. Other works assessed obfuscated code complexity as effectiveness measures: Goto\etal\cite{goto2000quantitative} used compiler syntax analysis; Visaggio\etal\cite{visaggio2013empirical} proposed code entropy.

However, De Sutter\etal\cite{desutter2024evaluation} indicate that software engineering metrics may not reliably indicate protection strength, can be difficult to compute on obfuscated binaries even with commercial disassemblers, and note that no consensus exists on appropriate metrics for evaluating protection potency. Additionally, metric-based approaches ignore protected asset characteristics such as run-time behaviour. By simulating attacks on specific protected applications, our method may obtain more realistic results given the modelled attacks are realistic.

Collberg\etal\cite{taxonomy} also introduced \emph{resilience} as an effectiveness metric. Talukder\etal\cite{Talukder2019} conjecture attacker effort strongly correlates with resilience, as resilience indicates how well obfuscations withstand automatic deobfuscation. They measure obfuscated program resilience using program slicing. Since program slicing identifies statements contributing to computing certain values, Talukder\etal compare this to human attacker program analysis approaches. Comparing slices for obfuscated versus unobfuscated program versions enables comparing different obfuscations and parameters.

Like our approach, Talukder\etal quantify attacker effort to determine optimal obfuscation configurations. Analogous to our memory scanning attack target, their program slicing attack narrows down  candidate data requiring inspection. However, our work differs in three aspects. First, we compare different obfuscations and different attack strategies against them, while Talukder\etal consider only a single attack type with one configuration. Second, unlike program slicing, memory scanning attacks are typically manual and interactive. Consequently, we quantify both post-attack work (remaining candidate locations) and manual effort required to reach that point (pruning steps needed). Lastly, we generate distributions accounting for attacker action randomness rather than single scores.

\subsection{Empirical assessment of software protections}

Multiple controlled experiments with human subjects have assessed software protection effectiveness. Participants typically perform reverse engineering attacks on target applications, with some attacking vanilla versions while others attack protected versions. Protection efficacy is assessed by comparing participant results---both attack success (\ie participants succeeding) and attack time (\ie completion time for successful participants). Different studies followed this format, introduced by Sutherland\etal~\cite{sutherland2006empirical}. Ceccato\etal~\cite{ceccato2009effectiveness} performed two experiments assessing obfuscation impact on understanding and modification tasks comparing vanilla and obfuscated decompiled Java code, with subsequent studies extending to other obfuscation techniques. Other studies with university students include Viticchié\etal~\cite{viticchie2016assessment} on data obfuscation and Ceccato\etal~\cite{viticchie20splitting} on code obfuscation. Ceccato\etal~\cite{icpc2017} involved professional hackers to assess protections in realistic attack scenarios and understand hacker attack techniques. The same authors extended this via a public challenge validating initial experiment findings~\cite{emse2018}.

Such empirical studies provide useful insights on protection effectiveness. However, they are typically limited in scale and frequency, being difficult to organise and conduct. They cannot provide adequate models for all relevant protection and attack strategy combinations. The simulation approach we validated helps fill these gaps.

 \section{Conclusions and Future Work}
\label{sec:conclusions}
In this work, we presented the first deployment of the methodology proposed by Faingnaert\etal for statistically modelling attacker effort in MATE attack scenarios. In particular, we instantiated this methodology for game resource localisation attacks, demonstrating that generating statistical models of attacker effort in an automated fashion is feasible. These models can be useful for defenders to understand the impact of MATE protection techniques on attacker effort. This is valuable decision support information for defenders seeking to choose the protection techniques that should be deployed to safeguard the assets in their application. \changed{We further complemented the simulation-based evaluation with an empirical validation involving human participants performing the same localisation task. The empirical results show that in most cases the simulated distributions closely match the distributions observed in practice.}

We plan to extend our work to assess more protection techniques, attack strategies, and other types of data assets, including game resources as well as data in other application domains, such as Digital Rights Management frameworks and software licence managers. \section*{Declaration of generative AI and AI-assisted technologies in the manuscript preparation process}
During the preparation of this work the authors used Anthropic Claude and OpenAI ChatGPT to help shortening the initial article draft, in order to limit the article main text to 10000 words, as required by the journal specifications. After using these tools, the authors reviewed and edited the content as needed and take full responsibility for the content of the published article.

\section*{Artifact Availability}
All code used during our experiments \changed{and the logs produced during the empirical experiment are} available at \url{https://github.com/alessandro-sanna/attacker_effort_estimation_experiments}. All obtained results are available at \url{https://doi.org/10.6084/m9.figshare.28578083}. We opted to not release the code for the \texttt{memhunt} tool used in the empirical experiments, since it can be used for offensive purposes; we will release its code upon reasonable requests to researchers with a legitimate scientific interest.

\section*{\changed{Compliance with Ethical Standards}}

\changed{\paragraph*{Human participants and research context} The empirical study involved seven adult participants enrolled in PhD programmes in cybersecurity at one of the authors' institutions. All participants were members of the participating research groups. The activity was conducted solely to evaluate whether the statistical models presented in this paper are representative of the outcomes obtained by real users performing the same resource localisation tasks with the provided tool. Participation was voluntary, and participants could stop the activity at any time without penalty. Participation, non-participation, and task performance had no effect on academic assessment, employment status, supervision, or any other institutional evaluation. No compensation was offered to participants.

\paragraph*{Informed participation and confidentiality} Participants were informed in advance about the aims and procedures of the experiment, the tools they were allowed to use, the restrictions on additional dynamic analysis, and the fact that the tool would log task-related actions for post-experiment analysis. The collected data were restricted to task results and task-related metadata necessary to evaluate the attack outcomes, such as the dumps taken, the pruning logic used, the number of remaining candidate locations, and whether the ground-truth location was retained. We did not collect personally identifiable information. Research analyses were performed only on de-identified and aggregate data, and no personal identifiers are stored, reported, or published. 

\paragraph*{Data protection} The experiment was designed according to data minimisation principles. Since no personally identifiable information was collected, stored, reported, or published, the research dataset used for analysis contains only de-identified task outcomes and aggregate statistics. Any processing of task-related data was performed in accordance with EU Regulation 2016/679 (General Data Protection Regulation), where applicable. 

\paragraph*{Ethics review} Because of the above, this study was deemed exempt from full ethics review.

\paragraph*{Conflict of interest} The authors declare no conflicts of interest.
}

\section*{Funding}
The research reported in this paper was in part funded by the Cybersecurity Research Program Flanders. This work was partially supported by project SERICS (PE00000014) and project SETA (PNRR M4.C2.1.1 PRIN 2022 PNRR, Cod. P202233M9Z, CUP F53D23009120001, Avviso D.D 1409 14.09.2022), both under the Italian NRRP MUR programme funded by the European Union - NextGenerationEU.

\bibliographystyle{splncs04}
\bibliography{references}

\begin{thebibliography}{10}
\providecommand{\url}[1]{\texttt{#1}}
\providecommand{\urlprefix}{URL }
\providecommand{\doi}[1]{https://doi.org/#1}

\bibitem{circulardebugging}
Abrath, B., Coppens, B., Nevolin, I., De~Sutter, B.: Resilient self-debugging software protection. In: 2020 IEEE European Symposium on Security and Privacy Workshops (EuroSPW). pp. 606--615. IEEE Computer Society (2020). \doi{10.1109/EuroSPW51379.2020.00088}

\bibitem{selfdebugging}
Abrath, B., Coppens, B., Volckaert, S., Wijnant, J., De~Sutter, B.: Tightly-coupled self-debugging software protection. In: Proc.\ of the 6th Workshop on Software Security, Protection, and Reverse Engineering. pp. 7:1--7:10. SSPREW '16, ACM (2016). \doi{10.1145/3015135.3015142}

\bibitem{anckaert2007obfuscation}
Anckaert, B., Madou, M., {De Sutter}, B., De~Bus, B., De~Bosschere, K., Preneel, B.: Program obfuscation: a quantitative approach. In: Proc. ACM Workshop on Quality of protection. pp. 15--20 (2007). \doi{10.1145/1314257.1314263}

\bibitem{Basile23}
Basile, C., {De Sutter}, B., Canavese, D., Regano, L., Coppens, B.: Design, implementation, and automation of a risk management approach for man-at-the-end software protection. Computers \& Security  \textbf{132},  103321 (2023). \doi{https://doi.org/10.1016/j.cose.2023.103321}

\bibitem{beran1977minimum}
Beran, R.: Minimum hellinger distance estimates for parametric models. The annals of Statistics pp. 445--463 (1977)

\bibitem{Cannell2013}
Cannell, J.: Obfuscation: Malware’s best friend (March 2013), \url{http://blog.malwarebytes.org/intelligence/2013/03/obfuscation-malwares-best-friend/}

\bibitem{game_hacking}
Cano, N.: Game hacking: developing autonomous bots for online games. No Starch Press (2016)

\bibitem{ceccato2009effectiveness}
Ceccato, M., Di~Penta, M., Nagra, J., Falcarin, P., Ricca, F., Torchiano, M., Tonella, P.: The effectiveness of source code obfuscation: An experimental assessment. In: IEEE 17th International Conference on Program Comprehension (ICPC). pp. 178--187 (May 2009). \doi{10.1109/ICPC.2009.5090041}

\bibitem{icpc2017}
Ceccato, M., Tonella, P., Basile, C., Coppens, B., {De Sutter}, B., Falcarin, P., Torchiano, M.: How professional hackers understand protected code while performing attack tasks. In: Proc.\ ICPC (2017). \doi{10.1109/ICPC.2017.2}

\bibitem{emse2018}
Ceccato, M., Tonella, P., Basile, C., Falcarin, P., Torchiano, M., Coppens, B., {De Sutter}, B.: Understanding the behaviour of hackers while performing attack tasks in a professional setting and in a public challenge. Empirical Software Engineering (EMSE)  \textbf{24},  240--286 (2019). \doi{10.1007/s10664-018-9625-6}

\bibitem{cheatengine}
{Cheat Engine} (2024), \url{https://www.cheatengine.org/}

\bibitem{tigress2025}
{Christian Collberg}: The {Tigress} {C} obfuscator (2025), \url{https://tigress.wtf/}

\bibitem{taxonomy}
Collberg, C., Thomborson, C., Low, D.: A taxonomy of obfuscating transformations. Tech. Rep.~148, University of Auckland (07 1997)

\bibitem{dagstuhl}
{De Sutter}, B., Collberg, C., Preda, M.D., Wyseur, B.: {Software Protection Decision Support and Evaluation Methodologies (Dagstuhl Seminar 19331)}. Dagstuhl Reports  \textbf{9}(8),  1--25 (2019). \doi{10.4230/DagRep.9.8.1}

\bibitem{desutter2024evaluation}
De~Sutter, B., Schrittwieser, S., Coppens, B., Kochberger, P.: Evaluation methodologies in software protection research. ACM Comput. Surv.  \textbf{57}(4) (Dec 2024). \doi{10.1145/3702314}

\bibitem{RNC}
Demissie, B.F., Ceccato, M., Tiella, R.: Assessment of data obfuscation with residue number coding. In: 2015 IEEE/ACM 1st International Workshop on Software Protection. pp. 38--44 (2015). \doi{10.1109/SPRO.2015.15}

\bibitem{k-hunt++}
Faingnaert, T., Van~Iseghem, W., De~Sutter, B.: K-hunt++: Improved dynamic cryptographic key extraction. In: Proceedings of the 2024 Workshop on Research on Offensive and Defensive Techniques in the Context of Man At The End (MATE) Attacks. p. 22–29. CheckMATE '24, Association for Computing Machinery, New York, NY, USA (2024). \doi{10.1145/3689934.3690818}, \url{https://doi.org/10.1145/3689934.3690818}

\bibitem{checkmate24}
Faingnaert, T., Zhang, T., Van~Iseghem, W., Everaert, G., Coppens, B., Collberg, C., De~Sutter, B.: Tools and models for software reverse engineering research. In: Proc.\ CheckMATE Workshop. p. 44–58 (2024). \doi{10.1145/3689934.3690817}

\bibitem{rr}
Feldman, S.I., Brown, C.B.: Igor: a system for program debugging via reversible execution. SIGPLAN Not.  \textbf{24}(1),  112–123 (Nov 1988). \doi{10.1145/69215.69226}

\bibitem{xor}
Fellin, R., Ceccato, M.: Experimental assessment of {XOR}-masking data obfuscation based on k-clique opaque constants. Journal of Systems and Software  \textbf{162},  110492 (2020). \doi{10.1016/j.jss.2019.110492}

\bibitem{Garner59}
Garner, H.L.: The residue number system. Electronic Computers, IRE Transactions on  \textbf{EC-8}(2),  140--147 (June 1959). \doi{10.1109/TEC.1959.5219515}

\bibitem{goto2000quantitative}
Goto, H., Mambo, M., Matsumura, K., Shizuya, H.: An approach to the objective and quantitative evaluation of tamper-resistant software. In: Third Int. Workshop on Information Security. pp. 82--96. Springer (2000). \doi{10.1007/3-540-44456-4\_7}

\bibitem{halstead1977elements}
Halstead, M.H.: Elements of Software Science (Operating and programming systems series). Elsevier Science Inc. (1977)

\bibitem{idapro}
{Hex-Rays}: {IDA Pro}. \url{https://hex-rays.com/ida-pro} (2025)

\bibitem{huangProtocolRE}
Huang, Y., Shu, H., Kang, F., Guang, Y.: Protocol reverse-engineering methods and tools: A survey. Computer Communications  \textbf{182},  238--254 (2022). \doi{https://doi.org/10.1016/j.comcom.2021.11.009}, \url{https://www.sciencedirect.com/science/article/pii/S0140366421004382}

\bibitem{rfc4648}
Josefsson, S.: Rfc 4648 - the base16, base32, and base64 data encodings (October 2006), \url{http://tools.ietf.org/html/rfc4648}

\bibitem{mccabe1976complexity}
McCabe, T.J.: A complexity measure. IEEE Transactions on software Engineering  \textbf{SE-2}(4),  308--320 (1976). \doi{10.1109/TSE.1976.233837}

\bibitem{collbergbook}
Nagra, J., Collberg, C.: Surreptitious Software: Obfuscation, Watermarking, and Tamperproofing for Software Protection. Pearson Education (2009)

\bibitem{ghidra}
{National Security Agency}: Ghidra (2025), \url{https://ghidra-sre.org/}

\bibitem{scanmem}
Scanmem (2024), \url{https://github.com/scanmem/scanmem}

\bibitem{survey2016}
Schrittwieser, S., Katzenbeisser, S., Kinder, J., Merzdovnik, G., Weippl, E.: Protecting software through obfuscation: Can it keep pace with progress in code analysis? ACM Comput. Surv.  \textbf{49}(1) (apr 2016). \doi{10.1145/2886012}

\bibitem{sutherland2006empirical}
Sutherland, I., Kalb, G.E., Blyth, A., Mulley, G.: An empirical examination of the reverse engineering process for binary files. Computers \& Security  \textbf{25}(3),  221--228 (2006). \doi{10.1016/j.cose.2005.11.002}

\bibitem{Talukder2019}
Talukder, M., Islam, S., Falcarin, P.: Analysis of obfuscated code with program slicing. In: 2019 International Conference on Cyber Security and Protection of Digital Services (Cyber Security). pp.~1--7. IEEE (Jun 2019). \doi{10.1109/cybersecpods.2019.8885094}

\bibitem{tangProtocolRE}
Tang, T., Lai, Y., Wang, Y.: Relational reasoning-based approach for network protocol reverse engineering. Computer Networks  \textbf{230},  109797 (2023). \doi{https://doi.org/10.1016/j.comnet.2023.109797}, \url{https://www.sciencedirect.com/science/article/pii/S1389128623002426}

\bibitem{ASLR}
{The Pax Team}: {PaX} address space layout randomization {(ASLR)}, \url{http://pax.grsecurity.net/docs/aslr.txt}

\bibitem{visaggio2013empirical}
Visaggio, C.A., Pagin, G.A., Canfora, G.: An empirical study of metric-based methods to detect obfuscated code. International Journal of Security \& Its Applications  \textbf{7}(2) (2013)

\bibitem{viticchie2016reactive}
Viticchi\'{e}, A., Basile, C., Avancini, A., Ceccato, M., Abrath, B., Coppens, B.: Reactive attestation: Automatic detection and reaction to software tampering attacks. In: Proceedings of the 2016 ACM Workshop on Software PROtection. p. 73–84. SPRO '16, ACM (2016). \doi{10.1145/2995306.2995315}

\bibitem{viticchie20splitting}
Viticchi{\'{e}}, A., Regano, L., Basile, C., Torchiano, M., Ceccato, M., Tonella, P.: Empirical assessment of the effort needed to attack programs protected with client/server code splitting. Empir. Softw. Eng.  \textbf{25}(1),  1--48 (2020). \doi{10.1007/s10664-019-09738-1}

\bibitem{viticchie2016assessment}
Viticchi{\'e}, A., Regano, L., Torchiano, M., Basile, C., Ceccato, M., Tonella, P., Tiella, R.: Assessment of source code obfuscation techniques. In: Int'l Working Conf.\ Source Code Analysis and Manipulation (SCAM). pp. 11--20. IEEE (2016). \doi{10.1109/SCAM.2016.17}

\bibitem{malwareMemory}
Willems, C., Freiling, F.C., Holz, T.: Using memory management to detect and extract illegitimate code for malware analysis. In: Proceedings of the 28th Annual Computer Security Applications Conference. p. 179–188. ACSAC '12, Association for Computing Machinery, New York, NY, USA (2012). \doi{10.1145/2420950.2420979}, \url{https://doi.org/10.1145/2420950.2420979}

\bibitem{RNCorig}
Zhu, W., Thomborson, C.: A provable scheme for homomorphic obfuscation in software security. In: The IASTED International Conference on Communication, Network and Information Security, CNIS. vol. Vol. 5. (2005)

\end{thebibliography}

\newpage
\begin{appendices}

\section{Experiment Encoding Details}
\label{app:obf_parameters}

\begin{table}[H]
    \centering
    \caption{Overview of the parameters of the encodings used in the experiments}
    \label{tab:experiments-parameters}
    \small
    \begin{tabular}{l|ll}
\hline
                        & \multicolumn{2}{c}{Parameters}                                                                                                    \\
Encoding                & SuperTux                                                          & AssaultCube                                                   \\
\hline
Base                    & -                                                                 & -                                                             \\
$+$                     & $O =$ \texttt{24}                                                 & $O =$ \texttt{24}                                             \\
$\xor$                  & $M = \texttt{0xABCD123}$                                          & $M = \texttt{0xABCD123}$                                      \\
$+\xor$                 & $M = \texttt{0xABCD123}$; $O = \texttt{17}$                       & $M = \texttt{0xABCD123}$; $O = \texttt{17}$                   \\
$\xor +$                & $M = \texttt{0xABCD123}$; $O = \texttt{17}$                       & $M = \texttt{0xABCD123}$; $O = \texttt{17}$                   \\
RNC                     & $m_1 = \texttt{89}$; $m_2 = \texttt{97}$; $m_3 = \texttt{93}$     & $m_1 = \texttt{2}$; $m_2 = \texttt{3}$; $m_3 = \texttt{5}$    \\
Dynamic $\xor$ - UoR    & $p_{u,r} = 300$                                                   & $p_{u,r} = 1500$                                              \\
Dynamic $\xor$ - UoW    & $p_{u,w} = 2$                                                    & $p_{u,w} = 2$                                                \\
\end{tabular} \end{table}

\section{Experiment Attack Details}

\begin{table}[H]
    \centering
    \caption{Overview of the \emph{dump collection}, \emph{dump selection}, and \emph{location pruning} strategies for the first set of experiments we performed.}
    \label{tab:experiments-1-overview}
    \small
    \resizebox{\linewidth}{!}{
        \begin{tabular}{l|llll}
\hline
Attack Strategy         & Dump Collection                                                               & Dump Selection                                                                        & Attack type   & Pruning logic     \\
\hline
Base                    & \hyperlink{sec:atk-strategies:subsec:dump-collection:par:paced}{Paced}    & \hyperlink{sec:atk-strategies:subsec:dumpselection:par:binned}{Binned}                & Greedy        & Base              \\
$+$                     & \hyperlink{sec:atk-strategies:subsec:dump-collection:par:paced}{Paced}    & \hyperlink{sec:atk-strategies:subsec:dumpselection:par:binned}{Binned}                & Greedy        & $+$               \\
$\xor$                  & \hyperlink{sec:atk-strategies:subsec:dump-collection:par:paced}{Paced}    & \hyperlink{sec:atk-strategies:subsec:dumpselection:par:binned}{Binned}                & Greedy        & $\xor$             \\
$+\xor$                 & \hyperlink{sec:atk-strategies:subsec:dump-collection:par:paced}{Paced}    & \hyperlink{sec:atk-strategies:subsec:dumpselection:par:incremental}{Incremental}      & Greedy        & $+\xor$            \\
$\xor +$                & \hyperlink{sec:atk-strategies:subsec:dump-collection:par:paced}{Paced}    & \hyperlink{sec:atk-strategies:subsec:dumpselection:par:incrmental}{Incremental}       & Greedy        & $\xor +$           \\
RNC                     & \hyperlink{sec:atk-strategies:subsec:dump-collection:par:paced}{Paced}    & \hyperlink{sec:atk-strategies:subsec:dumpselection:par:binned}{Binned}                & Greedy        & RNC               \\
Increase/Decrease       & \hyperlink{sec:atk-strategies:subsec:dump-collection:par:paced}{Paced}    & \hyperlink{sec:atk-strategies:subsec:dumpselection:par:fully-random}{Fully random}    & Greedy        & Increase/Decrease \\
Change/No Change        & \hyperlink{sec:atk-strategies:subsec:dump-collection:par:paced}{Paced}    & \hyperlink{sec:atk-strategies:subsec:dumpselection:par:fully-random}{Fully random}    & Greedy        & Change/No Change  \\
Change                  & \hyperlink{sec:atk-strategies:subsec:dump-collection:par:paced}{Paced}    & \hyperlink{sec:atk-strategies:subsec:dumpselection:par:fully-random}{Fully random}    & Greedy        & Change            \\                                
\end{tabular}
     }
\end{table}

\begin{table}[H]
    \centering
    \caption{Overview of the \emph{dump collection}, \emph{dump selection}, and \emph{location pruning} strategies for the second set of experiments we performed.}
    \label{tab:experiments-2-overview}
    \small
    \resizebox{\linewidth}{!}{
        \begin{tabular}{l|llll}
\hline
Attack Strategy         & Dump Collection                                                               & Dump Selection                                                                        & Attack type   & Pruning logic     \\
\hline
Change/No Change        & \hyperlink{sec:atk-strategies:subsec:dump-collection:par:fast}{Fast}      & \hyperlink{sec:atk-strategies:subsec:dumpselection:par:fully-random}{Fully random}    & Greedy        & Change/No Change  \\
Change                  & \hyperlink{sec:atk-strategies:subsec:dump-collection:par:fast}{Fast}      & \hyperlink{sec:atk-strategies:subsec:dumpselection:par:fully-random}{Fully random}    & Greedy        & Change            \\
Statistical $\xor$       & \hyperlink{sec:atk-strategies:subsec:dump-collection:par:fast}{Fast}      & \hyperlink{sec:atk-strategies:subsec:dumpselection:par:rapid}{Rapid}\tablefootnote{When simulating attacks on versions of the games with the dynamic $\xor$ UoR encoding, we used a maximum allowed interval between subsequent dumps $t$ of 5s, while we used t=12s for the UoW variant.} & Statistical   & $\xor$             \\
                                
\end{tabular}
     }
\end{table}

\section{Performed Experiments}
\label{app:experiment-combos}

\subsection{Overview of Experiments with Static Encodings}
\label{app:stat-experiments-combos}

\begin{table}[H]
    \centering
    \caption{Overview of the experiments we performed on SuperTux protected with static encodings. A filled dot ($\bullet$) indicates the experiment result is discussed in this paper. A hollow dot ($\circ$) indicates that we performed the experiment, but that we do not discuss them in this paper.}
    \label{tab:experiments-overview-st-static}
    \small
    \resizebox{\linewidth}{!}{
        \begin{tabular}{|c|>{\centering\arraybackslash}m{1.3cm}|>{\centering\arraybackslash}m{1.3cm}|
>{\centering\arraybackslash}m{1.3cm}|>{\centering\arraybackslash}m{1.3cm}|>{\centering\arraybackslash}m{1.3cm}|>{\centering\arraybackslash}m{1.3cm}|}
\hline
\backslashbox{\textbf{Attack}}{\textbf{Obfuscation}}    &  Base         & $+$           & $\xor$        & $+$-$\xor$    & $\xor$-$+$    & RNC           \\
\hline
Base                                                    & $\bullet$     & $\circ$       & $\circ$       & $\circ$       & $\circ$       & $\bullet$     \\ \hline
$+$                                                     & $\bullet$     & $\bullet$     & $\circ$       & $\circ$       & $\circ$       & $\bullet$     \\ \hline
$\xor$                                                  & $\bullet$     & $\circ$       & $\bullet$     & $\circ$       & $\circ$       & $\bullet$     \\ \hline
$+$-$\xor$                                              & $\bullet$     & $\circ$       & $\circ$       & $\bullet$     & $\bullet$     & $\bullet$     \\ \hline
$\xor$-$+$                                              & $\bullet$     & $\circ$       & $\circ$       & $\bullet$     & $\bullet$     & $\bullet$     \\ \hline
RNC                                                     & $\bullet$     & $\circ$       & $\circ$       & $\circ$       & $\circ$       & $\bullet$     \\ \hline
Increase/Decrease                                       & $\circ$       & $\circ$       & $\circ$       & $\circ$       & $\circ$       & $\circ$       \\ \hline
Change/No Change                                        & $\circ$       & $\circ$       & $\circ$       & $\circ$       & $\circ$       & $\circ$       \\ \hline
Change                                                  & $\circ$       & $\circ$       & $\circ$       & $\circ$       & $\circ$       & $\circ$       \\ \hline
\end{tabular}
     }
\end{table}

\begin{table}[H]
    \centering
    \caption{Overview of the experiments we performed on AssaultCube protected with static encodings. A filled dot ($\bullet$) indicates the experiment result is discussed in this paper. A hollow dot ($\circ$) indicates that we performed the experiment, but that we do not discuss them in this paper.}
    \label{tab:experiments-overview-ac-static}
    \small
    \resizebox{\linewidth}{!}{
        \begin{tabular}{|c|>{\centering\arraybackslash}m{1.3cm}|>{\centering\arraybackslash}m{1.3cm}|
>{\centering\arraybackslash}m{1.3cm}|>{\centering\arraybackslash}m{1.3cm}|>{\centering\arraybackslash}m{1.3cm}|>{\centering\arraybackslash}m{1.3cm}|}
\hline
\backslashbox{\textbf{Attack}}{\textbf{Obfuscation}}    &  Base         & $+$           & $\xor$        & $+$-$\xor$    & $\xor$-$+$    & RNC           \\
\hline
Base                                                    & $\bullet$     & $\circ$       & $\circ$       & $\circ$       & $\circ$       & $\bullet$     \\ \hline
$+$                                                     & $\bullet$     & $\bullet$     & $\circ$       & $\circ$       & $\circ$       & $\bullet$     \\ \hline
$\xor$                                                  & $\bullet$     & $\circ$       & $\bullet$     & $\circ$       & $\circ$       & $\bullet$     \\ \hline
$+$-$\xor$                                              & $\bullet$     & $\circ$       & $\circ$       & $\bullet$     & $\bullet$     & $\bullet$     \\ \hline
$\xor$-$+$                                              & $\bullet$     & $\circ$       & $\circ$       & $\bullet$     & $\bullet$     & $\bullet$     \\ \hline
RNC                                                     & $\bullet$     & $\circ$       & $\circ$       & $\circ$       & $\circ$       & $\bullet$     \\ \hline
Increase/Decrease                                       & $\circ$       & $\circ$       & $\circ$       & $\circ$       & $\circ$       & $\circ$       \\ \hline
Change/No Change                                        & $\bullet$     & $\bullet$     & $\bullet$     & $\bullet$     & $\bullet$     & $\bullet$     \\ \hline
Change                                                  & $\circ$       & $\circ$       & $\circ$       & $\circ$       & $\circ$       & $\circ$       \\ \hline
\end{tabular}
     }
\end{table}

\subsection{Overview of Experiments with Dynamic Encodings}
\label{app:dyn-experiments-combos}

\begin{table}[H]
    \centering
    \caption{Overview of the experiments we performed on SuperTux protected with dynamic encodings. A filled dot ($\bullet$) indicates the experiment result is discussed in this paper. A hollow dot ($\circ$) indicates that we performed the experiment, but that we do not discuss them in this paper.}
    \label{tab:experiments-overview-st-dynamic}
    \small
    \begin{tabular}{|c|>{\centering\arraybackslash}m{1.3cm}|>{\centering\arraybackslash}m{1.3cm}|}
\hline
\backslashbox{\textbf{Attack}}{\textbf{Obfuscation}}    & Dynamic $\xor$ UoW    & Dynamic $\xor$ UoR    \\
\hline
Change/No Change                                        & $\bullet$             & $\bullet$             \\ \hline
Change                                                  & $\bullet$             & $\bullet$             \\ \hline
Statistical $\xor$                                      & $\bullet$             & $\bullet$             \\ \hline
\end{tabular}
 \end{table}

\begin{table}[H]
    \centering
    \caption{Overview of the experiments we performed on AssaultCube protected with dynamic encodings. A filled dot ($\bullet$) indicates the experiment result is discussed in this paper. A hollow dot ($\circ$) indicates that we performed the experiment, but that we do not discuss them in this paper.}
    \label{tab:experiments-overview-ac-dynamic}
    \small
    \begin{tabular}{|c|>{\centering\arraybackslash}m{1.3cm}|>{\centering\arraybackslash}m{1.3cm}|}
\hline
\backslashbox{\textbf{Attack}}{\textbf{Obfuscation}}    & Dynamic $\xor$ UoW    & Dynamic $\xor$ UoR    \\
\hline
Change/No Change                                        & $\circ$               & $\circ$               \\ \hline
Change                                                  & $\circ$               & $\circ$               \\ \hline
Statistical $\xor$                                      & $\circ$               & $\circ$               \\ \hline
\end{tabular}
 \end{table}

\section{Empirical Experiment Encoding Details}
\label{app:empirical_obf_parameters}

\begin{table}[H]
    \centering
    \caption{Overview of the parameters of the encodings used in the empirical experiments}
    \label{tab:experiments-overview-st}
    \small
    \begin{tabular}{l|ll}
\hline
                        & \multicolumn{2}{c}{Parameters}                                                                                                     \\
Encoding                & SuperTux (version 1)                                              & SuperTux (version 2)                                           \\
\hline
Base                    & \multicolumn{2}{c}{-}                                                                                                                               \\
$+$                     & $O =$ \texttt{18}                                                 & $O =$ \texttt{8}                                               \\
$\xor$                  & $M = \texttt{2682338889}$                                         & $M = \texttt{2266022666}$                                      \\
$+\xor$                 & $M = \texttt{312889390}$; $O = \texttt{95}$                       & $M = \texttt{2918063068}$; $O = \texttt{30}$                   \\
$\xor +$                & $M = \texttt{3694758281}$; $O = \texttt{95}$                      & $M = \texttt{3980900406}$; $O = \texttt{37}$                   \\
RNC                     & $m_1 = \texttt{37}$; $m_2 = \texttt{107}$; $m_3 = \texttt{5}$     & $m_1 = \texttt{101}$; $m_2 = \texttt{7}$; $m_3 = \texttt{103}$ \\
\end{tabular} \end{table}

\end{appendices}

\end{document}